\documentclass{article}
\usepackage[latin9]{inputenc}
\usepackage{geometry}
\usepackage{amsmath}
\usepackage{graphicx}
\usepackage{setspace}
\makeatletter
\usepackage{dcolumn}% Align table columns on decimal point
\usepackage{bm}% bold math
\usepackage{amsfonts}
\usepackage{amsmath}
\usepackage{amssymb}

\usepackage[T1]{fontenc}
\date{} 

\makeatother

\begin{document}

\title{Understanding the computational difficulty of a binary-weight perceptron
and the advantage of input sparseness}

\author{$\text{Zedong Bi}^{1,2}$ and $\text{Changsong Zhou}^{1,2,3,*}$}

\maketitle
\center 1 Department of Physics, Centre for Nonlinear Studies and
Institute of Computational and Theoretical Studies, Hong Kong Baptist
University, Kowloon Tong, Hong Kong

\center 2 Research Centre, HKBU Institute of Research and Continuing
Education, Shenzhen, China

\center 3 Beijing Computational Science Research Center, Beijing,
China

\center {*} cszhou@hkbu.edu.hk

\raggedright 
\begin{abstract}
Limited precision of synaptic weights is a key aspect of both biological
and hardware implementation of neural networks. To assign low-precise
weights during learning is a non-trivial task, but may benefit from
representing to-be-learned items using sparse code. However, the
computational difficulty resulting from low weight precision and
the advantage of sparse coding remain not fully understood. Here,
we study a perceptron model, which associates binary (0 or 1) input
patterns with desired outputs using binary (0 or 1) weights, modeling
a single neuron receiving excitatory inputs. Two efficient perceptron
solvers (SBPI and rBP) usually find solutions in dense solution region.
We studied this dense-region-approaching process through a decimation
algorithm, where every time step, marginal probabilities of unfixed
weights were evaluated, then the most polarized weight was fixed
at its preferred value. We compared the decimation-fixing order of
weights with the dynamics of SBPI and rBP, and studied the structure
of solution subspace $\mathcal{S}$ where early-decimation-fixed
weights take their fixed values. We found that in SBPI and rBP, most
time steps are spent on determining values of late-decimation-fixed
weights. This algorithmic difficult point may result from strong
cross-correlation between late-decimation-fixed weights in $\mathcal{S}$,
and is related to solution condensation in $\mathcal{S}$ during
decimation. Input sparseness reduces time steps that SBPI and rBP
need to find solutions, due to reduction of cross-correlation between
late-decimation-fixed weights. When contraint density increases across
a transition point, the portion of weights whose values are difficult
to determine sharply increases, signaling difficult accessibility
of dense solution region. We propose that the computational difficulty
of binary-weight perceptron is related to a solution condensation
process during approaching the dense solution region. Our work highlights
the heterogeneity of learning dynamics of weights, which may help
understand axonal pruning in brain development, and inspire more
efficient algorithms to train neural networks.
\end{abstract}

\section{Introduction}

Information of neural networks is stored in synaptic weights. An
important feature of the synaptic weights in nature and simple hardware
implementation is low precision \cite{OConnor_2005,Montgomery_2004,Misra_2010}.
However, assigning suitable values to low-precision weights is a
non-trivial task, and is known to be intractable in the worst case
\cite{Rivest_1992,Amaldi_1991}. Artificial neural networks with
low-precision weights requires more epochs to train before performance
convergence \cite{Hubara_2018,Ott_2017}. Therefore, it is of great
interest to understand where the hardness comes from and how to avoid
it or use it.

Perceptron is a one-layer network that receives inputs from many
neurons and gives a single output. It serves as an elementary building
block of complex neural networks, and is also one of the basic structures
for learning and memory \cite{Engel_2001}. It is also a powerful
computational unit by itself, and has applications in, say, rule
inference \cite{Lage-Castellanos_2009} and data compression \cite{Hosaka_2002}.
Here we try to understand the computational difficulty induced by
low weight precision by studying perceptron whose weights take binary
values (0 or 1). We consider a classification task: given a set of
random input patterns, we want to adjust the synaptic weights, so
that the perceptron gives a desired output in response to each input
pattern. A synaptic weight configuration successfully associating
all these input-output pairs is a \textit{solution} of the perceptron.

Methods of statistical physics imply that most solutions of binary-weight
perceptron are isolated and computationally hard to find \cite{Huang_2014,Zdeborova_2008}.
In the seminal work of Ref. \cite{Baldassi_2015_b}, it is shown
that the weight configuration space of binary-weight perceptron has
a region where solutions densely aggregate. It is believed that solutions
found by algorithms usually belong to this dense solution region
\cite{Baldassi_2015_b,Baldassi_2018}, and efficient algorithms have
been designed by conducting weight configurations toward the dense
solution region \cite{Baldassi_2016b,Baldassi_2016c,Chaudhari_2017}.
This implies that the computational difficulty in solving binary-weight
perceptron, if any, should emerge during the process approaching
the dense solution region. This difficulty should be related to the
geometry structure of solution space around the dense solution region,
and has little to do with the isolated solutions. However, there
is a knowledge gap about how the structure of solution space around
the dense region influences algorithmic difficulty.

The ``difficulty\textquotedbl{} encountered by algorithms may have
two different meanings, depending on the constraint density $\alpha$
($\alpha=M/N$, with $M$ being the number of input-output pairs
to be associated and $N$ being the number of synapses). First is
\textit{difficult part} during efficient solving. When $\alpha$
takes small values, there exist algorithms (such as SBPI and rBP)
which can efficiently (i.e., typically take sub-exponential-to-$N$
time steps) solve the problem \cite{Baldassi_2007,Braunstein_2006}.
In this case, it is interesting to study what difficult part of the
solving process consumes most time, and think of methods to reduce
solving time. Second is \textit{exponential-time difficulty}. When
$\alpha$ takes large value, close to the theoretical capacity $\alpha_{c}^{\text{theo}}$
(which means the maximal $\alpha$ at which perceptron theoretically
has solutions), there are no known efficient solvers, so that people
have to wait for exponentially long time for theoretically existing
solutions. In this case, it is interesting to study what nature prevents
the existence of efficient solvers. As far as we know, no study has
been performed to address the difficult part during efficient solving
process. It has been suggested that exponential-time difficulty is
related to fragmentation of dense solution region at $\alpha_{U}$
when $\alpha\rightarrow\alpha_{c}^{\text{theo}}$ \cite{Baldassi_2016}.
However, it is unknown how this fragmentation influences the dynamics
of algorithms to prevent them from finding solutions.

A possible way to reduce the computational difficulty of binary-weight
perceptron is to represent the to-be-classified items using sparse
code in input patterns. The functional pros and cons of input sparseness
has been discussed in many contexts. On the one hand, input sparseness
allows algorithms to store more input-output (IO) associations into
perceptron \cite{Amit_1994,Barrett_2008,Legenstein_2008}, and facilitates
classification of input patterns \cite{Lin_2014}; on the other hand,
input sparseness also reduces the information content of input patterns
and reduces generalization performance of the trained network \cite{Spanne_2015,Barak_2013}.
Despite these possible cons, biological neural analogies of perceptron,
such as cerebellar Purkenje cells and insectile mushroom body output
neurons receive from a large number of cerebellar granule cells and
Keynon cells \cite{Brunel_2004,Huerta_2004}, which have low firing
probability. Theoretically, the advantage of input sparseness for
memory storage is mainly understood by calculating theoretical capacity
$\alpha_{c}^{\text{theo}}$. The result is that input sparseness
increases $\alpha_{c}^{\text{theo}}$ \cite{Brunel_2004,Clopath_2013}.
A point seldom discussed is how input sparsity influences the time
steps that an algorithm uses to find a solution, in the case that
the algorithm can ``efficiently\textquotedbl{} (i.e., take sub-exponential-to-$N$
time steps) solve the perceptron. Suppose in the case that solutions
can be efficiently found, sparser input requires more time steps
to find a solution, then the biological and engineering interest
of input sparseness should be significantly weakened. In this paper,
we will study this point above by investigating how input sparsity
influences the difficult part during efficient solving that consumes
most time steps.

One possible aspect to understand the computational difficulty of
binary-weight perceptron is cross-correlation of weights in solution
space. Strong cross-correlation breaks Bethe-Peierls approximation
\cite{Mezard_2009}, which supposes that the probability distribution
of the value of one weight in solution space is independent of the
values of the other weights. This approximation lies in the heart
of belief propagation \cite{Mezard_2009}, which is an ingredient
of or closely related to a number of efficient algorithms \cite{Braunstein_2006,Baldassi_2016b,Baldassi_2007,Baldassi_2015}.
From another aspect, strong cross-correlation implies that the value
preferences of a large number of weights are reconfigured in response
to the flipping of a single weight. This means that in a local search
algorithm, many subsequent rearrangements may be needed following
the modification of a single weight, which increases the difficulty
to find solutions \cite{Semerjian_2008}. Additionally, cross-correlation
may be related to the condensation of solutions \cite{Mezard_2009,Krzakala_2007},
which means that most solutions aggregate into a small number of
clusters.

In this paper, we study a perceptron model in which both inputs and
weights take binary values (0 or 1), modeling a single neuron receiving
excitatory inputs. To address the knowledge gap between structure
of solution space around dense solution region and algorithmic difficulty,
we propose to study decimation process, in which weights are successively
fixed. At each time step, marginal probabilities of unfixed weights
are computed using belief propagation or, in small-sized systems,
enumerated solutions, then the most polarized weight is fixed at
its preferred value. We will show that similarly as two efficient
algorithms (SBPI and rBP), decimation is also a process approaching
the dense solution region, and the order of weight-fixing in decimation
is comparable to the dynamics of these two efficient algorithms.
Additionally, decimation is also a simpler process than SBPI and
rBP, because a weight will not change its value again once gets fixed;
unlike in SBPI and rBP, where the preferences of weights are changing
continuously. Therefore, decimation is an ideal surrogate process
of SBPI and rBP for theoretical study. We will study the cross-correlation
between weights and the geometry of solution clusters in the solution
subspace made up of unfixed weights during decimation, thereby getting
insight into the reshaping of solution space in efficient algorithms
during approaching the dense solution region. With the help of decimation
process, we hope to shed light onto the links among solving dynamics
in efficient algorithms, algorithmic difficulty and structure of
solution space around the dense region.

This paper is organized as follows. After describing the model in
Section \ref{sec:Model}, we will investigate the capacity of the
model and time steps that SBPI and rBP need to obtain solutions at
different input sparsities in Section \ref{sec:advantage_sparse}.
In Sections \ref{sec:solving_nodus}-\ref{sec:geo_cluster}, we will
study difficult solving part and the advantage of input sparseness
with the help of decimation. In Sections \ref{sec:alpha_approach_alphac}
and \ref{sec:large_alpha_enum}, we will study exponential-time difficulty
by investigating the case when $\alpha\rightarrow\alpha_{c}^{\text{theo}}$.

\section{Model}

\label{sec:Model}

\begin{figure}
\includegraphics[scale=0.55]{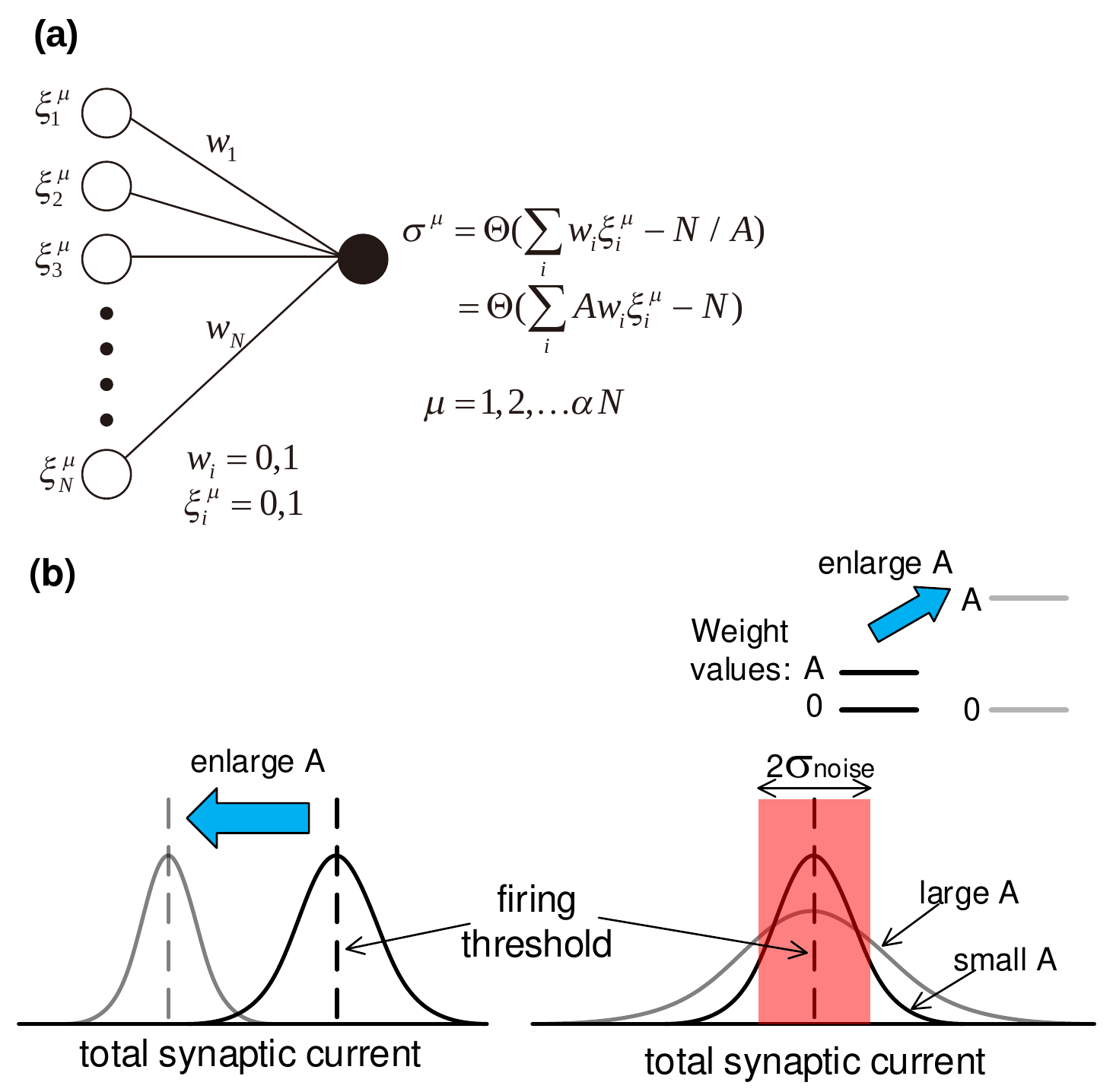}\protect\caption{Schematic of the perceptron model. (a) Structure of the network.
$N$ input units (open circles) feed directly to a single output
unit (solid circle). The task is that given $\alpha N$ input patterns
$\{\xi_{1}^{\mu},\cdots,\xi_{N}^{\mu}\}$ and desired outputs $\sigma^{\mu}$,
with $\mu=1,\cdots,\alpha N$, one needs to adjust the synaptic weights
$w_{i}$ so that the network gives the desired output in response
to each input pattern. All of $\xi_{i}^{\mu}$, $\sigma^{\mu}$ and
$w_{i}$ take binary (0 or 1) values. Input-output mapping function
is $\sigma^{\mu}=\Theta(\sum_{i}w_{i}\xi_{i}^{\mu}-N/A)$, in which
$A$ controls firing threshold $N/A$, or equivalently $\sigma^{\mu}=\Theta(\sum_{i}Aw_{i}\xi_{i}^{\mu}-N)$,
in which $A$ rescales weights $w_{i}$. (b) Two interpretations
of the advantage of large $A$. Left: $A$ can be regarded as the
controller of firing threshold. Total synaptic current (whose distribution
is represented by the curves) is on average approximately equal to
the firing threshold (dashed lines). When $A$ increases, firing
threshold decreases (gray dashed line), so that total synaptic current
also decreases (gray curve), reducing energy consumption. Right:
$A$ can also rescale the discrete values of weights. Input patterns
which induce total synaptic currents $I^{\mu}$s larger (or smaller)
than the firing threshold (dashed line) associate with output 1 (or
0). Under noise of strength $\sigma_{noise}$, perceptron may not
give correct output in response to an input pattern whose $I^{\mu}$
lies within $\sigma_{noise}$ from the threshold (red region). When
$A$ increases, distribution of $I^{\mu}$ becomes broader (gray
curve), so that less portion of $I^{\mu}$ lies in the red region,
which means that perceptron is more likely to give correct output
in response to a stored input pattern under noise. }
\label{fig:fig1} 
\end{figure}

We consider a perceptron model (\textbf{Fig.\ref{fig:fig1}a}) in
which a neuron receives $N$ binary inputs $\xi_{i}=\{1,0\}$ ($i=1,2,\cdots,N$),
which indicates whether the $i$th input neuron fires or not. The
synaptic weights are also binary $w_{i}=\{1,0\}$, modeling excitatory
synapses. The output of the perceptron is $\tau(\mathbf{w},\mathbf{\xi})=\Theta(\mathbf{w}\cdot\mathbf{\xi}-N/A)$,
which takes 1 or 0 depending on whether its total synaptic input
is larger than the firing threshold $N/A$, where $A>0$ is an adjustable
parameter. This neuron is then provided $\alpha N$ input patterns
${\xi^{\mu}}$ and desired outputs $\sigma^{\mu}$ with $\mu=1,2,\cdots,\alpha N$.
The task is to adjust the synaptic weights $\mathbf{w}$, so that
$\tau(\mathbf{w},\mathbf{\xi}^{\mu})=\sigma^{\mu}$ for all $\mu$s.
The probability that $\xi$ take 1 is input coding level $f_{in}$.
Input becomes sparser with smaller $f_{in}$. In our model, $f_{in}$
is fixed at a nonzero value when $N\rightarrow\infty$, so that the
number of $\xi_{i}$s that take 1 is of $\mathcal{O}(N)$ order.
$\sigma^{\mu}$ has equal probability to be 0 or 1.

Now let's discuss the meaning of $A$. For a well-trained perceptron,
it can be shown that the total synaptic current average over the
$\alpha N$ patterns $\langle\mathbf{w}\cdot\mathbf{\xi}^{\mu}\rangle_{\mu}$
is equal to the firing threshold $N/A$ up to order $\mathcal{O}(\sqrt{N})$
(see Appendix \ref{Append:ReplicaMethod}). As larger synaptic current
consumes more energy, energetic efficient perceptron should reduce
firing threshold by enlarging $A$ (\textbf{Fig.\ref{fig:fig1}b,
left}). Of course, if firing threshold is too low, neuron may have
high spontaneous activity due to membrane or synaptic noise \cite{Faisal_2008},
potentially consuming more energy. However, the advantage to enlarge
$A$ should be apparent as long as such spontaneous activity is rare.

The meaning of $A$ can also be interpreted in another way by noting
that the neuron output $\sigma=\Theta(\mathbf{w}\cdot\mathbf{\xi}-N/A)=\Theta(A\mathbf{w}\cdot\mathbf{\xi}-N)$,
where in the latter expression form, $A$ rescales the discrete values
that weights can take. This weight rescaling can increase successful
rate of memory retrieval under noise (\textbf{Fig.\ref{fig:fig1}b,
right}). Intuitively, because $\xi_{i}^{\mu}$s are supposed to be
independent with each other, the standard deviation of the distribution
of total synaptic current $I^{\mu}=A\mathbf{w}\cdot\mathbf{\xi}^{\mu}$
over $\mu$s is approximately $\sqrt{f_{in}(1-f_{in})\sum_{i}(Aw_{i})^{2}}\approx\sqrt{AN(1-f_{in})}$
(Appendix E), which increases with $A$; and the input patterns with
$I^{\mu}$ larger (or smaller) than the firing threshold $N$ associate
with output 1 (or 0). Under noise with strength $\sigma_{noise}$,
the perceptron may not give correct outputs in response to input
patterns whose $I^{\mu}$s lie within the range $[N-\sigma_{noise},N+\sigma_{noise}]$
(red region in \textbf{Fig.\ref{fig:fig1}b, right}). However, when
$A$ gets larger, the distribution of $I^{\mu}$s becomes broader
(gray curve in \textbf{Fig.\ref{fig:fig1}b, right}), so that the
portion of patterns whose $I^{\mu}$s lie within this range decreases,
which means that perceptron is more likely to output the associated
value in response to a stored input pattern under noise. We provide
numeric evidence to support this intuition in Supplemental Material
Section 1 and Supplementary Fig.1.

In a large body of literature (e.g. Ref. \cite{Gardner_1988,Brunel_2016}),
retrieval robustness is realized by requiring that total synaptic
input is above (below) firing threshold by a stability parameter
$\kappa>0$ when the desired output is active (inactive). Here we
effectively set $\kappa=0$. Introducing $\kappa>0$ ensures that
memory retrieval is always successful when the noise strength $\sigma_{noise}<\kappa$,
but enlarging $A$ increases the probability of successful retrieval
when $\sigma_{noise}>\kappa$.

Together, in our model, the requirements for low energy cost and
high robustness for memory retrieval can be unified in a single motivation:
enlarging $A$.

\section{Computational advantage of sparse input under large $A$}

\label{sec:advantage_sparse}

We evaluated the capability of our perceptron model in classification
task by investigating two efficient algorithmic solvers: stochastic
Belief-Propagation Inspired algorithm (SBPI) \cite{Baldassi_2007}
and reinforced Belief Propagation (rBP) \cite{Braunstein_2006}.
Both algorithms assign a hidden state $h_{i}$ to weight $w_{i}$,
update $h_{i}$ during solving, and take $w_{i}=0$ (or 1) when $h_{i}$
is negative (or positive), see Appendix \ref{Append:SBPI_rBP} for
more details of their implementations:

(1) SBPI is a biologically plausible on-line algorithm, in which
an input-output (IO) pair is presented at each time step. If the
presented IO pair is unassociated (i.e., when $\sum_{i}w_{i}\xi_{i}^{\mu}>N/A$
but $\sigma^{\mu}=0$ or when $\sum_{i}w_{i}\xi_{i}^{\mu}<N/A$ but
$\sigma^{\mu}=1$), then hidden state $h_{i}$ is updated as $h_{i}\leftarrow h_{i}+2\xi_{i}^{\mu}(2\sigma^{\mu}-1)$,
so that $w_{i}$ may be flipped towards the direction that facilitates
the association of the presented IO pair. If $\sigma^{\mu}=0$ and
the presented IO pair is associated, but the synaptic current $\sum_{i}w_{i}\xi_{i}^{\mu}$
is too close to the firing threshold $N/A$, or in other words, the
presented IO pair is just on the edge of association, then $h_{i}$
is also updated, with some probability, as $h_{i}\leftarrow h_{i}-2\xi_{i}^{\mu}$,
to increase the margin between $\sum_{i}w_{i}\xi_{i}^{\mu}$ and
$N/A$.

(2) rBP is a message-passing algorithm, which uses belief propagation
(BP) to update $h_{i}$. Specifically, 
\begin{equation}
h_{i\rightarrow\mu}^{t+1}=f[p(t)h_{i}^{t}+\phi_{i\rightarrow\mu}(\bigcup_{\nu\in\partial i\backslash\mu}\hat{h}_{\nu\rightarrow i}^{t})]+(1-f)h_{i\rightarrow\mu}^{t},\label{eq:rBP-SecIII-1}
\end{equation}
\begin{equation}
\hat{h}_{\mu\rightarrow i}^{t+1}=\hat{\phi}_{\mu\rightarrow i}(\bigcup_{j\in\partial\mu\backslash i}h_{j\rightarrow\mu}^{t}),\label{eq:rBP-SecIII-2}
\end{equation}
\begin{equation}
h_{i}^{t+1}=p(t)h_{i}^{t}+\phi_{i}(\bigcup_{\nu\in\partial i}\hat{h}_{\nu\rightarrow i}^{t}).\label{eq:rBP-SecIII-3}
\end{equation}
In the equations above, $\phi_{i\rightarrow\mu}(\cdot)$ and $\hat{\phi}_{\mu\rightarrow i}(\cdot)$
are BP update functions, $\phi_{i}(\cdot)$ is the BP estimate of
single-site field, $p(t)$ is a random number that takes $1$ with
probability $1-(\gamma)^{t}$ and 0 otherwise, and $f$ is a damping
factor. The meaning of eqs.\ref{eq:rBP-SecIII-1}-\ref{eq:rBP-SecIII-3}
is to use BP to evaluate single-site field $h_{i}$, and then in
the next time step, with probability $1-(\gamma)^{t}$ add $h_{i}$
as an external field to BP.

\begin{figure*}
\includegraphics[scale=0.6]{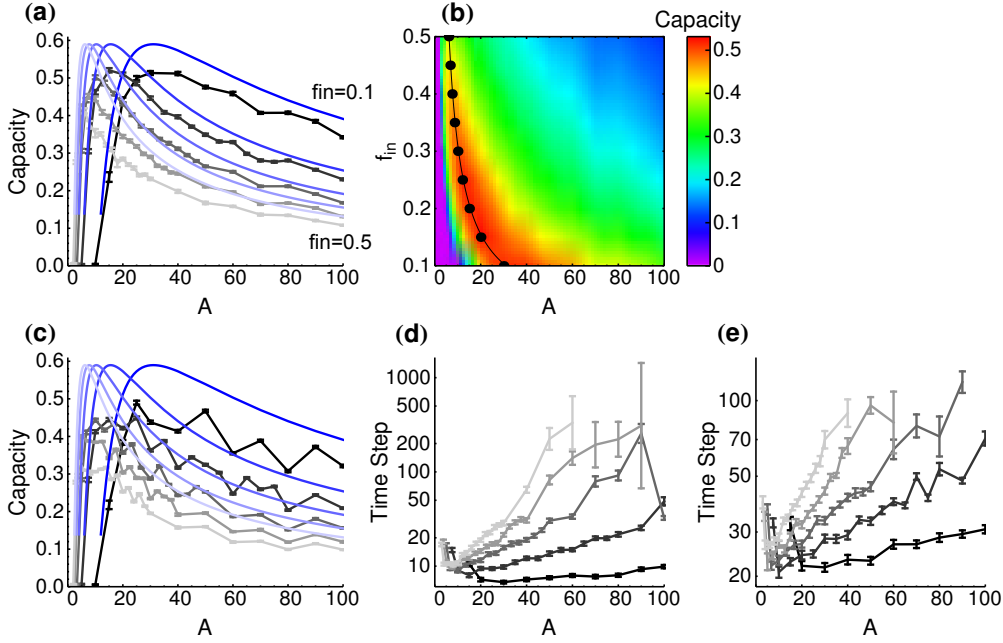}\protect\caption{Performance of SBPI and rBP. (a) Black lines: capacity of SBPI vs
$A$ when $f_{in}=0.1$, 0.2, 0.3, 0.4 and 0.5, represented by lines
with increasing transparency (the lines for $f_{in}=0.1$ and $f_{in}=0.5$
are indicated). Blue curves are theoretical capacity. (b) Density
plot of algorithmic capacity of SBPI. Black dots are the $A_{opt}$
values at which SBPI fulfills highest capacity at given $f_{in}$s.
Black curve is the $A_{opt}$ of the theoretical capacity. On the
left of the dots and the line is low-$A$ branch, which is not biologically
plausible and will not be considered in this paper; on the right
of the dots and the line is high-$A$ branch, at which capacity increases
with input sparseness at a given $A$. (c) The capacity of rBP (black
lines) compared with theoretical capacity (blue curves). (d) The
time steps to solve using SBPI when $\alpha=0.2$. (e) The time steps
to solve using rBP when $\alpha=0.2$. Each dot represents the minimal
time steps among when $\gamma=0.9$, 0.99 and 0.999 at the indicated
$A$ and $f_{in}$ values (see text for details). Detailed implementations
of SBPI and rBP are presented in Appendix \ref{Append:SBPI_rBP}.
$N=480$. Error bars represent standard deviation of the mean.}
\label{fig:performance_SBPI_rBP} 
\end{figure*}

We first studied the dependence of algorithmic capacity $\alpha_{c}^{\text{alg}}$
of SBPI and rBP on $f_{in}$ and $A$. $\alpha_{c}^{\text{alg}}N$
means the maximal number of IO pairs that the perceptron can associate
using an algorithm. We found that at a given $f_{in}$, $\alpha_{c}^{\text{alg}}$
increases with $A$ when $A$ is smaller than an optimal value $A_{opt}$,
and decreases with $A$ when $A>A_{opt}$ (\textbf{Fig.\ref{fig:performance_SBPI_rBP}a-c}).
So this $A_{opt}$, which decreases with $f_{in}$ (\textbf{Fig.\ref{fig:performance_SBPI_rBP}a-c}),
divides the function of $\alpha_{c}^{\text{alg}}$ with $A$ into
two branches: low-$A$ and high-$A$ branches. Because of the non-monotonicity
of $\alpha_{c}^{\text{alg}}$ with $A$, an $\alpha_{c}^{\text{alg}}$
usually corresponds with two $A$ values at a given $f_{in}$ (\textbf{Fig.\ref{fig:performance_SBPI_rBP}a-c}).
However, biologically and engineeringly good design should choose
to work at the larger $A$ value, because it implies lower energy
cost or higher retrieval robustness without compromising memory capacity.
So only the high-$A$ branch is of biological and engineering interest,
and will be considered in the following discussion. At a given $A$,
the dependence of $\alpha_{c}^{\text{alg}}$ on $f_{in}$ may not
be monotonic, but if we only look at high-$A$ branch, $\alpha_{c}^{\text{alg}}$
increases with input sparseness (\textbf{Fig.\ref{fig:performance_SBPI_rBP}a-c}).
This means that under the requirements of low energy consumption
and robust memory retrieval, input sparseness facilitates memory
capacity. From another aspect, the decrease of $\alpha_{c}^{\text{alg}}$
with $A$ at high-$A$ branch reflects the competition between capacity
with energy cost or retrieval robustness, but this competition can
be alleviated by increasing input sparseness: decreasing $f_{in}$
enables the same $\alpha_{c}^{\text{alg}}$ to be fulfilled using
a larger $A$ (\textbf{Fig.\ref{fig:performance_SBPI_rBP}a-c}).

To understand the change of $\alpha_{c}^{\text{alg}}$ with $A$
and $f_{in}$, we first calculated theoretical capacity $\alpha_{c}^{\text{theo}}$
using replica method (Appendix \ref{Append:ReplicaMethod}), which
means the maximal number of IO pairs that perceptron has a solution,
and provides an upper bound of $\alpha_{c}^{\text{alg}}$. We found
that the change of $\alpha_{c}^{\text{theo}}$ with $f_{in}$ and
$A$ follows similar profile as $\alpha_{c}^{\text{alg}}$ (\textbf{Fig.\ref{fig:performance_SBPI_rBP}a,c}):
the function of $\alpha_{c}^{\text{theo}}$ with $A$ also has two
branches; and at high-$A$ branch, $\alpha_{c}^{\text{theo}}$ increases
with $f_{in}$. We discuss the non-monotonicity of capacity with
$A$ and $f_{in}$ in Supplemental Material Section 2 and Supplementary
Fig.2, and we find that the existence of low-$A$ branch is because
of the upper boundedness of weights (i.e., $w_{i}\le1$) in our model.

However, only calculating $\alpha_{c}^{\text{theo}}$ is not sufficient
to understand the computational advantage of input sparseness, because
it is possible that algorithms need more time steps to find a solution
even though there are more solutions in the weight configuration
space. How easy it is to get solutions also matters: on the one hand,
algorithms usually have a upper limit of time step $T_{max}$, and
only the cases when solutions are found within $T_{max}$ time steps
report solving success and contribute to $\alpha_{c}^{\text{alg}}$;
on the other hand, solving perceptron in less time steps, which implies
quick learning, is itself of biological and engineering interest.

We investigated the time steps $T_{solve}$ that SBPI and rBP use
to solve under different input sparsities at a given $\alpha$. Under
SBPI, the change of $T_{solve}$ with $f_{in}$ and $A$ is largely
anti-correlated with the change of $\alpha_{c}^{\text{alg}}$: the
function of $T_{solve}$ with $A$ also has two branches, and at
high-$A$ branch, $T_{solve}$ decreases with input sparseness (\textbf{Fig.\ref{fig:performance_SBPI_rBP}d}).
Under rBP, however, the change of $T_{solve}$ with $f_{in}$ is
more subtle, dependent on the parameter $\gamma$ of rBP. At the
high-$A$ branch of $\alpha_{c}^{\text{alg}}$, at a fixed $\gamma$,
$T_{solve}$ may not decrease with input sparseness (Supplementary
Fig.3b); but if we also tune $\gamma$, defining $T_{solve}$ as
the minimal time steps that rBP achieves successful solving under
different $\gamma$s, then $T_{solve}$ follows similar profile as
that in SBPI (\textbf{Fig.\ref{fig:performance_SBPI_rBP}e}): $T_{solve}$
has two branches with $A$, and decreases with input sparseness at
high-$A$ branch. This means that we are, in principle, able to design
a new algorithm based on rBP, which cleverly chooses $\gamma$ for
quick solving based on, say, experience; and this clever-rBP algorithm
requires less time steps under sparser input at high-$A$ branch.
See Supplemental Material Section 3 and Supplementary Fig.3 for more
discussion on the rBP case.

Together, at large $A$ value, the computational advantage of input
sparseness is of two folds: (1) sparser input results in higher theoretical
capacity, which means that more IO pairs can potentially be associated
with sparser input; (2) at a given $\alpha$, solutions can be obtained
in fewer time steps under sparser input, which brings the theoretical
potentiality of input sparseness suggested by the first point into
algorithmic reality. The first point can be seen from the results
of replica method, and is consistent with the discovery in previous
studies \cite{Brunel_2004,Clopath_2013}. In Sections \ref{sec:solving_nodus}-\ref{sec:geo_cluster},
we will study which part of solving process consumes most time steps
and why input sparseness reduces time steps needed to find solutions
with the help of decimation process.

\section{Understanding the difficult part during perceptron solving through
decimation process}

\label{sec:solving_nodus}

To understand the difficult part in solving binary-weight perceptron
and the advantage of sparse input, we studied decimation process,
where we fixed a single weight in each time step. Specifically, at
the first time step, we evaluated the marginal probability $p_{i}$
that the $i$th weight took 1 value in solutions. Then we chose the
$j$th weight that had the strongest value polarization (i.e., $j=\arg_{i}\min\{p_{i},1-p_{i}\}$),
and fixed the $j$th weight at its preferred value $a_{j}^{\text{prefer}}$
(i.e., $a_{j}^{\text{prefer}}=0$ or 1 if $p_{j}<0.5$ or otherwise),
and eliminated the solutions with $w_{j}\neq a_{j}^{\text{prefer}}$
from consideration in further time steps. At the $k$th time step,
we evaluated $p_{i}$ of unfixed weights in the left solutions, then
fixed the most polarized unfixed weight and eliminated solutions
in the same way as above. By iteratively doing this, we could hopefully
fix all the weights, and got a solution of the perceptron problem.
The word ``decimation\textquotedbl{} means removing variable node
from factor graph \cite{Mezard_2002}.

When the marginal probabilities $p_{i}$ in each time step are computed
using belief propagation, the successive weight-fixing process above
has the name belief-propagation-guided decimation (BPD). In theoretical
interest, in small-sized systems, we can get the exact value of $p_{i}$
using enumerated solutions, which is called exact decimation.

The motivation why we studied decimation process is that BPD is heuristic
from similar idea with rBP and SBPI: BPD is equivalent to adding
an external field of infinite intensity to the fixed weights, while
rBP is a sort of smooth decimation in which each variable gets an
external field with intensity proportional to its polarization \cite{Braunstein_2006};
and SBPI is an on-line algorithm heuristic from rBP \cite{Baldassi_2007}.
So the dynamic processes of these three algorithms should be comparable
in some sense. Additionally, in decimation process, weight values
will not be changed again once fixed, so this process can be studied
in a more controlled manner, unlike in rBP and SBPI, where the preferences
of weights are changing continuously.

\begin{figure*}
\includegraphics[scale=0.6]{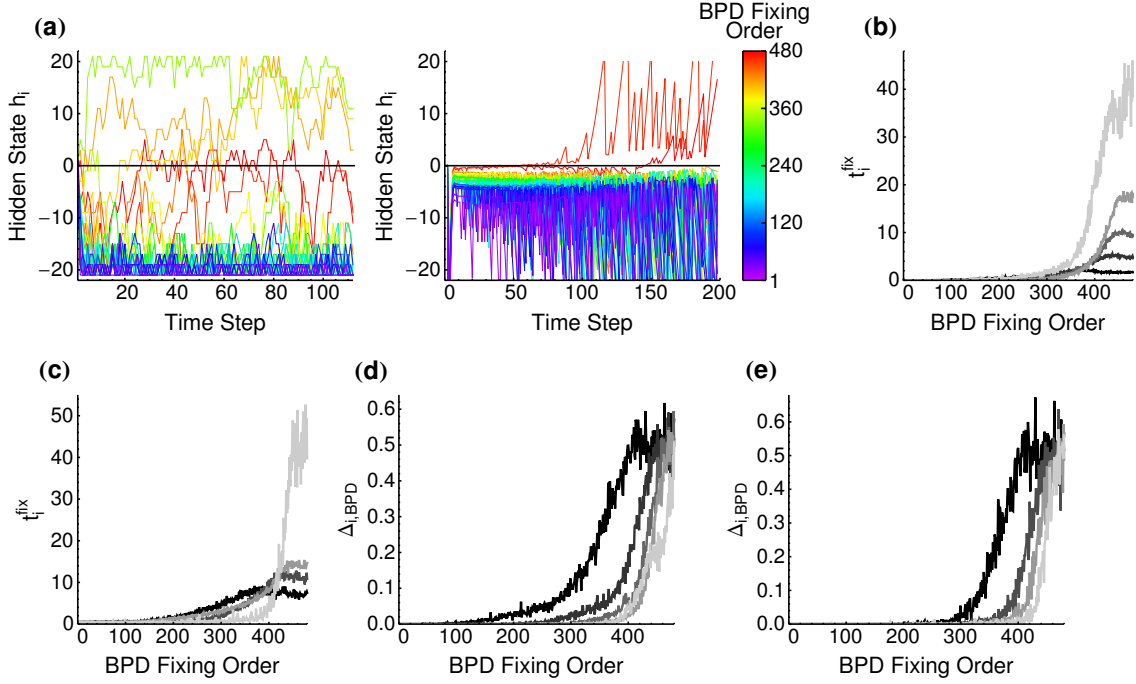}

\protect\caption{Comparing the dynamics of SBPI and rBP with weight-fixing order in
BPD. (a) Examples of the dynamics of hidden states in SBPI (left
panel) and rBP (right panel). Curve represents the change of hidden
state with time step. Color of curve represents the BPD-fixing order
of the weight that the hidden state belongs to. (b) Weight-fixing
time step $t_{i}^{\text{fix}}$ in SBPI as a function of weight-fixing
order in BPD. $t_{i}^{\text{fix}}$ is defined as the last time step
that hidden state $h_{i}$ changes its sign during SBPI. $f_{in}=0.1,0.2,0.3,0.4,0.5$,
represented by curves with decreasing blackness. (c) The same as
(b), but for rBP. $f_{in}=0.1,0.2,0.3,0.4$. We did not plot the
case when $f_{in}=0.5$, because rBP usually fails to find solutions
in this case. (d) The difference $\Delta_{i,\text{BPD}}$ (eq.\ref{eq:Delta_=00007Bi,BPD=00007D})
between the value of weight $w_{i}$ in a solution found by SBPI
from the value of $w_{i}$ in the solution found by BPD, as a function
of BPD-fixing order of $w_{i}$. (e) The same as (d), but for rBP.
$N=480$, $\alpha=0.2$, $A=40$. This $A$ value is in high-$A$
branch of theoretical capacity under the $f_{in}$ values we chose
(\textbf{Fig.\ref{fig:performance_SBPI_rBP}a-c}). Panels (b-e) average
over only trials in which SBPI or rBP succeeded to find solutions.
See Appendix \ref{Append:SBPI_rBP},\ref{Append:Miscellaneous} for
details of methods.}
\label{fig:SBPI_rBP_BPD} 
\end{figure*}

We compared the dynamics of SBPI and rBP with the weight-fixing order
in BPD. We found that for weights fixed early in BPD, the hidden
states in SBPI or rBP quickly deviate from zero in the first few
time step, and hardly change their signs in the subsequent solving
process; however, for weights fixed late in BPD, their hidden states
keep close to zero, prone to change their signs in the subsequent
solving process (\textbf{Fig.\ref{fig:SBPI_rBP_BPD}a}). We then
defined fixing time $t_{i}^{\text{fix}}$ of weight $w_{i}$ as the
last time step that the hidden state $h_{i}$ changes its sign during
SBPI or rBP. We found $t_{i}^{\text{fix}}\approx0$ for early-BPD-fixed
weights; while $t_{i}^{\text{fix}}$ is large for late-BPD-fixed
weights, larger under denser input (\textbf{Fig.\ref{fig:SBPI_rBP_BPD}b,c}).
This means that SBPI and rBP can assign values to early-BPD-fixed
weights in the first few time steps, while most time steps are spent
on determining the values of late-BPD-fixed weights; and spending
less time for these late-BPD-fixed weights is the key reason for
faster solving under sparser input.

We also calculated the difference of $w_{i}$ in solutions found
by SBPI or rBP from that found by BPD: 
\begin{equation}
\Delta_{i,\text{BPD}}=[\langle(w_{i}^{\mathbf{x}}-w_{i}^{\text{BPD}})^{2}\rangle_{\mathbf{x}}],\label{eq:Delta_=00007Bi,BPD=00007D}
\end{equation}
where $w_{i}^{\text{BPD}}$ is the value of $w_{i}$ in the solution
found by BPD, $w_{i}^{\mathbf{x}}$ is the value of $w_{i}$ in a
solution $\mathbf{x}$ found by SBPI or rBP, $\langle\cdot\rangle_{\mathbf{x}}$
means averaging over the solutions found by SBPI or rBP under different
seeds of random number generator, and $[\cdot]$ means quenched average
(i.e., average over different sets of IO pairs). We found that $\Delta_{i,\text{BPD}}\approx0$
for early-BPD-fixed weights (\textbf{Fig.\ref{fig:SBPI_rBP_BPD}d,e}),
indicating that $w_{i}^{\mathbf{x}}=w_{i}^{\text{BPD}}$ in most
cases; while $\Delta_{i,\text{BPD}}\approx0.5$ for late-BPD-fixed
weights (\textbf{Fig.\ref{fig:SBPI_rBP_BPD}d,e}), just like the
case when $w_{i}^{\mathbf{x}}$ takes 0 or 1 randomly. This result
indicates that these algorithms (BPD, SBPI and rBP) have hardly any
discrepancy in what values that early-BPD-fixed weights should take,
while the values of late-BPD-fixed weights are prone to stochasticity
during solving.

We tested the results above in large systems ($N=10000$), and found
similar results with those in \textbf{Fig.\ref{fig:SBPI_rBP_BPD}}
where $N=480$. Notably, the relative position $O_{BPD}/N$ (with
$O_{BPD}$ being BPD fixing order) at which $t_{i}^{\text{fix}}$
and $\Delta_{i,\text{BPD}}$ start to ramp up do not significantly
differ in $N=480$ and $N=10000$ cases (Supplementary Fig.5e-h).
This implies that the portion of weights that are difficult to determine
remains finite when $N\rightarrow\infty$.

Together, during SBPI and rBP, the values of early-BPD-fixed weights
are very easy to determine, and have little disagreement among algorithms;
most solving time steps are spent on determining the values of late-BPD-fixed
weights after the early-BPD-fixed weights are fixed to their little-disputed
values. Under sparser input, the fact that solutions can be found
in less time steps is mostly because less time steps are needed for
determining the values of late-BPD-fixed weights. In other words,
the difficult part during solving binary-weight perceptron and also
the advantage of input sparseness should originate from the subspace
of weight configuration space in which early-BPD-fixed weights are
fixed to their little-disputed values.

\section{Approaching dense solution region through decimation}

\label{sec:approach_dense_region}

\begin{figure*}
\includegraphics[scale=0.5]{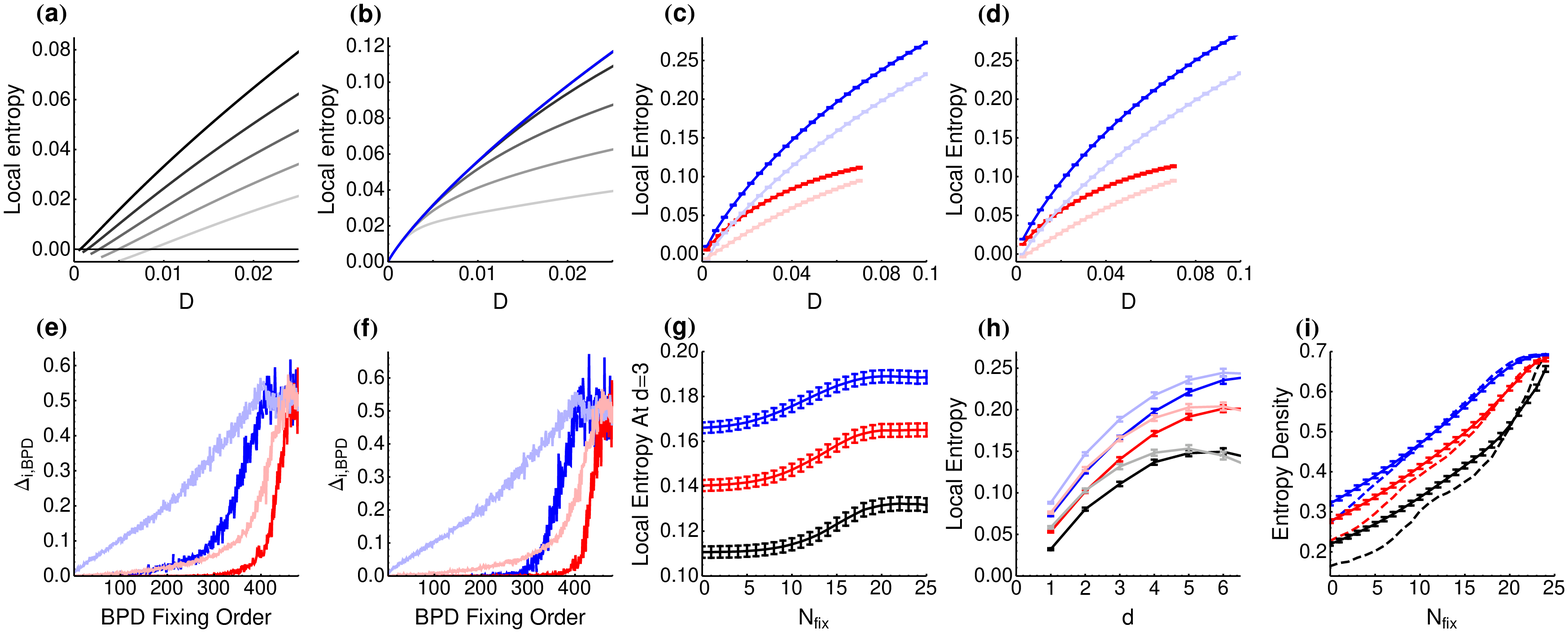}\protect\caption{BPD is a dense-solution-region approaching process. (a) Local entropy
at distance $D$ from typical solutions computed by replica method,
when $f_{in}=0.1,0.2,0.3,0.4,0.5$, representing by curves with decreasing
blackness. (b) Local entropy distance $D$ from a weight configuration
in the dense solution region computed by replica method. The blue
curve is the upper bound of local entropy, corresponding to the local
entropy when all weight configurations at distance $D$ are solutions.
The local entropy at $f_{in}=0.1$ is too close to the upper bound,
hard to be distinguished in the figure. (c) Local entropy computed
by belief propagation, around solutions found by SBPI when $f_{in}=0.1$
(blue) and 0.3 (red). The transparent curves represent the local
entropy around solutions reached after long-term random walk starting
from the solutions found by SBPI. (d) The same as (c), but for rBP.
(e) The difference $\Delta_{i,\text{BPD}}$ between the value of
weight $w_{i}$ in a solution $\mathbf{x}$ from the value of $w_{i}$
in the solution found by BPD, as a function of BPD-fixing order of
$w_{i}$. $\mathbf{x}$ is found by SBPI when $f_{in}=0.1$ (blue)
and 0.3 (red). The transparent curves represent the case when $\mathbf{x}$
is reached after long-term random walk starting from solutions found
by SBPI. (f) The same as (e), except for rBP. (g) Mean local entropy
(eq.\ref{eq:localEntropyInSolutionSubspace}) at Hamming distance
$d=3$ of the solutions in $\mathcal{S}(N_{\text{fix}})$ during
BPD, when $f_{in}=0.3$ (blue), 0.4 (red) and 0.5 (black). (h) Local
entropy at small Hamming distances average over from all solutions
(untransparent curves), and from the solution found by BPD after
fixing all weights (transparent curves). (i) Solution entropy density
in $\mathcal{S}(N_{\text{fix}})$ calculated by exact enumeration
(solid curves) and belief propagation (dashed curves) during BPD.
In the replica analysis in (a,b), $\alpha=0.2$, $A=40$. In (c-f),
$N=480$, $\alpha=0.2$, $A=40$. In small-sized systems in (g-i),
$N=25$, $\alpha=0.36$, $A=12$. We chose $f_{in}=0.1,0.2,0.3,0.4,0.5$
when $A=40$; and chose $f_{in}=0.3,0.4,0.5$ when $A=12$, so that
the value of $A$ lies at high-$A$ branch of theoretical capacity
(\textbf{Fig.\ref{fig:performance_SBPI_rBP}a-c}). Error bars represent
standard deviation of the mean. See Appendix \ref{Append:LocalEntropyBP},\ref{Append:ReplicaMethod},\ref{Append:Miscellaneous}
for details of methods. }
\label{fig:BPD_DenseSolutionCluster} 
\end{figure*}

Previous studies suggest that solutions of binary-weight perceptron
are typically isolated \cite{Huang_2014}, but there is a spatial
region in the weight configuration space where solutions densely
aggregate \cite{Baldassi_2015_b,Baldassi_2016}. It is believed that
solutions in the dense region have good generalization performance,
and are those found by efficient algorithms \cite{Baldassi_2015_b,Baldassi_2016b,Baldassi_2016c,Chaudhari_2017}.
Similar scenario also exists in our model. Using the replica method
introduced in Ref. \cite{Huang_2014,Baldassi_2015_b}, we calculated
the local entropy $F_{local}(D)$ (i.e., logarithm of solution number
at distance $D=d/N$ from a weight configuration) from a typical
solution or a configuration in the dense solution region (see Appendix
\ref{Append:ReplicaMethod} for details). Here, $d$ means Hamming
distance between two weight configurations, which is the number of
weights that take different values in the two weight configurations.
Consistent with the findings in Ref. \cite{Huang_2014,Baldassi_2015_b,Baldassi_2016b},
we found that $F_{local}(D)$ from a typical solution is negative
at small $D$ (\textbf{Fig.\ref{fig:BPD_DenseSolutionCluster}a}),
suggesting that solutions are typically isolated; however, from a
configuration in the dense solution region, at small $D$, $F_{local}(D)$
tends to its upper bound, which is the local entropy in the case
that all weight configurations are solutions (\textbf{Fig.\ref{fig:BPD_DenseSolutionCluster}b}),
suggesting extremely dense solution aggregation. Using belief propagation
(Appendix \ref{Append:LocalEntropyBP}), we found that the local
entropy from a solution $\mathbf{w}_{0}$ found by SBPI or rBP is
higher than that from a solution after long-term random walk (Appendix
\ref{Append:Miscellaneous}) starting from $\mathbf{w}_{0}$ (\textbf{Fig.\ref{fig:BPD_DenseSolutionCluster}c,d}),
suggesting that solutions found by SBPI or rBP are close to the dense
solution region.

Where does the dense solution region locate? To answer this question,
we calculated the difference of $w_{i}$ in a given solution $\mathbf{x}$
from that in the solution found by BPD: $\Delta_{i,\text{BPD}}(\mathbf{x})=(w_{i}^{\mathbf{x}}-w_{i}^{\text{BPD}})^{2}$.
We found that for solutions $\mathbf{w}_{0}$ found by SBPI or rBP,
$\Delta_{i,\text{BPD}}(\mathbf{w}_{0})$ is close to zero for $i$s
early fixed in BPD (\textbf{Fig.\ref{fig:BPD_DenseSolutionCluster}e,f}).
However, for solutions reached after long-term random walk starting
from $\mathbf{w}_{0}$, $\Delta_{i,\text{BPD}}$ is significantly
larger in these early-BPD-fixed weights (\textbf{Fig.\ref{fig:BPD_DenseSolutionCluster}e,f}).
This result, together with the calculation of local entropy (\textbf{Fig.\ref{fig:BPD_DenseSolutionCluster}c,d}),
suggests that solutions with small $\Delta_{i,\text{BPD}}$ in early-BPD-fixed
weights have high local entropy. In other words, the dense solution
region locates in the subspace where the early-BPD-fixed weights
take their fixed values. These early-fixed weights tend to have strong
polarization (i.e., small $\min\{p_{i}(1),1-p_{i}(1)\}$, with $p_{i}(1)$
being the probability that $w_{i}=1$ in solution space, see Supplementary
Fig.7e). Therefore, in a simple picture, solutions in the dense region
are those whose strong-polarized weights take their preferred values.

For the convenience of the following discussion, we define solution
subspace $\mathcal{S}(N_{\text{fix}})$ to be all the solutions in
which the first fixed $N_{\text{fix}}$ weights during a weight-fixing
process take their fixed values.

To better understand the dense-region approaching process during
BPD, we studied small-sized systems in which all solutions can be
exactly enumerated out. We calculated local entropy of the enumerated
solutions in $\mathcal{S}(N_{\text{fix}})$ during BPD: 
\begin{equation}
F_{local}(d,N_{\text{fix}})=\frac{1}{N}[\langle\Theta(\mathcal{N}(\mathbf{w},d))\log(\mathcal{N}(\mathbf{w},d))\rangle_{\mathbf{w}\in\mathcal{S}(N_{\text{fix}})}],\label{eq:localEntropyInSolutionSubspace}
\end{equation}
where $\Theta(x)$ is Heaviside step function which is 1 when $x>0$
and 0 otherwise, $\mathcal{N}(\mathbf{w},d)$ is the number of solutions
at Hamming distance $d$ from a configuration $\mathbf{w}$, $\langle\cdot\rangle_{\mathbf{w}\in\mathcal{S}(N_{\text{fix}})}$
means average over the solutions in $\mathcal{S}(N_{\text{fix}})$,
and $[\cdot]$ represents quenched average (i.e., average over sets
of IO pairs). We found that $F_{local}(d,N_{\text{fix}})$ at small
$d$ increases with $N_{\text{fix}}$ during BPD (\textbf{Fig.\ref{fig:BPD_DenseSolutionCluster}g}),
and the solutions found by BPD after fixing all weights have higher-than-average
local entropy at small $d$ (\textbf{Fig.\ref{fig:BPD_DenseSolutionCluster}h}).
These results further support BPD as a process approaching the dense
solution region.

As an intuitive understanding how BPD can be related with dense-solution-region
approaching, note that after we fix $w_{j}$ at its preferred value
$a_{j}^{\text{prefer}}$ at time step $N_{\text{fix}}$, we eliminate
the solutions with $w_{j}\neq a_{j}^{\text{prefer}}$ from $\mathcal{S}(N_{\text{fix}}-1)$.
By fixing the most polarized weights at its preferred value, we reduce
the number of unfixed weights $N_{\text{unfixed}}$ while eliminating
the least number of solutions. As a result, solution density in the
subspace of unfixed weights should continually increases during this
process. Consistent with this scenario, we found that solution entropy
density in the subspace of unfixed weights $\frac{1}{N_{\text{unfix}}}\log(|\mathcal{S}(N_{\text{fix}})|)$
(with $|\mathcal{S}(N_{\text{fix}})|$ denoting the set size of $\mathcal{S}(N_{\text{fix}})$)
continually increases during BPD (\textbf{Fig.\ref{fig:BPD_DenseSolutionCluster}i}).

We also tried exact decimation in small-sized systems, in which the
marginal probabilities used for weight fixing is calculated with
the enumerated solutions, and also found the increase of local entropy
and solution entropy in $\mathcal{S}(N_{\text{fix}})$ with the progress
of weight fixing (Supplementary Fig.6a-c). This means that the dense-region
approaching phenomenon observed in BPD is universal in decimation
algorithms, instead of an artifact introduced by belief propagation.

\section{Cross-correlation of unfixed weights in solution subspace $\mathcal{S}(N_{\text{fix}})$}

\label{sec:cross-correlation}

From the two subsections above, we see that the subspace of late-BPD-fixed
weights contains the dense solution region, and determining the values
of these late-BPD-fixed weights is the main difficulty in solving
binary-weight perceptron. To better understand this difficulty as
well as the advantage of input sparseness, we computed cross-correlation
within the subspace of solutions $\mathcal{S}(N_{\text{fix}})$ during
BPD: 
\begin{equation}
c_{ij}=\langle w_{i}w_{j}\rangle-\langle w_{i}\rangle\langle w_{j}\rangle,\label{eq:correlationDefinition}
\end{equation}
where $w_{i}$ and $w_{j}$ are two unfixed weights, and $\langle\cdot\rangle$
means average over solutions in $\mathcal{S}(N_{\text{fix}})$.

\begin{figure*}
\includegraphics[scale=0.6]{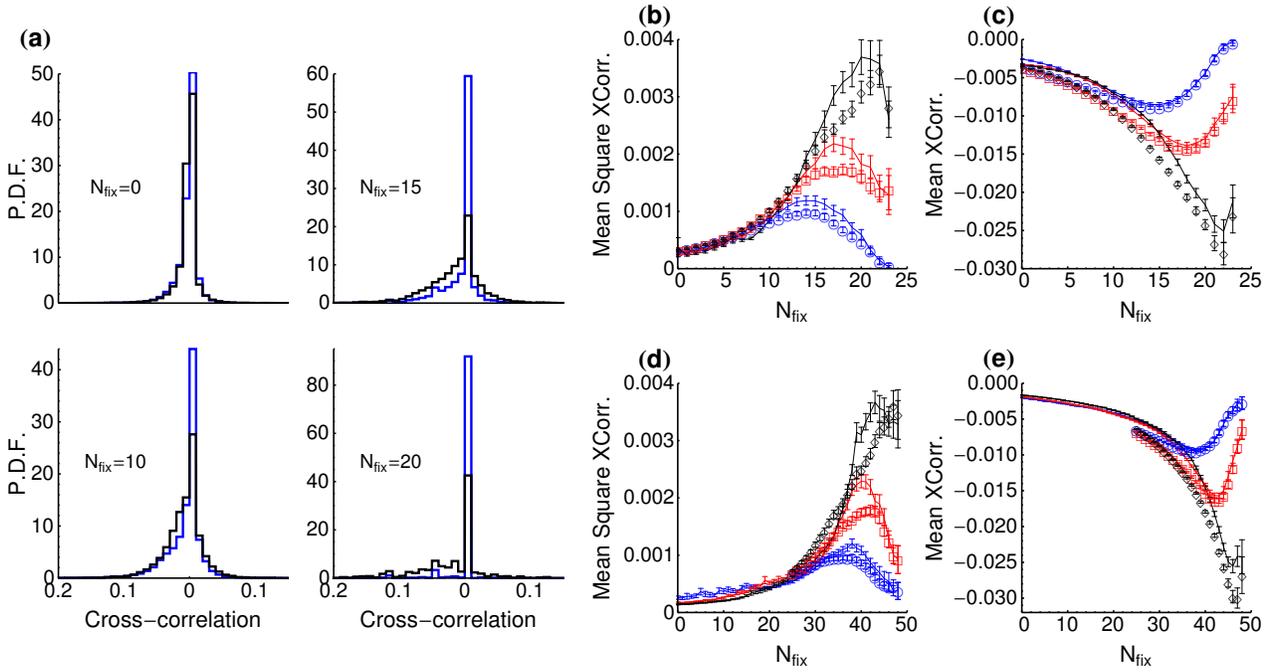}

\protect\caption{\label{fig:fig7} Cross-correlation between unfixed weights in $\mathcal{S}(N_{\text{fix}})$
during BPD. (a) Probability distribution function (P.D.F.) of cross-correlations
at four different time steps of BPD for $f_{in}=0.3$ (blue) and
0.5 (black) in systems of size $N=25$. (b) Mean square cross-correlation
$\overline{\text{XCorr}^{2}}$ during BPD, when $f_{in}=0.3$ (blue),
0.4 (red), and 0.5 (black) in systems of size $N=25$. Error bars
connected by lines represent results computed by belief propagation;
plot markers (circles, squares, diamonds) represent results computed
by exactly enumerated solutions in $\mathcal{S}(N_{\text{fix}})$.
(c) Mean cross-correlation $\overline{\text{XCorr}}$ during BPD,
$N=25$. (d,e) The same as (b,c), except for $N=50$. Results for
$N=100$ and $N=200$ are given in Supplementary Fig.7a-d. $\alpha=0.36$,
$A=12$. See Appendix \ref{Append:XCorrBP} for details of methods.
\label{fig:XCorr}}
\end{figure*}

Before decimation, $c_{ij}$ distributes narrowly around zero. With
the progress of BPD, the distribution of $c_{ij}$ becomes broader
(\textbf{Fig.\ref{fig:XCorr}a}). Consistently, the mean square cross-correlations
\begin{equation}
\overline{\text{XCorr}^{2}}=\frac{1}{N_{\text{unfix}}(N_{\text{unfix}}-1)}[\sum_{i\neq j}c_{ij}^{2}]\label{eq:XCorr2}
\end{equation}
increases with BPD progress (\textbf{Fig.\ref{fig:XCorr}b,d}, Supplementary
Fig.7a,c). Cross-correlations are negative-dominated: their mean
value 
\begin{equation}
\overline{\text{XCorr}}=\frac{1}{N_{\text{unfix}}(N_{\text{unfix}}-1)}[\sum_{i\neq j}c_{ij}]
\end{equation}
is negative and decreases with BPD progress (\textbf{Fig.\ref{fig:XCorr}c,e},
Supplementary Fig.7b,d). This indicates strong cross-correlation
between late-BPD-fixed weights. This strong cross-correlation, which
breaks Bethe-Peierls approximation, is probably the reason for the
difficulty in assigning values to late-BPD-fixed weights, and why
SBPI and rBP spend so many times steps on assigning values to late-BPD-fixed
weights (\textbf{Fig.\ref{fig:SBPI_rBP_BPD}b,c}). At the late stage
of BPD, with sparser input, $\overline{\text{XCorr}^{2}}$ is smaller
and $\overline{\text{XCorr}}$ is less negative (\textbf{Fig.\ref{fig:XCorr}b-e},
Supplementary Fig.7a-d). The reduction of cross-correlation between
late-BPD-fixed weights may be the reason for quicker solving under
sparser input.

Notably, at the early stage of BPD, both $\overline{\text{XCorr}^{2}}$
and $|\overline{\text{XCorr}}|$ are small, and they do not change
much with input sparsities (\textbf{Fig.\ref{fig:XCorr}b-e}, Supplementary
Fig.7a-d). This implies that the difficulty in solving binary-weight
perceptron and also the computational advantage of input sparseness
come from the structure of solution space near the dense solution
region, and cannot be unveiled using equilibrium analysis where all
solutions have equal statistical contribution.

We also investigated $\overline{\text{XCorr}^{2}}$ and $\overline{\text{XCorr}}$
during exact decimation in small-sized systems, and found similar
profile how $\overline{\text{XCorr}^{2}}$ and $\overline{\text{XCorr}}$
change with $N_{\text{fix}}$ and $f_{in}$ (Supplementary Fig.6d,e).

\section{Understanding the weight cross-correlation in $\mathcal{S}(N_{\text{fix}})$
through geometry of solution clusters }

\label{sec:geo_cluster}

In this subsection, we will try to understand the cross-correlation
between unfixed weights in $\mathcal{S}(N_{\text{fix}})$ from the
geometry of solution clusters.

A solution cluster in $\mathcal{S}(N_{\text{fix}})$ is defined as
a set of solutions that can be connected by flipping a single unfixed
weight. A cluster is used to represent a pure state, which is a set
of weight configurations that are separated from other configurations
by infinite free-energy barrier. In other words, from a configuration
in a pure state, the system needs infinitely long time to access
a configuration outside of this pure state under natural thermodynamics.
Strictly speaking, it is not known how to adapt the definition of
pure states to instances of finite size. Here we adopt the suggestion
in Ref. \cite{Ardelius_2008,Obuchi_2009}, and numerically identify
a pure state in $\mathcal{S}(N_{\text{fix}})$ as a solution cluster.

Under 1-step replica symmetry breaking (1RSB) ansatz, the overlap
between solutions in the same cluster (or different clusters) is
a delta function locates at $q_{\text{same}}$ (or $q_{\text{diff}}$).
In this case, mean square cross-correlation can be related with solution
overlap in the limit $N_{\text{unfix}}\rightarrow\infty$ as (Supplemental
Material Section 5) 
\begin{equation}
\overline{\text{XCorr}^{2}}=\frac{1}{N_{\text{unfix}}^{2}}[\sum_{i,j}c_{ij}^{2}]=\frac{1}{3}x_{*}(1-x_{*})(q_{\text{same}}-q_{\text{diff}})^{2},\label{eq:CorrGeo}
\end{equation}
where $x_{*}$ is Parisi parameter, which is the probability that
two solutions belong to different clusters. Numerically, $x_{*}$
is defined as 
\begin{equation}
x_{*}=1-[\sum_{\gamma}(\frac{N_{\gamma}}{N_{solution}})^{2}],\label{eq:Parisi}
\end{equation}
where $N_{\gamma}$ is the number of solutions in pure state $\gamma$,
and $N_{solution}$ is the total number of solutions. Using the replica
method introduced in Ref. \cite{Obuchi_2009}, it can be shown that
in the full solution space $\mathcal{S}(N_{\text{fix}}=0)$ of infinite-sized
systems, $x_{*}=1$ and $q_{\text{same}}=q_{\text{diff}}$ (see Appendix
\ref{Append:GeometryOfPureStatesFullSpace} for discussion and Appendix
Fig.\ref{fig:geoLargeNFig} for numeric support), which from eq.\ref{eq:CorrGeo}
implies zero cross-correlation. Possibly because of this zero cross-correlation,
belief propagation can give entropy landscape closely matching that
predicted by replica method in the solution space before decimation
\cite{Huang_2013}.

In the following context, we will try to understand the cross-correlation
between unfixed weights through eq.\ref{eq:CorrGeo}, by investigating
1-weight-flip clusters in $\mathcal{S}(N_{\text{fix}})$ using enumerated
solutions of small-sized systems. Strictly speaking, eq.\ref{eq:CorrGeo}
is valid only when $N_{\text{unfix}}\rightarrow\infty$. Here, by
investigating small-sized systems, we hope to get some understanding
on the change of cross-correlation during decimation observed in
finite-sized systems (\textbf{Fig.\ref{fig:XCorr}}), and also the
possible phase transition ($\overline{\text{XCorr}^{2}}$ transits
from zero to nonzero) during decimation in infinite-sized systems.

\begin{figure}
\includegraphics[scale=0.6]{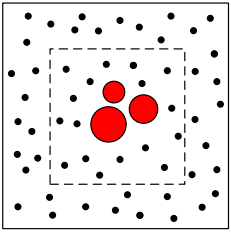}

\protect\caption{Schematic for solution condensation around dense solution region.
The full solution space (inside the solid box) contains many small
clusters (black dots) and a small number of large clusters (red circles).
The dense solution region lies in a subspace (inside the dashed box)
where most solutions belong to large clusters. In this subspace,
solutions become condensed in large clusters, which reduces Parisi
parameter $x_{*}$ from 1, increasing cross-correlation through eq.\ref{eq:CorrGeo}. }
\label{fig:solutionCondensationSchem}
\end{figure}

\begin{figure*}
\includegraphics[scale=0.6]{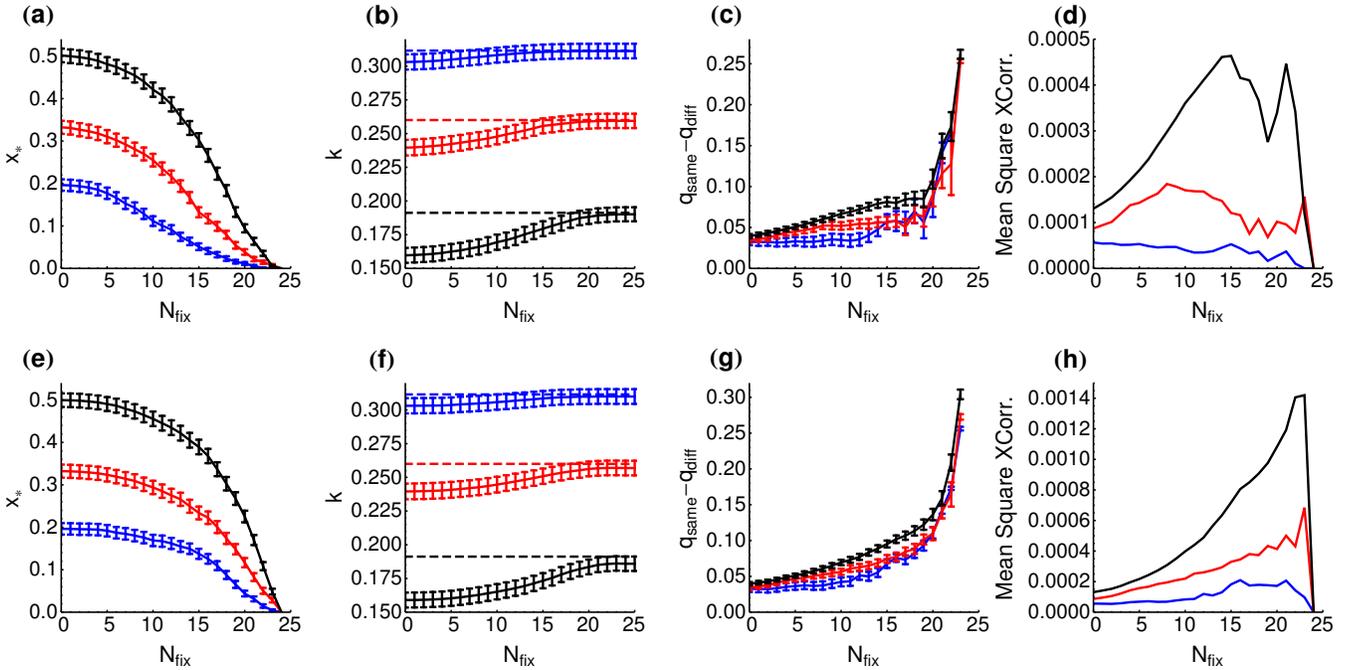}

\protect\caption{Understanding cross-correlation between unfixed weights in $\mathcal{S}(N_{\text{fix}})$
through geometry of clusters. (a) Parisi parameter $x_{*}$ (eq.\ref{eq:Parisi})
with the progress of BPD, when $f_{in}=0.3$ (blue), 0.4 (red) and
0.5 (black). (b) $k(N_{\text{fix}})$ (eq.\ref{eq:kIndex}) during
BPD. Dashed lines represent the entropy density of the largest cluster
in the full solution space before BPD. (c) $q_{\text{same}}-q_{\text{diff}}$
during BPD. (d) Mean square cross-correlation $\overline{\text{XCorr}^{2}}$
predicted by eq.\ref{eq:CorrGeo}, using the values of $x_{*}$ and
$q_{\text{same}}-q_{\text{diff}}$ plotted in (a,c). (e-h) The same
as (a-d), except for exact decimation, where marginal probabilities
used for weight fixing are computed using enumerated solutions. $N=25$,
$\alpha=0.36$, $A=12$. }
\label{fig:geometryPureState} 
\end{figure*}

We found that $x_{*}$ decreases with $N_{\text{fix}}$ during decimation
(\textbf{Fig.\ref{fig:geometryPureState}a,e}), which implies a solution
condensation process. One possible scenario is as follows. Before
decimation, the solution space contains sub-exponential number of
large clusters with exponentially many solutions in each, and exponentially
many small clusters with sub-exponential number of solutions in each
\cite{Obuchi_2009} (Appendix \ref{Append:GeometryOfPureStatesFullSpace}).
The dense solution region presumably lies in a region where most
solutions belong to large clusters. Therefore, in the solution subspace
$\mathcal{S}(N_{\text{fix}})$ at the late stage of decimation, solutions
may be more condensed in large clusters (\textbf{Fig.\ref{fig:solutionCondensationSchem}}).
With this solution condensation, $x_{*}$ decreases from 1, because
two randomly chosen solutions become more likely to locate in the
same large pure state. The decrease of $x_{*}$ from 1 increases
weight cross-correlation through eq.\ref{eq:CorrGeo}.

We tested a hypothesis in the discussion above: most solutions around
the dense solution region come from large clusters. To this end,
we calculated a $k$ index, defined as 
\begin{equation}
k(N_{\text{fix}})=\frac{1}{N}[\frac{\sum_{\mathbf{w}\in\mathcal{S}(N_{\text{fix}})}\log(N_{\gamma\subset\mathcal{S}(N_{\text{fix}}=0)}(\mathbf{w}))}{\sum_{\mathbf{w}\in\mathcal{S}(N_{\text{fix}})}1}],\label{eq:kIndex}
\end{equation}
where $N_{\gamma\subset\mathcal{S}(N_{\text{fix}}=0)}(\mathbf{w})$
means the number of solutions in cluster $\gamma$ that contains
solution $\mathbf{w}$, and the subscript $\gamma\subset\mathcal{S}(N_{\text{fix}}=0)$
means that $\gamma$ is a cluster defined in the full solution space
$\mathcal{S}(N_{\text{fix}}=0)$ before weight fixing. Therefore,
$k(N_{\text{fix}})$ means the average entropy of the clusters in
$\mathcal{S}(N_{\text{fix}}=0)$ that solutions in $\mathcal{S}(N_{\text{fix}})$
belong to. We found that $k(N_{\text{fix}})$ increases with $N_{\text{fix}}$
(\textbf{Fig.\ref{fig:geometryPureState}b,f}); and that $k(N_{\text{fix}}=N)$,
which is the entropy of the cluster that contains the solution obtained
after fixing all weights, is close to the entropy of the largest
cluster in $\mathcal{S}(N_{\text{fix}}=0)$ (\textbf{Fig.\ref{fig:geometryPureState}b,f}).
These results support the scenario depicted in \textbf{Fig.\ref{fig:solutionCondensationSchem}}:
solutions in the neighborhood of the dense solution region mostly
belong to large clusters.

We found that $x_{*}<1$ before decimation (i.e., when $N_{\text{fix}}=0$),
see \textbf{Fig.\ref{fig:geometryPureState}a,e}. This is probably
a finite-size effect: replica method in Ref.\cite{Obuchi_2009} (Appendix
\ref{Append:GeometryOfPureStatesFullSpace}) predicts that $x_{*}=1$
before decimation; and numerically, we found that $x_{*}$ before
decimation approaches to 1 when $N$ gets large (Appendix Fig.\ref{fig:geoLargeNFig}).

We found that $q_{\text{same}}-q_{\text{diff}}$ tends to increase
with decimation progress, and decrease with input sparseness (\textbf{Fig.\ref{fig:geometryPureState}c,g}).
This means that around the dense solution region, the difference
of overlaps within and between clusters is larger than that in full
solution space, and gets smaller with sparser input. This point contributes
to the increase of weight cross-correlation during decimation, and
also the reduction of weight cross-correlation under sparse input
at the late stage of decimation through eq.\ref{eq:CorrGeo}.

We compared $\overline{\text{XCorr}^{2}}$ calculated from eq.\ref{eq:CorrGeo}
with that directly calculated by eq.\ref{eq:XCorr2}. We found that
eq.\ref{eq:CorrGeo} cannot reproduce the value of $\overline{\text{XCorr}^{2}}$,
but it can manifest some features of the profile how $\overline{\text{XCorr}^{2}}$
changes with $N_{\text{fix}}$ and $f_{in}$ (\textbf{Fig.\ref{fig:geometryPureState}d,h}):
(1) $\overline{\text{XCorr}^{2}}$ increases with $f_{in}$; (2)
$\overline{\text{XCorr}^{2}}$ increases with $N_{\text{fix}}$,
especially for BPD when $f_{in}$ is large and for exact decimation.
This suggests that eq.\ref{eq:CorrGeo} provides a promising aspect
to understand the strong weight cross-correlation at the late stage
of decimation, and the reduction of weight cross-correlation under
sparse input.

Possible reasons why eq.\ref{eq:CorrGeo} cannot reproduce the value
of $\overline{\text{XCorr}^{2}}$ include: (1) finite size of $N_{\text{unfix}}$,
(2) failure of 1RSB ansatz, (3) improper numeric definition of pure
state. The first and second points undermine the conditions to establish
eq.\ref{eq:CorrGeo}. As for the third point, as mentioned in Ref.
\cite{Ardelius_2008}, a shortcoming of defining pure state using
1-weight flip is that it cannot describe entropic barrier. For example,
suppose there are two subsets of solutions. Both subsets are densely
intra-connected by 1-weight flip, but there is only a single long
1-weight flip routine connecting the two subsets. According to the
1-weight-flip definition of pure state, these two subsets should
belong to the same pure state. However, it is very hard to access
a subset starting from the other through this single routine using
random-walk dynamics when $N$ is large. So according to the physical
definition of pure state mentioned at the beginning of this subsection,
these two subsets should belong to different pure states.

\section{The case when $\alpha\rightarrow\alpha_{c}^{\text{theo}}$}

\label{sec:alpha_approach_alphac}

In previous sections, we studied the cases when $\alpha$ is not
so close to theoretical capacity $\alpha_{c}^{\text{theo}}$, so
that SBPI and rBP can efficiently solve the perceptron problem. In
this section, we will study the case when $\alpha\rightarrow\alpha_{c}^{\text{theo}}$,
so that SBPI and rBP are no longer able to solve the problem in reasonable
time. This implies that in this case, people have to wait for exponential-to-$N$
time steps for theoretically existing solutions. Here, we will try
to understand the origin of this exponential-time difficulty by investigating
the dynamics of SBPI and rBP when $\alpha\rightarrow\alpha_{c}^{\text{theo}}$,
keeping $f_{in}=0.5$.

\begin{figure*}
\includegraphics[scale=0.6]{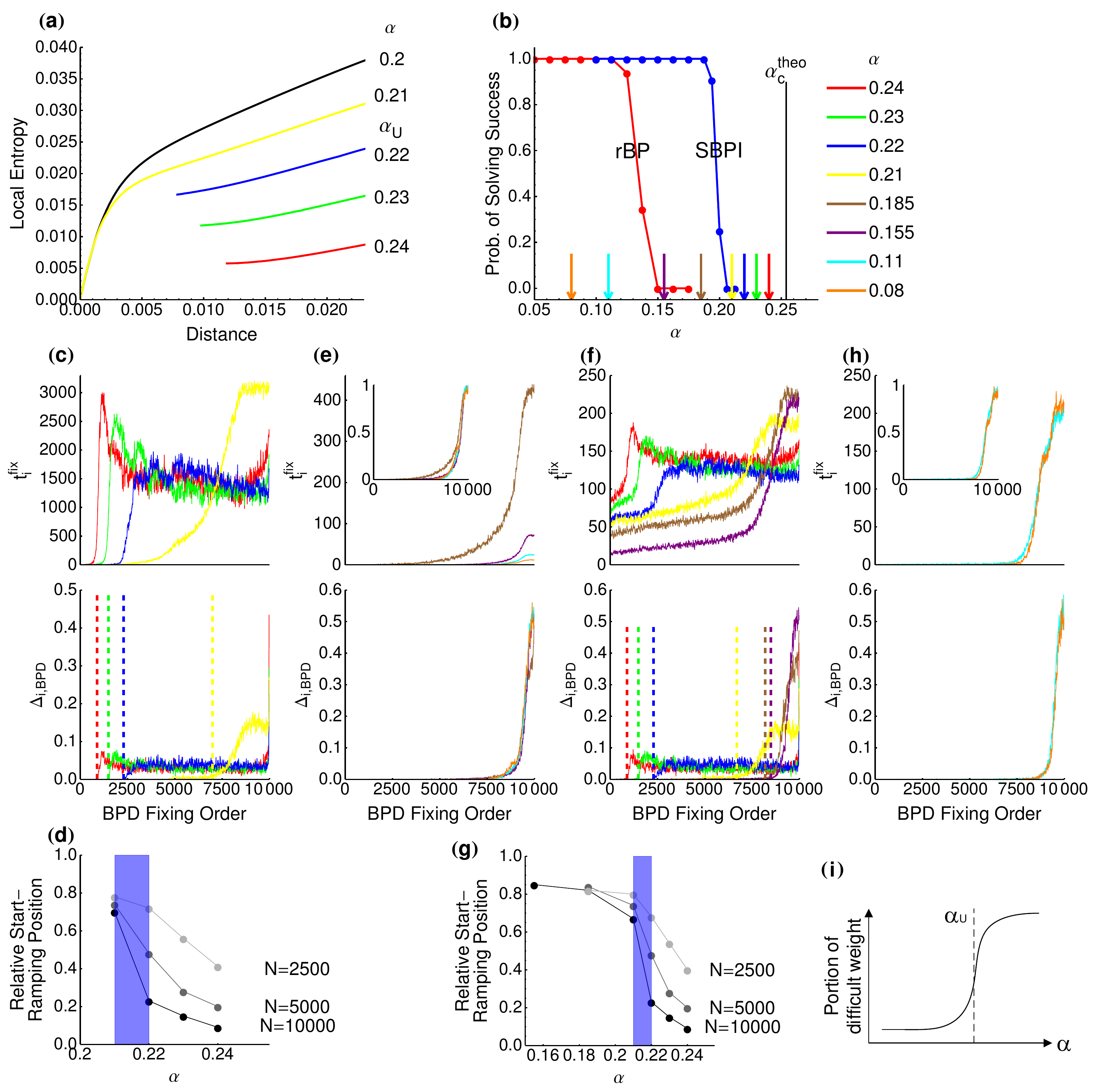}

\protect\caption{Performance of SBPI and rBP when $\alpha\rightarrow\alpha_{c}^{\text{theo}}$.
(a) Large-deviation local entropy as a function of distance. A phase
transition happens at $\alpha_{U}\in(0.21,0.22)$. (b) Probability
of successful solving of SBPI (blue) and rBP (red). Black vertical
line is theoretical capacity $\alpha_{c}^{\text{theo}}$. Colored
arrows indicate the $\alpha$ values that we investigate in the following
panels. $N=10000$. (c) Upper: Weight-fixing time step $t_{i}^{\text{fix}}$
in SBPI as a function of BPD weight-fixing order. Lower: The difference
$\Delta_{i,\text{BPD}}$ between the value of weight $w_{i}$ that
SBPI ends up from the value of $w_{i}$ that BPD ends up, as a function
of BPD-fixing order of $w_{i}$. In panel (c), $\alpha$ takes the
values at which SBPI fails to solve, represented by colors of lines
indicated in panel (b). Vertical lines in the lower panel indicate
the positions when curves start to ramp, determined by eye-looking.
(d) Start-ramping positions of $\Delta_{i,\text{BPD}}$ relative
to $N$. The $N=10000$ line is plotted according to vertical lines
in the lower panel of (c). See Supplementary Fig.9 for correspondences
of panel (c) in $N=5000$ and $N=2500$ cases. Blue shading indicates
the interval $(0.21,0.22)$ that $\alpha_{U}$ lies in, indicated
by panel (a). (e) The same as panel (c), except that $\alpha$ takes
the values at which SBPI succeeds to solve. In the inset of the upper
panel, $t_{i}^{\text{fix}}$ is rescaled with maximum 1. (f) The
case when rBP fails to solve. (g) The same as panel (d), except for
rBP. (h) The case when rBP succeeds to solve. (i) Schematic of the
conclusion from the previous panels: when $\alpha$ increases, the
portion of weights whose values are difficult to determine keeps
almost unchanged when $\alpha<\alpha_{U}$, but increases sharply
near $\alpha_{U}$. $N=10000$ by default, $A=40$, $f_{in}=0.5$. }
\label{fig:alphaApproachCapacityFig} 
\end{figure*}

As the first step, we studied the reshaping of dense solution region
with $\alpha$ by calculating large-deviation local entropy $F_{local}(D)$,
which quantifies the number of solutions at distance $D$ from a
configuration in the dense solution region. Similarly as in previous
studies \cite{Baldassi_2016b,Baldassi_2016}, we found a transition
point $\alpha_{U}$ of $F_{local}(D)$ (\textbf{Fig.\ref{fig:alphaApproachCapacityFig}a}):
when $\alpha<\alpha_{U}$, saddle point equations always have solutions,
so that we can calculate $F_{local}(D)$ at $D$ close to zero; when
$\alpha>\alpha_{U}$, however, saddle point equations stop having
solutions at some value of $D>0$. As suggested in Ref.\cite{Baldassi_2016b,Baldassi_2016},
$F_{local}(D)$ may be no longer monotonic when $\alpha>\alpha_{U}$,
which implies that the dense solution region fragments into separate
regions. After noting that no known solvers have algorithmic capacity
larger than $\alpha_{U}$, Ref.\cite{Baldassi_2016} suggests that
$\alpha_{U}$ signals a transition between easy and hard solving
phase.

We have shown that SBPI and rBP spend most time steps on determining
late-decimation-fixed weights (\textbf{Fig.\ref{fig:SBPI_rBP_BPD}}),
and with weight fixing, the subspace of unfixed weights gets shrinking
to the dense solution region (\textbf{Fig.\ref{fig:BPD_DenseSolutionCluster}}).
Therefore, if the structural transition of the dense region at $\alpha_{U}$
is indeed responsible for the difficult solving when $\alpha>\alpha_{U}$,
there should be some observable effect on late-decimation-fixed weights
around $\alpha_{U}$.

To study this possible effect, we investigated the dynamics of SBPI
and rBP at a few $\alpha$ values (\textbf{Fig.\ref{fig:alphaApproachCapacityFig}}).
With the increase of $\alpha$, the probabilities that rBP and SBPI
can solve the problem decrease (\textbf{Fig.\ref{fig:alphaApproachCapacityFig}b}).
No matter whether SBPI or rBP can solve successfully, we defined
$t_{i}^{\text{fix}}$ as the last time step that the hidden state
$h_{i}$ of weight $w_{i}$ changed its sign, and $\Delta_{i,\text{BPD}}$
as the difference of the value of $w_{i}$ that SBPI or rBP ended
up from the value that BPD ended up. We considered the following
two cases. Case I: when $\alpha$ took small values so that SBPI
(or rBP) succeeded to solve with high probability, we investigated
$t_{i}^{\text{fix}}$ and $\Delta_{i,\text{BPD}}$ in the trials
when solving succeeded. Case II: when $\alpha$ took large values
so that SBPI (or rBP) failed to solve with high probability, we investigated
$t_{i}^{\text{fix}}$ and $\Delta_{i,\text{BPD}}$ in the trials
when solving failed. We studied the functions $t_{i}^{\text{fix}}(O_{\text{BPD}})$
and $\Delta_{i,\text{BPD}}(O_{\text{BPD}})$, where $O_{\text{BPD}}$
is BPD fixing order, with particular interest in the relative position
$O_{\text{BPD}}/N$ where these two functions start to ramp. $O_{\text{BPD}}/N$
estimates the portion of weights whose values can be easily determined
during algorithmic solving.

Interestingly, we found that in Case I, the functions of $\Delta_{i,\text{BPD}}(O_{\text{BPD}})$
at different $\alpha$ values almost overlap (\textbf{Fig.\ref{fig:alphaApproachCapacityFig}e,h,
lower panels}); and after normalizing the maximum of $t_{i}^{\text{fix}}$
to 1, functions $t_{i}^{\text{fix}}(O_{\text{BPD}})$ at different
$\alpha$s also almost overlap (\textbf{Fig.\ref{fig:alphaApproachCapacityFig}e,h,
upper panels}). This suggests that the portion of weights whose values
are easy to determine does not change much with $\alpha$ at small
values of $\alpha$ when perceptron problem can be easily solved.
In Case II, we recognized a sharp decrease of start-ramping position
around $\alpha_{U}$ (\textbf{Fig.\ref{fig:alphaApproachCapacityFig}c,f});
and this decrease gets shaper with larger $N$ (\textbf{Fig.\ref{fig:alphaApproachCapacityFig}d,g}).
Note that we only plotted start-ramping positions of $\Delta_{i,\text{BPD}}$
with $\alpha$ in \textbf{Fig.\ref{fig:alphaApproachCapacityFig}d,g},
because $\Delta_{i,\text{BPD}}$ has more clear-cut start-ramping
positions. The start-ramping positions of $t_{i}^{\text{fix}}$ are
largely comparable to those of $\Delta_{i,\text{BPD}}$ (\textbf{Fig.\ref{fig:alphaApproachCapacityFig}c,f}).
Collectively, it seems that the portion of weights whose values are
difficult to determine does not change much with $\alpha$ when $\alpha<\alpha_{U}$,
but gets sharply increased around $\alpha_{U}$ (\textbf{Fig.\ref{fig:alphaApproachCapacityFig}i}).
This sharp increase of difficult weights may be responsible for the
difficult solving when $\alpha>\alpha_{U}$.

\section{The geometry of solution clusters when $\alpha\rightarrow\alpha_{c}^{\text{theo}}$}

\label{sec:large_alpha_enum}

To get more understanding on the exponential-time difficulty encountered
by algorithms when $\alpha\rightarrow\alpha_{c}^{\text{theo}}$,
we performed exact decimation in small-sized systems, and studied
the geometry of solution clusters connected by single weight flipping.

\begin{figure*}
\includegraphics[scale=0.55]{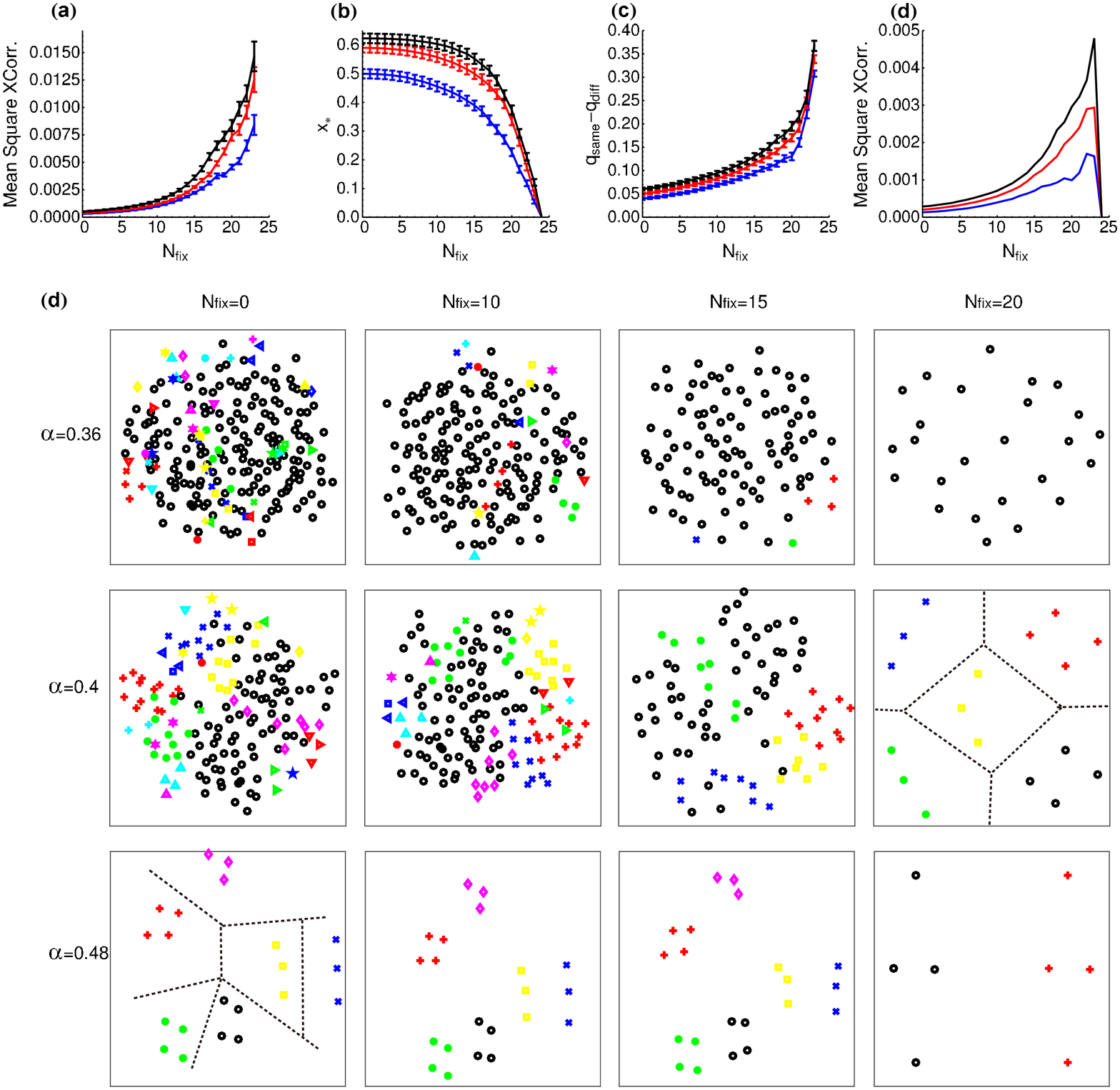}

\protect\caption{Studying geometry of clusters when $\alpha\rightarrow\alpha_{c}^{\text{theo}}$
using exact enumeration. (a) Mean square cross-correlation $\overline{\text{XCorr}^{2}}$
during exact decimation, when $\alpha=0.36$ (blue), 0.4 (red) and
0.48 (black). (b) Parisi parameter $x_{*}$ during exact decimation.
(c) $q_{\text{same}}-q_{\text{diff}}$ during exact decimation. (d)
Mean square cross-correlation $\overline{\text{XCorr}^{2}}$ predicted
by eq.\ref{eq:CorrGeo}, using the values of $x_{*}$ and $q_{\text{same}}-q_{\text{diff}}$
plotted in (b,c). (e) Visualization of solution clusters during exact
decimation at different $\alpha$ values, after mapping solutions
into 2-dimensional space using multi-dimensional scaling. Markers
of the same shape and color represent solutions in the same cluster.
Black circles represent the largest cluster. In some subplots, dashed
lines are plotted to guide eyes, showing that different clusters
are well spatially separated. $N=25$, $A=12$. Theoretical capacity
$\alpha_{c}^{\text{theo}}=0.503$. See Appendix \ref{Append:Miscellaneous}
for details of multi-dimensional scaling. }
\label{fig:Enum_MDS_Fig} 
\end{figure*}

As the first step, similarly as in \textbf{Fig.\ref{fig:geometryPureState}e-h},
we studied cross-correlation in the solution subspace $\mathcal{S}(N_{\text{fix}})$
after fixing $N_{\text{fix}}$ weights during exact decimation. We
found that mean square cross-correlation $\overline{\text{XCorr}^{2}}$
increases with both $N_{\text{fix}}$ and $\alpha$ (\textbf{Fig.\ref{fig:Enum_MDS_Fig}a}).
We found that Parisi parameter $x_{*}$ decreases with $N_{\text{fix}}$
(\textbf{Fig.\ref{fig:Enum_MDS_Fig}b}), suggesting a solution condensation
process during decimation, and that $q_{\text{same}}-q_{\text{diff}}$
increases with $N_{\text{fix}}$ and $\alpha$ (\textbf{Fig.\ref{fig:Enum_MDS_Fig}c}).
The $\overline{\text{XCorr}^{2}}$ value calculated by eq.\ref{eq:CorrGeo}
can reproduce the profile how $\overline{\text{XCorr}^{2}}$ changes
with $N_{\text{fix}}$ and $\alpha$ (\textbf{Fig.\ref{fig:Enum_MDS_Fig}d}).
Notably, $x_{*}$ increases with $\alpha$ (\textbf{Fig.\ref{fig:Enum_MDS_Fig}b}),
which, when $x_{*}$ is close to 1, reduces $\overline{\text{XCorr}^{2}}$
from eq.\ref{eq:CorrGeo}. Therefore, the observed increase of $\overline{\text{XCorr}^{2}}$
with $\alpha$ (\textbf{Fig.\ref{fig:Enum_MDS_Fig}d}) comes from
the increase of $q_{\text{same}}-q_{\text{diff}}$. This increase
of $\overline{\text{XCorr}^{2}}$ with $\alpha$ should at least
contribute to the exponential-time difficulty when $\alpha\rightarrow\alpha_{c}^{\text{theo}}$.

We then visualized the geometry of solution clusters using multi-dimensional
scaling (MDS). MDS maps high-dimensional data points into 2-dimensional
space while preserving inter-point distances to the most extent.
Using MDS, we hope to get some intuitive understanding on the spatial
reorganization of solution clusters with decimation and $\alpha$.

In \textbf{Fig.\ref{fig:Enum_MDS_Fig}e}, we illustrate the solution
clusters during the decimation processes of three perceptron instances
at three $\alpha$ values. Each of these instances was chosen so
that its solution number took the median value in a number of random
instances at the corresponding $\alpha$. In this way, the illustrated
instances are hopefully representative to manifest the reorganization
of solution clusters with decimation and $\alpha$.

When $\alpha$ takes small value, in the full solution space before
decimation (upper left corner of \textbf{Fig.\ref{fig:Enum_MDS_Fig}e}),
a large cluster dominates. It is hard to spatially separate this
large cluster from the other small clusters by depicting boundaries
between them, which suggests that the large cluster spatially intertwines
with small clusters. Using the replica method introduced in Ref.\cite{Obuchi_2009}
(see Appendix \ref{Append:GeometryOfPureStatesFullSpace}), we know
that $q_{\text{same}}=q_{\text{diff}}$ before decimation, and that
$q_{\text{same}}$ is contributed by overlaps between solutions in
the same large cluster. Here we show that the phenomenon $q_{\text{same}}=q_{\text{diff}}$
is possibly because that the large cluster is highly spatially dispersed,
filling the same spatial region together with small clusters, so
that two solutions in the large cluster or in different clusters
have almost the same overlap. With decimation, the number of clusters
get reduced, consistent with the solution condensation scenario.
At the late stage of decimation, different clusters get easier to
spatially separate (right column of \textbf{Fig.\ref{fig:Enum_MDS_Fig}e}),
consistent with the observation that $q_{\text{same}}-q_{\text{diff}}$
increases with decimation (\textbf{Fig.\ref{fig:Enum_MDS_Fig}c}).

With the increase of $\alpha$, solution clusters get easier to separate
even in the full solution space before decimation (lower left corner
of \textbf{Fig.\ref{fig:Enum_MDS_Fig}e}). However, if we accept
that the result $q_{\text{same}}=q_{\text{diff}}$ is valid at all
$\alpha$ values, large clusters should also disperse even when $\alpha\rightarrow\alpha_{c}$,
so the observed easy separation should be a finite-size effect. A
conjecture that is consistent with the current knowledge is that
when $\alpha>\alpha_{U}$, spatial separation of clusters happens
at very early stage of decimation, due to the fragmentation of the
dense solution region. This spatial separation induces high $q_{\text{same}}-q_{\text{diff}}$,
resulting in high cross-correlation through eq.\ref{eq:CorrGeo}
after decimating only a few weights.

\section{Conclusion and Discussion}

\label{sec:discussion}

In this paper, we tried to understand the computational difficulty
of binary-weight perceptron and the advantage of input sparseness
in classification task. The novelty of our work is we regard decimation
process as reference to study the dynamics of two efficient solvers
(SBPI and rBP) and use decimation process to study the reshaping
of the structure of solution space during approaching the dense solution
region. We studied two types of algorithmic difficulty: (1) difficult
part during efficient solving, with particular interest in the advantage
of input sparseness and (2) exponential-time difficulty, which happens
when $\alpha\rightarrow\alpha_{c}^{\text{theo}}$. We found that
in the case of successful solving, most time steps in SBPI and rBP
are spent on determining the values of the weights late fixed in
decimation, which is the difficult solving part. Under biologically
plausible requirements for low energy consumption and high robustness
of memory retrieval, perceptron can be solved in fewer time steps
with sparser input, because the values of late-decimation-fixed weights
become easier to determine. We then tried to understand the difficulty
in assigning late-decimation-fixed weights and the advantage of input
sparseness from the aspect of cross-correlation of unfixed weights
during decimation. We found that this cross-correlation increases
with decimation progress and decreases with input sparseness. We
then studied the geometry structure of clusters in the subspace of
unfixed weights during decimation. We proposed that the change of
cross-correlation with decimation progress and input sparsity may
be related to solution condensation in the subspace of unfixed weights
during decimation, and also the fact that the difference $q_{\text{same}}-q_{\text{diff}}$
between overlaps of solutions in the same and different clusters
increases with decimation progress and decreases with input sparsity.
To get understanding of exponential-time difficulty, we compared
the dynamics of SBPI and rBP with weight-fixing order in BPD when
$\alpha\rightarrow\alpha_{c}^{\text{theo}}$, and found that the
portion of weights whose values are difficult to determine sharply
increases around $\alpha_{U}$. By studying small-sized systems using
exact enumeration, we suggest that this large portion of difficult
weights is because that spatial separation of solution clusters happens
at very early stage of decimation, due to the fragmentation of the
dense solution region around $\alpha_{U}$.

Binary weight perceptron belongs to constraint satisfaction problems
(CSP). At present, we still do not have a full understanding on the
computational difficulty of CSP. In this paper, we suggest that solution
condensation during decimation is an important aspect to understand
the difficulty in solving CSP. This viewpoint is supported by Ref.\cite{Ricci-Tersenghi_2009},
where a uniform BP decimation process was studied in XORSAT and SAT
problems. In uniform BP decimation, the variable to be fixed at each
time step is randomly chosen, unlikely in the BPD studied in this
paper, in which the to-be-fixed variable in each time step is the
most biased variable. Uniform BP decimation, in the ideal case, will
end up with a uniformly chosen solution from the whole solution set,
so it is unsuitable for studying the process approaching the dense
solution region. However, statistical properties during uniform BP
decimation can be computed using cavity method, which are helpful
for understanding the computational difficulty during decimation.
It was found that the $\alpha$ value at which uniform BP decimation
fails to find solution coincides well with the $\alpha$ value at
which condensation occurs during decimation \cite{Ricci-Tersenghi_2009}.
In the context of binary perceptron, another observation that supports
the importance of solution condensation for understanding the exponential-time
difficulty when $\alpha>\alpha_{U}$ is the close-to-zero external
entropy in the large-deviation calculation of the dense solution
region \cite{Baldassi_2016b}. External entropy measures the logarithm
of the number of configurations that can be defined as the center
of the dense region. Therefore, zero external entropy means that
there are only sub-exponential number of configurations can be regarded
as the center of the dense region. Therefore, when $\alpha>\alpha_{U}$,
the dense region should fragment into sub-exponential pieces, each
of which surrounds a center. In this case, the scenario of solution
condensation into sub-exponential clusters naturally occurs in the
subspace around the dense region.

It is tempting to understand exponential-time difficulty and difficult
solving part during efficient solving from a unified viewpoint. From
eq.\ref{eq:CorrGeo}, condensation (i.e., $x_{*}<1$) implies finite
cross-correlation in large $N$ limit. If finite cross-correlation
implies exponential solving time, it is reasonable to hypothesize
that strong but $o(1)$ order of cross-correlation implies difficult
solving part, at which solving takes sub-exponential, but many time
steps. In Supplementary Fig.8, we studied the scaling of mean square
cross-correlation ($\overline{\text{XCorr}^{2}}$) with $N$ during
decimation at a low $\alpha$ value at which BPD can find solution
with high probability. We found that although $\overline{\text{XCorr}^{2}}$
gets high at late stage of decimation, it tends to zero when $N$
gets large (Supplementary Fig.8a-c). In \textbf{Fig.\ref{fig:geometryPureState}},
we observed that $x_{*}<1$ and $q_{\text{same}}-q_{\text{diff}}>0$
during decimation. If $\overline{\text{XCorr}^{2}}\rightarrow0$
when $N\rightarrow\infty$, then at least one of these two observations
should be a finite-size effect. Our numerical results support both
$x_{*}\rightarrow1$ and $q_{\text{same}}-q_{\text{diff}}\rightarrow0$
with $N\rightarrow\infty$ (Supplementary Fig.8d-i). The scaling
speeds of $x_{*}$ and $q_{\text{same}}-q_{\text{diff}}$ with $N$
control the time steps needed to find solutions when solving is efficient.

One influential viewpoint is that exponential-time difficulty comes
from frozen variables in clusters \cite{Zdeborova_2007,Zdeborova_2008,Zdeborova_2008b},
which are variables that take the same value in solutions in the
same cluster. However, when solving CSPs in which all solutions are
isolated, or in other words, all variables are frozen in every cluster,
BPD can still evaluate the preferences of variables at the early
stage, while hardness emerges during BPD progress \cite{Zdeborova_2008}.
This implies that the freezing in the full solution space does not
hamper variable assignment at the beginning of BPD, while difficulty
comes from the subspace of unfixed variables during BPD. In other
words, freezing may not be the origin of computational difficulty,
although it leads to this difficulty. From the solution condensation
scenario depicted in this paper, we presume that the difficulty during
BPD in Ref. \cite{Zdeborova_2008} may be related to a solution elimination
process: the number of isolated solutions in the subspace of unfixed
variables decreases during BPD, and difficulty arises when this number
becomes finite so that $x_{*}<1$, indicating solutions are condensed
into a small number of clusters; the isolation of solutions in Ref.
\cite{Zdeborova_2008} results in large $q_{\text{same}}-q_{\text{diff}}$,
which makes large cross-correlation from eq.\ref{eq:CorrGeo}. In
other words, the number of un-eliminated solutions decreases with
decimation, until, if solving is successful, only one solution is
obtained after fixing all variables. So during decimation, there
should be some steps at which the solution number is finite. Because
of the isolation nature of the solutions in the model of Ref.\cite{Zdeborova_2008},
the number of clusters at these steps is also finite, resulting in
finite cross-correlation. Therefore, it is hard to tell whether the
solving failure observed in Ref.\cite{Zdeborova_2008} results from
freezing or from condensation during decimation.

The idea that solution condensation causes exponential-time difficulty
was also evaluated in Ref.\cite{Zdeborova_2007}. It was found that
algorithms can efficiently found solutions even when $\alpha$ is
larger than the condensation transition point $\alpha_{c}$, so that
the solution space is dominated by sub-exponential number of large
pure states. This observation argues against the relevance of condensation
with computational difficulty. However, in our view, for a fair comparison
between condensation point and the point of algorithmic failure,
one should use the same sampling measure. Specifically, an algorithm
may be regarded as a configuration sampler. On the one hand, during
on-going solving process, it continually samples configurations with
positive energy. On the other hand, the algorithm stops once it encounters
a solution (i.e., zero-energy configuration), so a solution is sampled
at each successful solving trial. However, the probability measure
that a configuration is sampled by the algorithm may not coincide
with the Boltzmann measure widely used in statistical calculation.
So it is of no surprise that the condensation transition point $\alpha_{c}$
based on Boltzmann measure cannot well describe the algorithmic failure.
In other words, the energetic landscape calculated with Boltzmann
measure may be different from the energetic landscape encountered
by the algorithm. A good example of this discrepancy is binary-weight
perceptron studied in this paper, in which typical solutions are
isolated \cite{Huang_2014}, but algorithms tend to find solutions
in the dense region \cite{Baldassi_2015_b}. In the model of Ref.\cite{Zdeborova_2007},
one possible scenario is that when $\alpha>\alpha_{c}$, algorithms
tend to sample configurations near the largest pure state, so that
this single pure state dominates the subspace where algorithms usually
sample, which makes algorithmic solving simple. Moreover, as we mentioned
previously, uniform BP decimation algorithm samples all solutions
uniformly; and the failure of this algorithm is well predicted by
condensation during decimation calculated using Boltzmann measure
\cite{Ricci-Tersenghi_2009}.

The dense solution region is called ``sub-dominant\textquotedbl{}
in Ref. \cite{Baldassi_2015_b}, because solutions in the dense region
are so rare and cannot be unveiled using equilibrium analysis. We
found solutions in the dense region are those whose strong-polarized
weights take their preferred values, which may provide an aspect
to understand the scarcity of solutions in the dense region: in solutions
of the dense region, all strong polarized weights take their preferred
values, while in typical solutions, some strong polarized weights
violate their preferences (manifested by \textbf{Fig.\ref{fig:BPD_DenseSolutionCluster}e,f});
typical solutions are much more than solutions in dense region because
there are combinatorially many ways to choose the weights that violate
their preferences. As a simple example, suppose a perceptron with
5 weights has six solutions: 00000,10000,01000,00100,00010 and 00001.
Here, each digit indicates the value of a weight (e.g., 10000 means
that the value of the first weight is 1, while the other four weights
are zero). In this case, each weight has probability 0.833 to be
at 0, so according to our finding (i.e., dense solution region is
made up solutions in which strong polarized weights take their preferred
values), the dense solution region should be the solution in which
all these weights take 0 (i.e., 00000). Consistently, it is easy
to check that 00000 has 5 neighbors at Hamming distance 1, while
all the other solutions have only one neighbor at Hamming distance
1, so 00000 has larger local entropy. However, we see that there
is only one solution 00000 in which all weights take their preferred
value 0, while in all the other five solutions, there is a weight
that violates its preference and takes 1.

Our result suggests that during training neural networks, the learning
dynamics of different weights are heterogeneous: the values of some
weights can be quickly determined in the first few time steps, while
most time steps are spent on determining the values of some other
weights (\textbf{Fig.\ref{fig:SBPI_rBP_BPD}b,c}). This result has
implications in both fields of neuroscience and machine learning.
During the development of embryo and infant, synapses are vastly
eliminated \cite{Piochon_2016}. People may wonder why the brain
removes synapses that could be used to store information, long before
the brain becomes functionally competent. Our result suggests that
some synapses may be assigned at zero efficacy at the very early
learning stage of the animal, unlikely to be changed in the subsequent
learning. In this case, it is structurally and spatially economical
to remove these silent synapses as soon as possible. Additionally,
we presume that this heterogeneous dynamics of weights may also exist
during training artificial neural networks. We can evaluate the confidence
that a weight stays at a value using hidden states as in SBPI (also
see Ref. \cite{Hubara_2018,Ott_2017}), and fix the high-confident
weights at their values in a few initial time steps, excluding them
from further updating. This scheme can significantly reduce the computational
overhead during training. Notably, in both biological and artificial
contexts, this early-weight-fixing strategy may not compromise the
generalization performance of the neural network after training,
because the dense solution region lies in the subspace where these
early-fixed weights take the values they are fixed at.

\section*{acknowledgments}

We thank Haiping Huang for constructive comments. This work was partially
supported by Hong Kong Baptist University Strategic Development Fund,
RGC (Grant No. 12200217) and NSFC (Grants No.11275027). %\end{acknowledgments}

% Specify following sections are appendices. Use \appendix* if there% only one appendix.%\onecolumngrid

\appendix
%dummy comment inserted by tex2lyx to ensure that this paragraph is not empty

\section{Implementations of SBPI and rBP \label{Append:SBPI_rBP}}

In \textbf{Fig.\ref{fig:performance_SBPI_rBP}}, the capacity of
an algorithm was estimated by adding the number of input-output associations
one by one, until the algorithm failed to find a solution within
a given time step. Below we briefly introduce the two efficient algorithms
(SBPI and rBP) we studied in this paper.

\subsection*{SBPI}

We used the SBPI01 algorithm introduced in Ref. \cite{Baldassi_2007}.
In this algorithm, each synapse has a hidden variable $h_{i}$ that
takes odd integer values between $-K$ and $K$. $w_{i}=1$ if $h_{i}>0$,
and $w_{i}=0$ if $h_{i}<0$. Suppose an input-output (IO) pair $\mu$
is presented, the hidden states are updated in the following two
cases. Firstly, when this IO pair is unassociated, $h_{i}$ is updated
as $h_{i}\leftarrow h_{i}+2\xi_{i}^{\mu}(2\sigma^{\mu}-1)$. Secondly,
when this IO pair is associated, but $\sigma^{\mu}=0$ and $0<(1-2\sigma^{\mu})(\mathbf{w}\cdot\mathbf{\xi}^{\mu}-N/A)\le1$,
then with probability $p_{s}$, $h_{i}\leftarrow h_{i}-2\xi_{i}^{\mu}$.
Solving was stopped when a solution was found or after going sweep
of all IO pairs in random sequential order for $T_{max}=4000$ times.
$p_{s}=0.3$, $K=[21\sqrt{\frac{N}{480}}]$, with $[x]$ being the
integer nearest to $x$.

\subsection*{rBP}

The rBP updating rule \cite{Braunstein_2006} is 
\begin{equation}
h_{i\rightarrow\mu}^{t+1}=f[p(t)h_{i}^{t}+\sum_{\nu\in\partial i\backslash\mu}\hat{h}_{\nu\rightarrow i}^{t}]+(1-f)h_{i\rightarrow\mu}^{t},\quad\hat{h}_{\mu\rightarrow i}=\log\frac{H(s^{\mu}\frac{\theta-1-a_{\mu\rightarrow i}}{\sigma_{\mu\rightarrow i}})}{H(s^{\mu}\frac{\theta-a_{\mu\rightarrow i}}{\sigma_{\mu\rightarrow i}})},\label{eq:rBP-1}
\end{equation}
with 
\begin{equation}
s^{\mu}=2\sigma^{\mu}-1,\quad\theta=N/A\label{eq:rBP-2}
\end{equation}
\begin{equation}
a_{\mu\rightarrow i}=\sum_{j\neq i}\frac{1}{1+e^{-h_{j\rightarrow\mu}}},\quad\sigma_{\mu\rightarrow i}^{2}=\sum_{j\neq i}\frac{e^{-h_{j\rightarrow\mu}}}{(1+e^{-h_{j\rightarrow\mu}})^{2}}.\label{eq:rBP-3}
\end{equation}
The values of weights $\mathbf{w}$ are determined by the sign of
single-site fields 
\begin{equation}
h_{i}^{t+1}=p(t)h_{i}^{t}+\sum_{\nu\in\partial i}\hat{h}_{\nu\rightarrow i}^{t},\label{eq:rBP-4}
\end{equation}
so that $w_{i}=1$ if $h_{i}>0$ and $w_{i}=0$ otherwise. The message-passing
equations eq.\ref{eq:rBP-1} are iterated in a factor graph in which
the $i$th variable node (representing $w_{i}$) and the $\mu$th
factor node (representing the $\mu$th IO pairs) are linked if $\xi_{i}^{\mu}=1$.
In eqs.\ref{eq:rBP-1} and \ref{eq:rBP-4}, $p(t)$ is a random number
that takes $1$ with probability $1-(\gamma)^{t}$ and 0 otherwise.
We also added damping through $f=0.05$ in eq.\ref{eq:rBP-1}, which
is necessary to avoid divergence of the algorithm before finding
a solution. In practice, we constrained $h_{i\rightarrow\mu}$ between
$[-8,8]$, so that if $h_{i\rightarrow\mu}>8$ (or $h_{i\rightarrow\mu}<-8$)
by the first equation of eq.\ref{eq:rBP-1}, we set $h_{i\rightarrow\mu}=8$
(or $h_{i\rightarrow\mu}=-8$). We found adding this constraint can
improve the performance of the algorithm. Solving was stopped when
a solution was found or after $T_{max}=4000$ time steps. In \textbf{Fig.\ref{fig:performance_SBPI_rBP}c},
$\gamma=0.99$; in \textbf{Fig.\ref{fig:performance_SBPI_rBP}e},
at a given pair of $A$ and $f_{in}$ values, we chose the minimal
time steps that rBP used to solve among when $\gamma=0.9,0.99$ and
0.999. We found that at the parameter values we chose ($\alpha=0.2$,
$A=40$ and $N=480$), when $f_{in}=0.1$, 0.2 and 0.3, rBP can successfully
solve the perceptron with high probability in either the three $\gamma$
values, and it takes minimal time steps to solve when $\gamma=0.9$;
when $f_{in}=0.4$, however, rBP has low probability to solve the
problem if $\gamma=0.9$, and it takes less time steps to solve when
$\gamma=0.99$ than when $\gamma=0.999$ (Supplementary Fig.3). Because
of this, in \textbf{Fig.\ref{fig:SBPI_rBP_BPD}} and \textbf{Fig.\ref{fig:BPD_DenseSolutionCluster}},
$\gamma=0.9$ when $f_{in}=0.1$, 0.2 and 0.3, and $\gamma=0.99$
when $f_{in}=0.4$. When studying the cases when $N=10000$, 5000
or 2500 in \textbf{Fig.\ref{fig:alphaApproachCapacityFig}} and Supplementary
Fig.9, we kept $\gamma=0.99$, $f=0.002$. Lookup tables were used
to efficiently evaluate the H function in eq.\ref{eq:rBP-1} during
computation.

\section{Belief-propagation-guided decimation}

\label{Append:BPD}

The idea of belief-propagation-guided decimation (BPD) is to use
belief propagation (BP) to evaluate the marginal probabilities of
unfixed weights, and then fix the most polarized weight at its preferred
value; then run BP again to fix another weight, until all weights
are fixed. At a step of BPD, $\tau^{\mu}=\Theta(\sum_{i}w_{i}\xi_{i}^{\mu}-N/A)$
may become fixed at 1 or 0, whatever the values of the unfixed weights.
If $\tau^{\mu}=\sigma^{\mu}$, then the $\mu$th IO pair is successfully
associated and eliminated from the factor graph; if $\tau^{\mu}\neq\sigma^{\mu}$,
however, it means that BPD fails to find a solution of the perceptron
problem and is terminated.

BP equations are 
\begin{equation}
p_{i\rightarrow\mu}(w_{i})\propto\prod_{\nu\in\partial i\backslash\mu}\hat{p}_{\nu\rightarrow i}(w_{i})\label{eq:BP_1}
\end{equation}
\begin{equation}
\hat{p}_{\mu\rightarrow i}(w_{i})\propto\sum_{\{w_{j}\}_{j\in\partial\mu\backslash i}}\Theta(s^{\mu}(w_{i}+\sum_{j\in\partial\mu\backslash i}w_{j}+N_{\partial\mu}^{\text{fix}}-\theta))\prod_{j\in\partial\mu\backslash i}p_{j\rightarrow\mu}(w_{j}),\label{eq:BP_2}
\end{equation}
where as in eq.\ref{eq:rBP-2}, $s^{\mu}=2\sigma^{\mu}-1$ and $\theta=N/A$.
These BP equations run on a factor graph where unfixed weights (represented
by $i,j$) and IO pairs (represented by $\mu,\nu$) are variable
and factor nodes respectively, and there is a link between the $i$th
variable node and the $\mu$th factor node if $\xi_{i}^{\mu}=1$.
$N_{\partial\mu}^{\text{fix}}$ in eq.\ref{eq:BP_2} means the number
of fixed weights $w_{k}$ with $\xi_{k}^{\mu}=1$. After getting
a fixed point of BP, marginal probabilities of unfixed weights can
be estimated as 
\begin{equation}
p_{i}(w_{i})\propto\prod_{\mu\in\partial i}\hat{p}_{\mu\rightarrow i}(w_{i}).\label{eq:BP_marginal}
\end{equation}

After introducing $h_{i\rightarrow\mu}=\log\frac{p_{i\rightarrow\mu}(1)}{p_{i\rightarrow\mu}(0)}$
and $\hat{h}_{\mu\rightarrow i}=\log\frac{\hat{p}_{\mu\rightarrow i}(1)}{\hat{p}_{\mu\rightarrow i}(0)}$,
and correspondingly $p_{i\rightarrow\mu}(1)=\frac{1}{1+e^{-h_{i\rightarrow\mu}}}$
and $\hat{p}_{\mu\rightarrow i}(1)=\frac{1}{1+e^{-\hat{h}_{\mu\rightarrow i}}}$,
BP equations can be simplified as 
\begin{equation}
h_{i\rightarrow\mu}=\sum_{\nu\in\partial i\backslash\mu}\hat{h}_{\nu\rightarrow i}\label{eq:BP_h}
\end{equation}
\begin{equation}
\hat{h}_{\mu\rightarrow i}=\log\frac{\sum_{\{w_{j}\}_{j\in\partial\mu\backslash i}}\Theta(s^{\mu}(1+\sum_{j\in\partial\mu\backslash i}w_{j}+N_{\partial\mu}^{\text{fix}}-\theta))\prod_{j\in\partial\mu\backslash i}p_{j\rightarrow\mu}(w_{j})}{\sum_{\{w_{j}\}_{j\in\partial\mu\backslash i}}\Theta(s^{\mu}(\sum_{j\in\partial\mu\backslash i}w_{j}+N_{\partial\mu}^{\text{fix}}-\theta))\prod_{j\in\partial\mu\backslash i}p_{j\rightarrow\mu}(w_{j})}.\label{eq:BP2}
\end{equation}

When the number $|\partial\mu|$ of unfixed variable nodes around
factor node $\mu$ is large, eq.\ref{eq:BP2} can be efficiently
calculated using Gaussian approximation: 
\begin{equation}
\hat{h}_{\mu\rightarrow i}=\log\frac{H(s^{\mu}\frac{\theta-1-N_{\partial\mu}^{\text{fix}}-a_{\mu\rightarrow i}}{\sigma_{\mu\rightarrow i}})}{H(s^{\mu}\frac{\theta-N_{\partial\mu}^{\text{fix}}-a_{\mu\rightarrow i}}{\sigma_{\mu\rightarrow i}})},
\end{equation}
where $a_{\mu\rightarrow i}=\sum_{j\in\partial\mu\backslash i}\frac{1}{1+e^{-h_{j\rightarrow\mu}}}$,
$\sigma_{\mu\rightarrow i}^{2}=\sum_{j\in\partial\mu\backslash i}\frac{e^{-h_{j\rightarrow\mu}}}{(1+e^{-h_{j\rightarrow\mu}})^{2}}$.

In practice, for factor nodes with $|\partial\mu|>7$, we used the
Gaussian approximation above, while for factor nodes with $|\partial\mu|\le6$,
we used exact enumeration to calculate eq.\ref{eq:BP2}. We found
that this small-degree-enumeration strategy results in better convergence
of BP than pure Gaussian strategy. We also added damping to eq.\ref{eq:BP_h}:
\begin{equation}
h_{i\rightarrow\mu}^{t+1}=f\sum_{\nu\in\partial i\backslash\mu}\hat{h}_{\nu\rightarrow i}^{t}+(1-f)h_{i\rightarrow\mu}^{t}.
\end{equation}
We chose $f=0.02$ when $A=12$; $f=0.05$ when $A=40$ and $N=480$;
and $f=0.002$ when $A=40$ and $N=10000$, 5000 or 2500.

Numerically, in \textbf{Fig.\ref{fig:SBPI_rBP_BPD}} where $N=480$,
we initialized BP messages randomly at the beginning of each decimation
step. We defined convergence of BP as the case when the change of
BP entropy density $F_{BP}$ is smaller than $10^{-7}$ in adjacent
iteration step. In the case when BP failed to converge, we estimated
the iteration step that most close to the fixed point from the iteration
dynamics of $F_{BP}$ using a number of heuristics, and calculated
the marginal probabilities using the BP messages in that time step
using eq.\ref{eq:BP_marginal}. See Supplemental Material Section
4 for more details. In \textbf{Fig.\ref{fig:alphaApproachCapacityFig}}
and Supplementary Fig.9 where $N=10000$, 5000 or 2500, to speed
up computation, we initialized BP messages as the BP messages obtained
from the last decimation step if the last decimation step converged.
We defined convergence as the case when the change of $\frac{1}{N}\sum_{i=1}^{N}(h_{i}^{t+1}-h_{i}^{t})^{2}$
in adjacent iteration steps is smaller than $6.5\times10^{-5}$,
where $h_{i}^{t}$ is single-site BP field in the $t$th iteration
step. We found these two strategies do not give significantly different
results when $N=480$.

\section{Evaluating cross-correlation using belief propagation}

\label{Append:XCorrBP}

The cross-correlation between $w_{i}$ and $w_{j}=0,1$ can be calculated
using 
\begin{equation}
c_{ij}=\langle w_{i}w_{j}\rangle-\langle w_{i}\rangle\langle w_{j}\rangle=p(w_{i}=1)p(w_{j}=1|w_{i}=1)-p(w_{i}=1)p(w_{j}=1),\label{eq:BP_corr}
\end{equation}
where $p(w_{i}=1)$ and $p(w_{j}=1)$ can be evaluated using BP through
eq.\ref{eq:BP_marginal}. To evaluate $p(w_{j}=1|w_{i}=1)$, we fixed
$w_{i}=1$, run BP, got the fixed point, and calculated the marginal
probabilities of the other weights.

In \textbf{Fig.\ref{fig:XCorr}b-d} and Supplementary Fig.7, we evaluated
mean cross-correlation and mean square cross-correlation using 
\begin{equation}
\overline{\text{XCorr}}=[\frac{1}{|\mathcal{A}|(N_{\text{unfixed}}-1)}\sum_{i\in\mathcal{A}}\sum_{j\neq i}c_{ij}],\quad\overline{\text{XCorr}^{2}}=[\frac{1}{|\mathcal{A}|(N_{\text{unfixed}}-1)}\sum_{i\in\mathcal{A}}\sum_{j\neq i}c_{ij}^{2}],\label{eq:BP_meanAndMeanSquareCorr}
\end{equation}
where $[\cdot]$ represents quenched average (i.e., average over
IO pairs $\{\xi_{i}^{\mu},\sigma^{\mu}\}$), $\mathcal{A}$ is the
set of values of $i$, and $N_{\text{unfixed}}$ is the number of
unfixed weights during BPD. When $N_{\text{unfixed}}\le50$, $\mathcal{A}$
contains all the unfixed weights; when $N_{\text{unfixed}}>50$,
$\mathcal{A}$ contains 50 randomly chosen unfixed weights.

If BP for evaluating unconditioned marginal probabilities (i.e.,
$p(w_{i}=1),p(w_{j}=1)$ in eq.\ref{eq:BP_corr}) did not converge,
we did not continue to evaluate conditional probabilities, and excluded
this case from quenched average. Sometimes, BP for unconditioned
probabilities converged, but BP for conditional probabilities $p(w_{j}=1|w_{i}=1)$
did not converge. We found that this usually happened when $p(w_{i}=1)$
was small, so the ill-convergence of BP when we fixed $w_{i}=1$
may reflect the unlikelihood that $w_{i}$ takes 1, or in other words,
$w_{i}=0$ in most solutions. Therefore, we set $c_{ij}=0$ for all
$j\neq i$ in this case. We also tried the operation which excludes
from $\mathcal{A}$ the $i$s whose fixation to 1 lead to non-convergence
of BP, and found that the results were not significantly different.

\section{Evaluating local entropy using belief propagation}

\label{Append:LocalEntropyBP}

We followed Ref.\cite{Huang_2013} to evaluate the local entropy
around a given weight configuration $\tilde{\mathbf{w}}$ using belief
propagation. The partition function we calculated is 
\begin{equation}
Z=\sum_{\mathbf{w}}\prod_{\mu}\Theta(s^{\mu}(\sum_{j}w_{j}\xi_{j}^{\mu}-N/A))\prod_{i}\exp(x(w_{i}-\tilde{w}_{i})^{2}),
\end{equation}
with $x$ being a coupling factor controlling the distance of $\mathbf{w}$
from $\tilde{\mathbf{w}}$.

\section{Replica-method analysis}

\label{Append:ReplicaMethod}

The replica methods we used in \textbf{Fig.\ref{fig:performance_SBPI_rBP}a-c}
and \textbf{Fig.\ref{fig:BPD_DenseSolutionCluster}a,b} closely follows
the standard approaches presented in Ref. \cite{Brunel_2016,Huang_2014,Baldassi_2016}.
Here we only list out the free entropy density used to calculate
theoretical capacity, local entropy around typical solutions and
local entropy around configurations in the dense solution region.
Details how to introduce replicas to calculate the quenched average
of free entropy density are seen in Ref. \cite{Brunel_2016,Huang_2014,Baldassi_2016}.

When calculating theoretical capacity, we introduce partition function
\begin{equation}
Z(\xi,s)=\sum_{\mathbf{w}}\mathbb{X}_{\xi,s}(\mathbf{w}),
\end{equation}
with 
\begin{equation}
\mathbb{X}_{\xi,s}(\mathbf{w})=\prod_{\mu=1}^{\alpha N}\Theta[s^{\mu}(\sum_{i=1}^{N}w_{i}\xi_{i}^{\mu}-N/A)].
\end{equation}
$Z(\xi,s)$ counts the number of solutions given a set of IO pairs
$(\xi,s)$. Using replica method, we can calculate free entropy density
\begin{equation}
F=\frac{1}{N}\langle\log Z(\xi,s)\rangle_{\xi,s},
\end{equation}
with $\langle\cdot\rangle_{\xi,s}$ indicating quenched average of
sets of IO pairs. Theoretical capacity plotted in \textbf{Fig.\ref{fig:performance_SBPI_rBP}a-c}
is the $\alpha$ value at which $F=0$. During replica calculation,
an order parameter $M$ is defined as $M=\frac{1}{\sqrt{N}}\sum_{i}w_{i}-\frac{\sqrt{N}}{Af_{in}}$.
This means that in a well trained perceptron, the average total synaptic
current $\langle\mathbf{w}\cdot\xi^{\mu}\rangle_{\mu}=\sum_{i}w_{i}f_{in}$
deviates from the firing threshold $N/A$ up to $\mathcal{O}(\sqrt{N})$
order. In the classification task we considered, output $\sigma^{\mu}$
has equal probability to be 0 and 1. In this case, $M$ has saddle
point value 0, so $\sum_{i}w_{i}f_{in}=N/A$. The expression $\sqrt{f_{in}(1-f_{in})\sum_{i}(Aw_{i})^{2}}\approx\sqrt{AN(1-f_{in})}$
in Section \ref{sec:Model} can be derived using the facts that $w_{i}=\{0,1\}$
and $\sum_{i}Aw_{i}\approx N/f_{in}$.

Local entropy around typical solution plotted in \textbf{Fig.\ref{fig:BPD_DenseSolutionCluster}a}
has the name Franz-Parisi potential at zero temperature. It is defined
as 
\begin{equation}
F_{FP}(D)=\frac{1}{N}\langle\frac{\sum_{\tilde{\mathbf{w}}}\mathbb{X}_{\xi,s}(\tilde{\mathbf{w}})\ln(\mathcal{N}_{\xi,s}(\tilde{\mathbf{w}},D))}{\sum_{\tilde{\mathbf{w}}}\mathbb{X}_{\xi,s}(\tilde{\mathbf{w}})}\rangle_{\xi,s},
\end{equation}
where $\mathcal{N}_{\xi,s}(\tilde{\mathbf{w}},D)=\sum_{\mathbf{w}}\mathbb{X}_{\xi,s}(\mathbf{w})\delta[\frac{1}{N}\sum_{i}(w_{i}-\tilde{w}_{i})^{2}-D]$
means the number of solutions at distance $D=\frac{1}{N}\sum_{i}(w_{i}-\tilde{w}_{i})^{2}$
from reference configuration $\tilde{\mathbf{w}}$. The meaning of
$F_{FP}(D)$ is the mean free entropy density at distance $D$ from
a solution.

The local entropy around a weight configuration in the dense solution
region is calculated using large-deviation free entropy density 
\begin{equation}
F_{LD}(y,D)=\frac{1}{Ny}\langle\ln(\sum_{\tilde{\mathbf{w}}}\mathcal{N}(\tilde{\mathbf{w}},D)^{y})\rangle_{\xi,s}.
\end{equation}
In \textbf{Fig.\ref{fig:BPD_DenseSolutionCluster}b}, we calculated
the value of $F_{LD}(y,D)$ in the limit $y\rightarrow\infty$ under
1RSB ansatz, following the replica method introduced in Ref.\cite{Baldassi_2016}.

\section{The geometry of pure states in full solution space }

\label{Append:GeometryOfPureStatesFullSpace}

In this section, we discuss the geometry of pure states in the full
solution space before weight fixing, based on the replica method
introduced in Ref. \cite{Obuchi_2009}.

The structure of solution pure states is closely related to 1RSB
generalized free entropy density: 
\begin{equation}
g(x)=\frac{1}{N}[\log(\sum_{\gamma}Z_{\gamma}^{x})]\label{eq:1RSBFreeEntropyDensity}
\end{equation}
where $Z_{\gamma}$ is the solution number in the $\gamma$th pure
state, and $x$ is Parisi parameter. Following similar argument as
Ref. \cite{Obuchi_2009}, it can be shown that $g(x)$ is a piecewise
linear function: 
\begin{equation}
g(x)=\begin{cases}
x\frac{\partial\phi_{\text{1RSB1}}(n,x)}{\partial n}|_{n=0}=\frac{\mathrm{d}\phi_{\text{RS1}}(n)}{\mathrm{d}n}|_{n=0}, & x\le1\\
x\frac{\partial\phi_{\text{1RSB2}}(n,x)}{\partial n}|_{n=0}=x\frac{\mathrm{d}\phi_{\text{RS1}}(n)}{\mathrm{d}n}|_{n=0}, & x>1
\end{cases}\label{eq:g(x)Calculate}
\end{equation}
where $\phi_{\text{1RSB1}}(n,x)$ and $\phi_{\text{1RSB2}}(n,x)$
are two saddle-point solutions of the generating function $\phi(n)=\frac{1}{N}\log[Z^{n}]$
of partition function $Z$ under 1RSB ansatz, and $\phi_{\text{RS1}}(n)$
is a solution under RS ansatz, with the relation $\phi_{\text{1RSB1}}(n,x)=\phi_{\text{RS1}}(n/x)$
and $\phi_{\text{1RSB2}}(n,x)=\phi_{\text{RS1}}(n)$. The overlaps
$q_{\text{same}}$ and $q_{\text{diff}}$ in $\phi_{\text{1RSB1}}(n,x)$
and $\phi_{\text{1RSB2}}(n,x)$ respectively take the following values:
\begin{equation}
\text{1RSB1 }:\quad q_{\text{same}}=1/(f_{in}A),\quad q_{\text{diff}}=q\label{eq:1RSB1_overlap}
\end{equation}
\begin{equation}
\text{1RSB2 }:\quad q_{\text{same}}=q,\quad q_{\text{diff}}=q\label{eq:1RSB2_overlap}
\end{equation}
where $q$ is the saddle-point value of overlap in RS solution $\phi_{\text{RS1}}(n)$
when $n\rightarrow0$.

Now let us discuss the implication about solution-space structure
from these results. Firstly, from eq.\ref{eq:g(x)Calculate}, it
can be shown \cite{Obuchi_2009} that the complexity function $\Sigma(s)$
(which is the entropy density of pure states that contain $e^{Ns}$
number of solutions) is a line connecting $(0,\phi_{\text{RS1}}'(0))$
and $(\phi_{\text{RS1}}'(0),0)$ (where we have denoted $\phi_{\text{RS1}}'(0)=\frac{\mathrm{d}\phi_{\text{RS1}}(n)}{\mathrm{d}n}|_{n=0}$).
As discussed in Ref. \cite{Obuchi_2009}, this linear function may
be a convex hull of a convex-downward function connecting these two
points, whose convex-downward part between these two points is undetectable
by mean-field theory. This suggests that the solution space is dominated
by $\mathcal{O}(\exp(N\phi_{\text{RS1}}'(0)))$ number of small pure
states with sub-exponential number of solutions in each, and by sub-exponential
number of large clusters with $\mathcal{O}(\exp(N\phi_{\text{RS1}}'(0)))$
number of solutions in each. Additionally, the slope of function
$\Sigma(s)$ at $\Sigma=0$ is $-1$, which means that Parisi parameter
in real systems $x_{*}=1$ (see eq.\ref{eq:CorrGeo}) when $N$ is
large. Note that according to the results above, the numbers of solutions
in large and small pure states are the same to the leading exponential
order (i.e., $\mathcal{O}(\exp(N\phi_{\text{RS1}}'(0))$). However,
the result $x_{*}=1$ means that the probability of two random chosen
solutions lie in the same pure state is 0, which implies that the
number of solutions in small pure states dominates over that in the
large ones in sub-exponential order. Using exact enumeration, we
show that $x_{*}$ tends to 1 when $N$ gets large (Fig.\ref{fig:geoLargeNFig}a),
in support of the result $x_{*}=1$ in the large $N$ limit.

\setcounter{figure}{0} \global\long\def\thefigure{\thesection\arabic{figure}}
 
\begin{figure*}
\includegraphics[scale=0.6]{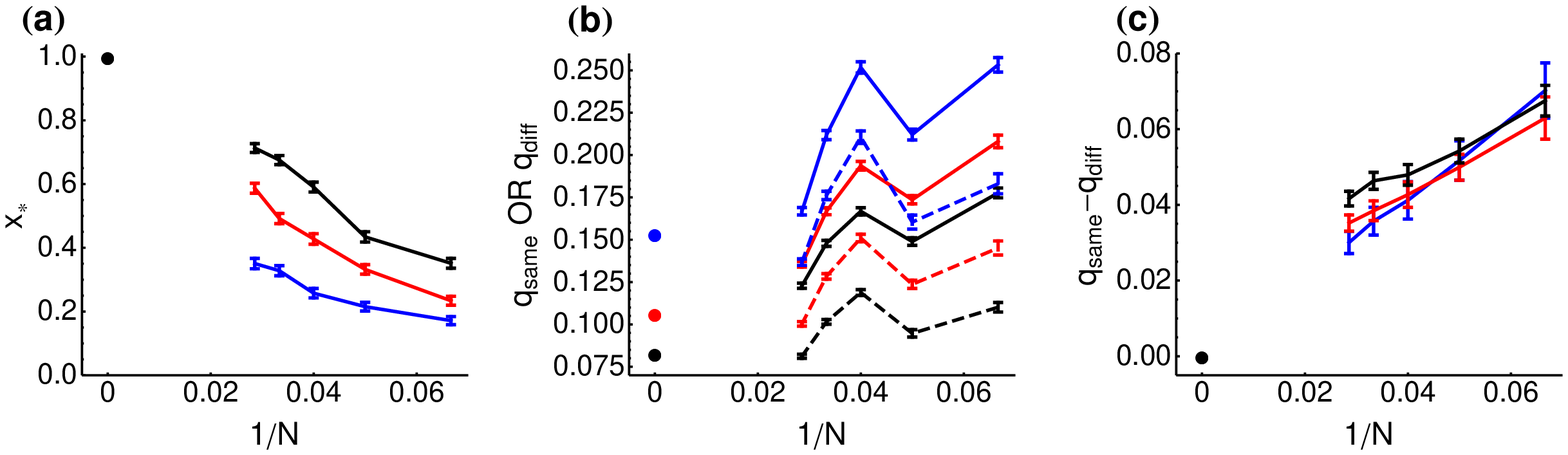} \protect\caption{Numeric support to the analytic results of $x_{*}$, $q_{\text{same}}$
and $q_{\text{diff}}$. (a) $x_{*}$ as a function of $1/N$, when
$f_{in}=0.3$ (blue), 0.4 (red) and 0.5 (black). The black dot is
the result predicted by replica method at $N\rightarrow\infty$.
(b) Solid lines: $q_{\text{same}}$; dashed lines: $q_{\text{diff}}$
. Dots of different colors indicate their analytic values at different
$f_{in}$s. (c) $q_{\text{same}}-q_{\text{diff}}$. The black dot
is the analytic result. $N=15$,20,25,30,35. $\alpha=0.4$, $A=12$.
Error bars indicate standard deviation of the mean. }
\label{fig:geoLargeNFig} 
\end{figure*}

Now let's discuss the overlaps $q_{\text{same}}$ and $q_{\text{diff}}$.
First, let's investigate the pure states that dominate in the summation
of eq.\ref{eq:1RSBFreeEntropyDensity}: 
\begin{equation}
\sum_{\gamma}Z_{\gamma}^{x}\approx N_{\text{small}}Z_{\text{small}}^{x}+N_{\text{large}}Z_{\text{large}}^{x}
\end{equation}
\begin{equation}
\sim\mathcal{O}(e^{N\phi_{\text{RS1}}'(0)})+\mathcal{O}(e^{xN\phi_{\text{RS1}}'(0)}),
\end{equation}
where $N_{\text{small}}$ and $N_{\text{large}}$ are the numbers
of small and large pure states respectively, and $Z_{\text{small}}$
and $Z_{\text{large}}$ are the numbers of solutions in a small and
large pure state respectively. From our previous discussion, $N_{\text{small}}\sim Z_{\text{large}}\sim\mathcal{O}(\exp(N\phi_{\text{RS1}}'(0))$,
while both $Z_{\text{small}}$ and $N_{\text{large}}$ are of sub-exponential
order. When $x<1$, small pure states (i.e., the term $N_{\text{small}}Z_{\text{small}}^{x}$)
dominate in $\sum_{\gamma}Z_{\gamma}^{x}$. Because $g(x)$ uses
$\phi_{\text{1RSB1}}$ to calculate when $x<1$ (eq.\ref{eq:g(x)Calculate}),
the overlaps $q_{\text{same}}$ and $q_{\text{diff}}$ of $\phi_{\text{1RSB1}}$
(eq.\ref{eq:1RSB1_overlap}) should correspond to the structure of
small clusters. In other words, two different solutions in the same
small pure state have overlap $1/(f_{in}A)$, and two solutions in
different small pure states have overlap $q$. Similarly, when $x>1$,
large pure states (i.e., the term $N_{\text{small}}Z_{\text{large}}^{x}$)
dominate in $\sum_{\gamma}Z_{\gamma}^{x}$, so $q_{\text{same}}$
and $q_{\text{diff}}$ of $\phi_{\text{1RSB2}}$ (eq.\ref{eq:1RSB2_overlap})
should correspond to the structure of large pure states. In other
words, overlaps in the same large pure state or in different large
pure states should both be $q$.

At $x=1$, which is the value $x_{*}$ that $x$ should take in ``real\textquotedbl{}
systems, both $\phi_{\text{1RSB1}}$ and $\phi_{\text{1RSB2}}$ predict
$q_{\text{diff}}=q$. There is discrepancy of $q_{\text{same}}$
predicted by $\phi_{\text{1RSB1}}$ and $\phi_{\text{1RSB2}}$ at
$x=1$. To get a reasonable $q_{\text{same}}$ value, note that the
number of solution pairs in the same small pure states is of $\mathcal{O}(e^{N\phi_{\text{RS1}}'(0)})$
order, while the number of solution pairs in the same large pure
states is of $\mathcal{O}(e^{2N\phi_{\text{RS1}}'(0)})$ order. Therefore,
$q_{\text{same}}$ at $x=1$ should be dominated by the solution
pairs in large pure states, which means $q_{\text{same}}=q$. Together,
$q_{\text{same}}=q_{\text{diff}}=q$. Using exact enumeration, we
show that both $q_{\text{same}}$ and $q_{\text{diff}}$ stay around
$q$, and $q_{\text{same}}-q_{\text{diff}}$ tends to 0 when $N$
gets large (Fig.\ref{fig:geoLargeNFig}b,c), in support of the result
$q_{\text{same}}=q_{\text{diff}}=q$ in the large $N$ limit.

\section{Miscellaneous}

\label{Append:Miscellaneous}

The method of enumeration was used both for listing out all the solutions
of small-sized systems and for calculating eq.\ref{eq:BP2} when
$|\partial\mu|\le6$. The basic idea to speed up enumeration is to
sweep all possible weight configurations in an order so that adjacent
configurations have minimal Hamming distance. In practice, configurations
were swept in the order of Gray code \cite{Krauth_1989b}.

The random walk in \textbf{Fig.\ref{fig:SBPI_rBP_BPD}} and \textbf{Fig.\ref{fig:BPD_DenseSolutionCluster}}
was done as follows. Starting from a solution, we tried to flip a
chosen weight, and accepted this flip if the configuration after
flipping was still a solution. Weights were chosen in random sequential
order. In \textbf{Fig.\ref{fig:SBPI_rBP_BPD}} and \textbf{Fig.\ref{fig:BPD_DenseSolutionCluster}},
we started random walk from a solution found by SBPI or rBP, and
swept all weights 10000 times.

Note the choice of parameter $A$ in numeric experiments. In our
model, $A$ does not scale when $N\rightarrow\infty$. Therefore,
if we set $A$ a large value, comparable to or larger than $N$,
how well the numeric results can represent the large $N$ case is
questionable. In all the figures, we chose $A=40$, or 12 when studying
systems of size $N\ge480$, or $N=25$, so that $A$ is well smaller
than $N$ in all cases.

Multi-dimensional scaling in \textbf{Fig.\ref{fig:Enum_MDS_Fig}e}
was performed using mdscale routine of MATLAB, with Sammon's mapping
criterion.


\bibliographystyle{ieeetr}\bibliographystyle{ieeetr}
\bibliography{reference}

\begin{thebibliography}{10}

\bibitem{OConnor_2005}
D.~H. O'Connor, G.~M. Wittenberg, and S.~S.-H. Wang, ``Graded bidirectional
  synaptic plasticity is composed of switch-like unitary events,'' {\em Proc.
  Natl. Acad. Sci. U. S. A.}, vol.~102, no.~27, pp.~9679--9684, 2005.

\bibitem{Montgomery_2004}
J.~M. Montgomery and D.~V. Madison, ``Discrete synaptic states define a major
  mechanism of synapse plasticity,'' {\em Trends Neurosci.}, vol.~27, no.~12,
  pp.~744--750, 2004.

\bibitem{Misra_2010}
J.~Misra and I.~Saha, ``Artificial neural networks in hardware: A survey of two
  decades of progress,'' {\em Neurocomputing}, vol.~74, pp.~239--255, 2010.

\bibitem{Rivest_1992}
R.~L. Rivest and A.~L. Blum, ``Training a 3-node neural network is
  {NP}-complete,'' {\em Neural Networks}, vol.~5, pp.~117--127, 1992.

\bibitem{Amaldi_1991}
E.~Amaldi, ``On the complexity of training perceptrons,'' in {\em Artificial
  Neural Networks} (O.~Simula, T.~Kohonen, K.~Makisara, and J.~Kangas, eds.),
  pp.~55--60, Elsevier, 1991.

\bibitem{Hubara_2018}
I.~Hubara, M.~Courbariaux, D.~Soudry, R.~El-Yaniv, and Y.~Bengio, ``Quantized
  neural networks: Training neural networks with low precision weights and
  activations,'' {\em J. Mach. Learn. Res.}, vol.~18, pp.~1--30, 2018.

\bibitem{Ott_2017}
J.~Ott, Z.~Lin, Y.~Zhang, S.-C. Liu, and Y.~Bengio, ``Recurrent neural networks
  with limited numerical precision,'' {\em arXiv:1608.06902}, 2017.

\bibitem{Engel_2001}
A.~Engel and C.~V. den Broeck, {\em Statistical Mechanics of Learning}.
\newblock Cambridge, England: Cambridge University Press, 2001.

\bibitem{Lage-Castellanos_2009}
A.~Lage-Castellanos, A.~Pagnani, and M.~Weigt, ``Statistical mechanics of
  sparse generalization and graphical model selection,'' {\em J. Stat. Mech.},
  vol.~2009, p.~P10009, 2009.

\bibitem{Hosaka_2002}
T.~Hosaka, Y.~Kabashima, and H.~Nishimori, ``Statistical mechanics of lossy
  data compression using a nonmonotonic perceptron,'' {\em Phys. Rev. E},
  vol.~66, p.~066126, 2002.

\bibitem{Huang_2014}
H.~Huang and Y.~Kabashima, ``Origin of the computational hardness for learning
  with binary synapses,'' {\em Phys. Rev. E}, vol.~90, p.~052813, 2014.

\bibitem{Zdeborova_2008}
L.~Zdeborov{\'a} and M.~M{\'e}zard, ``Locked constraint satisfaction
  problems,'' {\em Phys. Rev. Lett.}, vol.~101, p.~078702, 2008.

\bibitem{Baldassi_2015_b}
C.~Baldassi, A.~Ingrosso, C.~Lucibello, L.~Saglietti, and R.~Zecchina,
  ``Subdominant dense clusters allow for simple learning and high computational
  performance in neural networks with discrete synapses,'' {\em Phys. Rev.
  Lett.}, vol.~115, p.~128101, 2015.

\bibitem{Baldassi_2018}
C.~Baldassi, F.~Gerace, H.~J. Kappen, C.~Lucibello, L.~Saglietti,
  E.~Tartaglione, and R.~Zecchina, ``Role of synaptic stochasticity in training
  low-precision neural networks,'' {\em Phys. Rev. Lett.}, vol.~120, p.~268103,
  2018.

\bibitem{Baldassi_2016b}
C.~Baldassi, A.~Ingrosso, C.~Lucibello, L.~Saglietti, and R.~Zecchina, ``Local
  entropy as a measure for sampling solutions in constraint satisfaction
  problems,'' {\em J. Stat. Mech. Theory Exp.}, vol.~2016, p.~023301, 2016.

\bibitem{Baldassi_2016c}
C.~Baldassi, C.~Borgs, J.~T. Chayes, A.~Ingrosso, C.~Lucibello, L.~Saglietti,
  and R.~Zecchina, ``Unreasonable effectiveness of learning neural networks:
  From accessible states and robust ensembles to basic algorithmic schemes,''
  {\em Proc. Natl. Acad. Sci. U.S.A.}, vol.~113, pp.~E7655--E7662, 2016.

\bibitem{Chaudhari_2017}
P.~Chaudhari, A.~Choromanska, S.~Soatto, Y.~LeCun, C.~Baldassi, C.~Borgs,
  J.~Chayes, L.~Sagun, and R.~Zecchina, ``Entropy-{SGD}: Biasing gradient
  descent into wide valleys,'' in {\em International Conference on Learning
  Representations (ICLR)}, 2017.

\bibitem{Baldassi_2007}
C.~Baldassi, A.~Braunstein, N.~Brunel, and R.~Zecchina, ``Efficient supervised
  learning in networks with binary synapses,'' {\em Proc. Natl. Acad. Sci. U.
  S. A.}, vol.~104, no.~26, pp.~11079--11084, 2007.

\bibitem{Braunstein_2006}
A.~Braunstein and R.~Zecchina, ``Learning by message passing in networks of
  discrete synapses,'' {\em Phys. Rev. Lett.}, vol.~96, p.~030201, 2006.

\bibitem{Baldassi_2016}
C.~Baldassi, F.~Gerace, C.~Lucibello, L.~Saglietti, and R.~Zecchina, ``Learning
  may need only a few bits of synaptic precision,'' {\em Phys. Rev. E},
  vol.~93, p.~052313, 2016.

\bibitem{Amit_1994}
D.~J. Amit and S.~Fusi, ``Learning in neural networks with material synapses,''
  {\em Neural Comput.}, vol.~6, pp.~957--982, 1994.

\bibitem{Barrett_2008}
A.~B. Barrett and M.~C.~W. van Rossum, ``Optimal learning rules for discrete
  synapses,'' {\em PLoS Comput. Biol.}, vol.~4, no.~11, p.~e1000230, 2008.

\bibitem{Legenstein_2008}
R.~Legenstein and W.~Maass, ``On the classification capability of
  sign-constrained perceptrons,'' {\em Neural Comput.}, vol.~20, pp.~288--309,
  2008.

\bibitem{Lin_2014}
A.~C. Lin, A.~M. Bygrave, A.~de~Calignon, T.~Lee, and G.~Miesenb{\"o}ck,
  ``Sparse, decorrelated odor coding in the mushroom body enhances learned odor
  discrimination,'' {\em Nat. Neurosci.}, vol.~17, pp.~559--568, 2014.

\bibitem{Spanne_2015}
A.~Spanne and H.~J{\"o}rntell, ``Questioning the role of sparse coding in the
  brain,'' {\em Trends in Neurosci.}, vol.~38, no.~7, pp.~417--427, 2015.

\bibitem{Barak_2013}
O.~Barak, M.~Rigotti, and S.~Fusi, ``The sparseness of mixed selectivity
  neurons controls the generalization-discrimination trade-off,'' {\em J.
  Neurosci.}, vol.~33, no.~9, pp.~3844--3856, 2013.

\bibitem{Brunel_2004}
N.~Brunel, V.~Hakim, P.~Isope, J.~P. Nadal, and B.~Barbour, ``Optimal
  information storage and the distribution of synaptic weights: Perceptron
  versus {P}urkinje cell,'' {\em Neuron}, vol.~43, pp.~745--757, 2004.

\bibitem{Huerta_2004}
R.~Huerta, ``Learning classification in the olfactory system of insects,'' {\em
  Neural Comput.}, vol.~16, pp.~1601--1640, 2004.

\bibitem{Clopath_2013}
C.~Clopath and N.~Brunel, ``Optimal properties of analog perceptrons with
  excitatory weights,'' {\em PLoS Comput Biol}, vol.~9, no.~2, p.~e1002919,
  2013.

\bibitem{Mezard_2009}
M.~M{\'e}zard and A.~Montanari, {\em Information, Physics, and Computation}.
\newblock Oxford, England: Oxford University Press, 2009.

\bibitem{Baldassi_2015}
C.~Baldassi and A.~Braunstein, ``A max-sum algorithm for training discrete
  neural networks,'' {\em J. Stat. Mech.}, vol.~P08008, p.~052313, 2015.

\bibitem{Semerjian_2008}
G.~Semerjian, ``On the freezing of variables in random constraint satisfaction
  problems,'' {\em J. Stat. Phys.}, vol.~130, pp.~251--293, 2008.

\bibitem{Krzakala_2007}
F.~Krz{\c a}ka{\l}a, A.~Montanari, F.~Ricci-Tersenghie, G.~Semerjianc, and
  L.~Zdeborov{\'a}, ``Gibbs states and the set of solutions of random
  constraint satisfaction problems,'' {\em Proc. Natl. Acad. Sci. U. S. A.},
  vol.~104, pp.~10318--10323, 2007.

\bibitem{Faisal_2008}
A.~A. Faisal, L.~P.~J. Selen, and D.~M. Wolpert, ``Noise in the nervous
  system,'' {\em Nat. Rev. Neurosci.}, vol.~9, pp.~292--303, 2008.

\bibitem{Gardner_1988}
E.~Gardner, ``The space of interactions in neural network models,'' {\em J.
  Phys. A}, vol.~21, pp.~257--270, 1988.

\bibitem{Brunel_2016}
N.~Brunel, ``Is cortical connectivity optimized for storing information?,''
  {\em Nat. Neurosci.}, vol.~19, no.~5, pp.~749--755, 2016.

\bibitem{Mezard_2002}
M.~M{\'e}zard, G.~Parisi, and R.~Zecchina, ``Analytic and algorithmic solution
  of random satisfiability problems,'' {\em Science}, vol.~297, pp.~812--815,
  2002.

\bibitem{Ardelius_2008}
J.~Ardelius and L.~Zdeborov{\'a}, ``Exhaustive enumeration unveils clustering
  and freezing in the random 3-satisfiability problem,'' {\em Phys. Rev. E},
  vol.~78, p.~040101, 2008.

\bibitem{Obuchi_2009}
T.~Obuchi and Y.~Kabashima, ``Weight space structure and analysis using a
  finite replica number in the {I}sing perceptron,'' {\em J. Stat. Mech.},
  vol.~2009, p.~P12014, 2009.

\bibitem{Huang_2013}
H.~Huang, K.~Y.~M. Wong, and Y.~Kabashima, ``Entropy landscape of solutions in
  the binary perceptron problem,'' {\em J. Phys. A: Math. Theor.}, vol.~46,
  p.~375002, 2013.

\bibitem{Ricci-Tersenghi_2009}
F.~Ricci-Tersenghi and G.~Semerjian, ``On the cavity method for decimated
  random constraint satisfaction problems and the analysis of belief
  propagation guided decimation algorithms,'' {\em J. Stat. Mech.}, vol.~2009,
  p.~P09001, 2009.

\bibitem{Zdeborova_2007}
L.~Zdeborov{\'a} and F.~Krz{\c a}ka{\l}a, ``Phase transitions in the coloring
  of random graphs,'' {\em Phys. Rev. E}, vol.~76, p.~031131, 2007.

\bibitem{Zdeborova_2008b}
L.~Zdeborov{\'a} and M.~M{\'e}zard, ``Constraint satisfaction problems with
  isolated solutions are hard,'' {\em J. Stat. Mech.}, vol.~2008, p.~P12004,
  2008.

\bibitem{Piochon_2016}
C.~Piochon, M.~Kano, and C.~Hansel, ``{LTD}-like molecular pathways in
  developmental synaptic pruning,'' {\em Nat. Neurosci.}, vol.~19,
  pp.~1299--1310, 2016.

\bibitem{Krauth_1989b}
W.~Krauth and M.~Opper, ``Critical storage capacity of the ${J}=\pm 1$ neural
  network,'' {\em J. Phys. A: Math. Gen.}, vol.~22, pp.~L519--L523, 1989.

\end{thebibliography}


\begin{thebibliography}{1}

\bibitem{Brunel_2016}
N.~Brunel, ``Is cortical connectivity optimized for storing information?,''
  {\em Nat. Neurosci.}, vol.~19, no.~5, pp.~749--755, 2016.

\bibitem{Baldassi_2007}
C.~Baldassi, A.~Braunstein, N.~Brunel, and R.~Zecchina, ``Efficient supervised
  learning in networks with binary synapses,'' {\em Proc. Natl. Acad. Sci. U.
  S. A.}, vol.~104, no.~26, pp.~11079--11084, 2007.

\bibitem{Mezard_1987}
M.~M{\'e}zard, G.~Parisi, and M.~A. Virasoro, {\em Spin glass theory and
  beyond}.
\newblock Singapore: World Scientific, 1987.

\bibitem{Mezard_1984}
M.~M{\'e}zard, G.~Parisi, N.~Sourlas, G.~Toulouse, and M.~Virasoro, ``Replica
  symmetry breaking and the nature of the spin glass phase,'' {\em J. Phys.},
  vol.~45, pp.~843--854, 1984.

\end{thebibliography}

\end{document}

\title{Understanding the computational difficulty of a binary-weight perceptron
and the advantage of input sparseness: Supplemental Material}

\author{Zedong Bi, Changsong Zhou}

\maketitle
\renewcommand{\figurename}{Supplementary Figure }
\renewcommand{\theequation}{S\arabic{equation}}

\section{Large $A$ improves robustness of memory retrieval under noise}

\begin{figure}[tbph]
\includegraphics[scale=0.8]{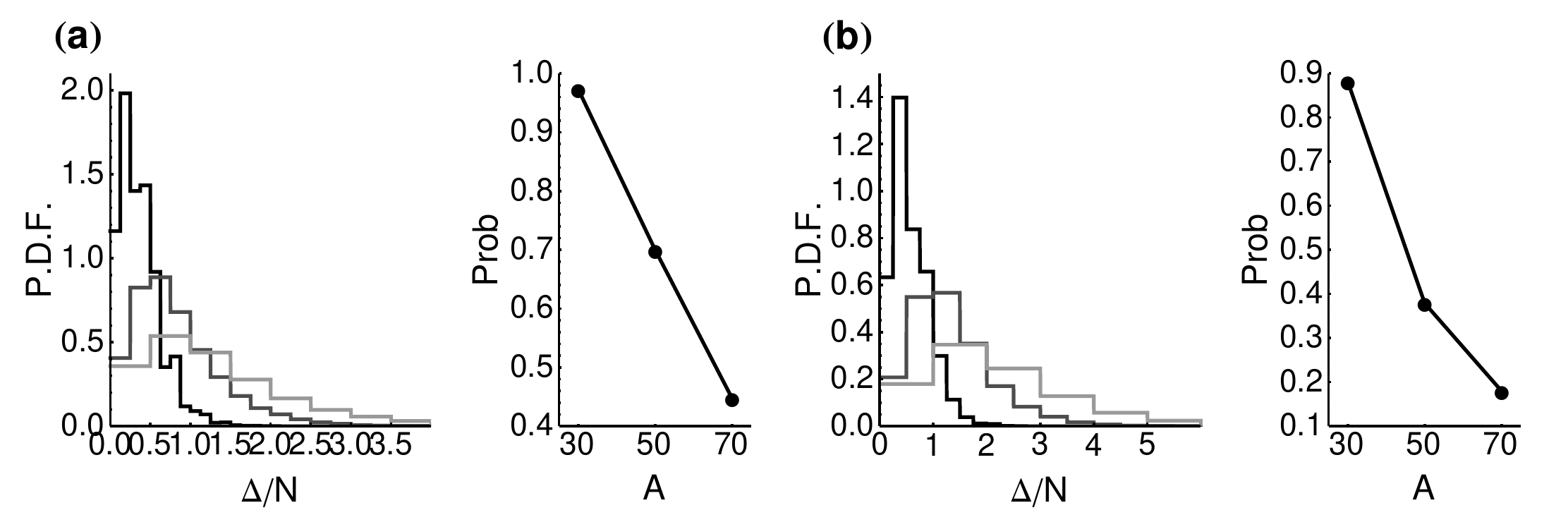}

\protect\caption{\textbf{\label{fig:LargeAImprovesRobustness}Large $A$ improves
robustness of memory retrieval under noise. }(a) Left panel: The
probability distribution function (P.D.F.) of $\Delta^{\mu}/N$ over
$\mu$ for solutions found by SBPI. $A=30$, 50 and 70, indicated
by curves with decreasing blackness. Note that the distribution becomes
broader with the increase of $A$. Right panel: The probability of
$\Delta^{\mu}/N<1$ when $A=30$, 50 and 70. Note that this probability
decreases with $A$. $\alpha=0.4$, $f_{in}=0.1$, $N=1000$. (b)
The same as (a), except for rBP. $\alpha=0.35$, $f_{in}=0.1$, $N=500$.
In the left panels of (a,b), the probability distributions were plotted
after solving 200 perceptron problems (a problem means to find a
solution of $\alpha N$ input-output pairs). In the right panels
of (a,b), error bars that represent s.e.m. over these 200 problems
are smaller than the plot markers. }

\end{figure}

In the main text, we have intuitively explained why large $A$ facilitates
robustness of memory retrieval under noise when interpreting $A$
as weight rescaler (\textbf{Fig.1b} in the main text), here we provide
numeric evidence to support this intuitive explanation. We solved
perceptron problems at different $A$ values using SBPI and rBP (see
Appendix A for their implementations) and then calculated the stability
parameter $\Delta^{\mu}=(2\sigma^{\mu}-1)(A\mathbf{w}_{0}\cdot\xi^{\mu}-N)$
for each input-output pair $\mu$ at the found solution $\mathbf{w}_{0}$.
This stability parameter means that the perceptron is able to output
the desired value $\sigma^{\mu}$ in response to the input pattern
$\xi^{\mu}$ when the strength of the noise of total synaptic current
is smaller than $\Delta^{\mu}$. We found that with the increase
of $A$, the distribution of $\Delta^{\mu}$ becomes broader, so
that the probability that $\Delta^{\mu}$ is smaller than a given
value also becomes smaller (\textbf{Supplementary Fig.\ref{fig:LargeAImprovesRobustness}}).
This suggests that perceptron is more likely to give correct output
in response to a stored input pattern with larger $A$ value. 

\section{Understanding the low-$A$ branch of capacity}

From \textbf{Fig.2a-c} in the main text, the dependence of algorithmic
and theoretical capacity with $A$ and $f_{in}$ is not monotonic.
At high-$A$ branch, capacity decreases with $A$ and $f_{in}$;
at low-$A$ branch, capacity increases with $A$ and $f_{in}$. While
our study mostly focuses on high-$A$ branch, here we discuss why
low-$A$ branch exists in our model. 

We propose that the existence of low-$A$ branch is because of the
upper boundedness of weights (i.e., $w_{i}\le1$) in our model. To
understand this, let us interpret $A$ as the threshold controller
so that the neuronal output is $\Theta(\mathbf{w}\cdot\xi-N/A)$.
When $A$ is too small (smaller than $1/f_{in}$), the threshold
$N/A$ becomes so large that the synaptic current $\mathbf{w}\cdot\xi<N/A$
even when $w_{i}=1$ for all $i$s. In this case, a single desired
output $\sigma^{\mu}=1$ makes the perceptron problem have no solution.
Therefore, capacity is zero when $A\le1/f_{in}$, and gets increased
when $A$ increases from $1/f_{in}$. Similar discussion applies
for the case when input is too sparse ($f_{in}<1/A$): in this case
the synaptic current $\mathbf{w}\cdot\xi<N/A$ even when $w_{i}=1$
for all $i$s. Therefore, capacity gets increased when $f_{in}$
increases from $1/A$. This argument explains why in low-$A$ branch
capacity increases with $A$ and $f_{in}$. 

\begin{figure}[tbph]
\includegraphics[scale=0.8]{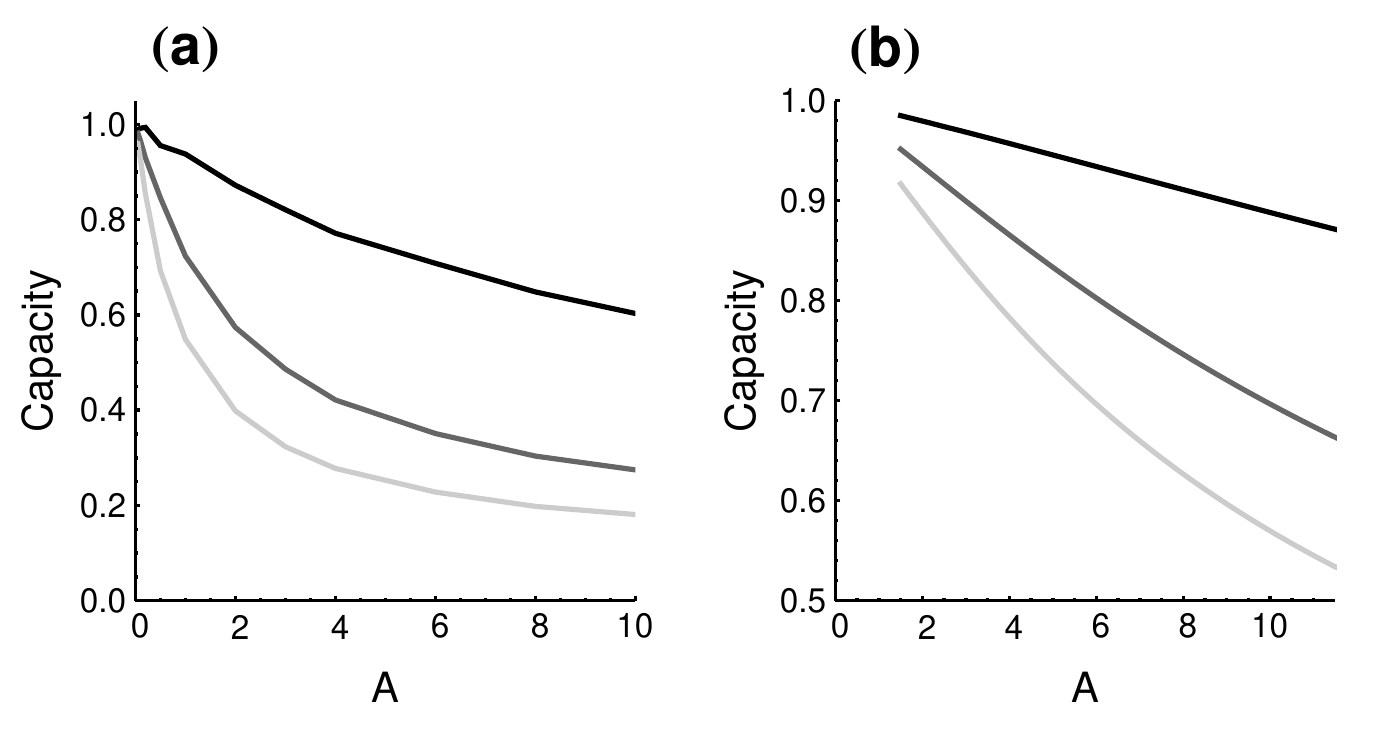}

\protect\caption{\label{fig:UnboundedWeight}\textbf{Capacity decreases with $A$
after removing the upper bound of $w_{i}$.} (a) Numeric capacity
of the extended-CP algorithm, when $f_{in}=0.1$, 0.3 and 0.5 represented
by curves with decreasing blackness. $T_{max}=8000$, $K=2$, $N=1000$.
(b) Theoretical capacity with $A$. In the numeric and analytic calculations
in panels (a,b), we chose $w_{i}=0,1,2,\cdots,W$, with $W$ being
a large enough value so that the probability that $w_{i}=W$ is too
low and the change of $W$ does not significantly influence the results. }

\end{figure}

In support of the discussion above, we considered a perceptron model
where $w_{i}=0,1,2,\cdots$, removing the upper bound of weights.
We calculated the theoretical capacity of this model using replica
method, and also investigated its algorithmic capacity using an extended-CP
algorithm (introduced at the end of this section). We found that
both theoretical and algorithmic capacity monotonically decrease
with $A$ and $f_{in}$ (\textbf{Supplementary Fig.\ref{fig:UnboundedWeight}});
and when $A\rightarrow0$, both capacities monotonically increase
to 1 (\textbf{Supplementary Fig.\ref{fig:UnboundedWeight}}), which
is the capacity when weights take continuous values \cite{Brunel_2016}.
This result suggests that the existence of low-$A$ branch in the
model used in the main text, which causes the non-monotonicity of
capacity with $A$ and $f_{in}$, is due to the upper boundedness
of weights. 

The extended-CP algorithm we used above is extended from Clip Perceptron
algorithm \cite{Baldassi_2007}. In extended-CP, each weight $w_{i}$
has a hidden variable $h_{i}$ which takes intergers between $-K$
and $K$. When an unassociated input-output (IO) pair $\mu$ is presented,
$h_{i}$ is updated as $\ensuremath{h_{i}\leftarrow h_{i}+\xi_{i}^{\mu}(2\sigma^{\mu}-1)}$.
When $h_{i}<-K$, $w_{i}$ is reduced by 1 until it reach the lower
bound 0, and $h_{i}$ is reset to $K$. When $h_{i}>K$, $w_{i}$
is increased by 1 and $h_{i}$ is reset to $-K$. Solving was stopped
when a solution was found or after going sweep of all IO pairs $T_{max}$
times. 

\section{The performance of rBP at different $\gamma$s}

The performance of rBP at different $\gamma$s is shown in \textbf{Supplementary
Fig.\ref{fig:rBPPerformance}}. With the increase of $\gamma$, the
probability of successful solving increases (\textbf{Supplementary
Fig.\ref{fig:rBPPerformance}a}), while the time steps $T_{solve}$
needed to solve in the case of successful solving also increases
(\textbf{Supplementary Fig.\ref{fig:rBPPerformance}b}). Additionally,
at a given $\gamma$ value, $T_{solve}$ increases with $f_{in}$
when $\gamma$ is small, but decreases with $f_{in}$ when $\gamma$
is large (\textbf{Supplementary Fig.\ref{fig:rBPPerformance}b}). 

\begin{figure}[tbph]
\includegraphics{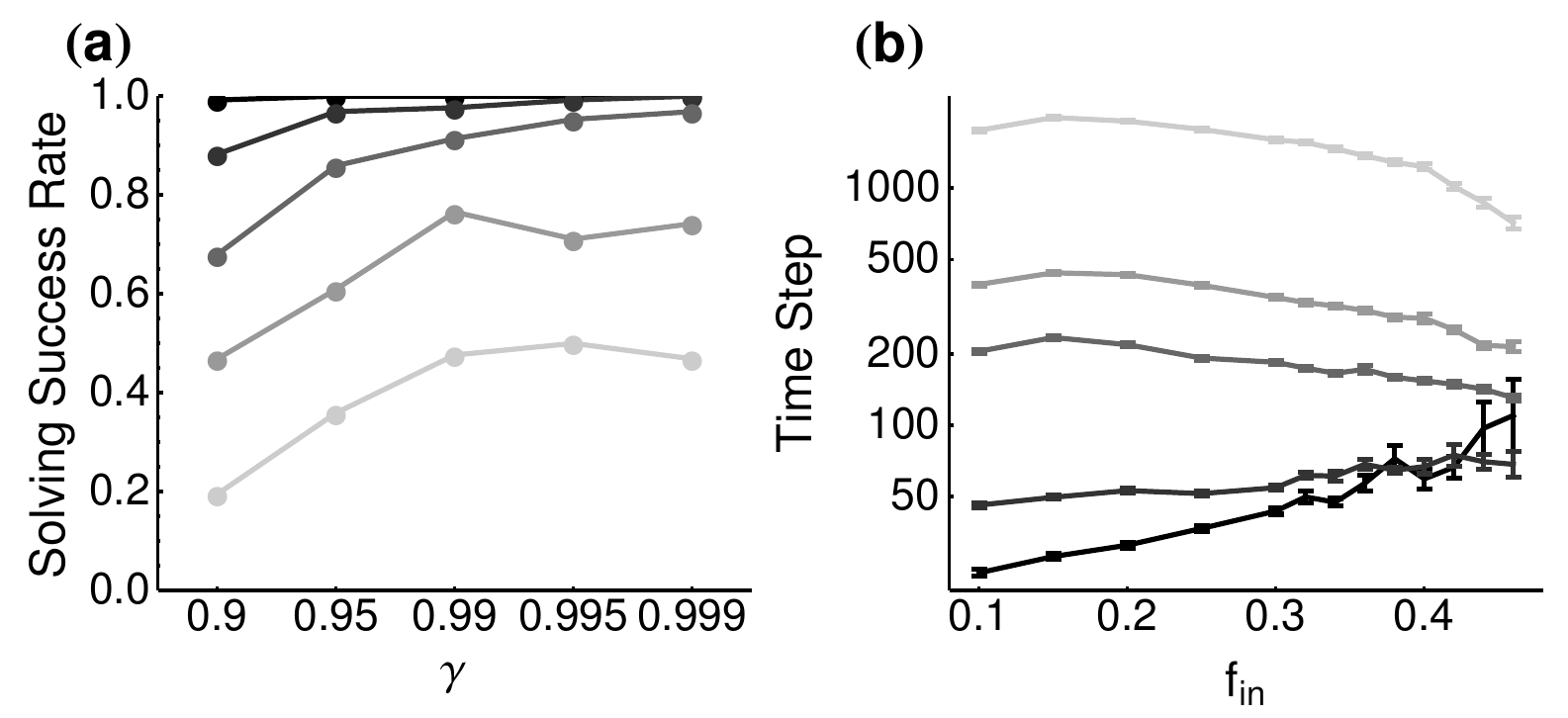}\protect\caption{\textbf{The performance of rBP at different $\gamma$s.\label{fig:rBPPerformance}
}(a) The rate of successful solving at different $\gamma$s, when
$f_{in}=0.25$, 0.32, 0.36, 0.4, 0.44, indicated by lines with decreasing
blackness. (b) Time steps needed in the case of successful solving
as a function of $f_{in}$, when $\gamma=0.9$, 0.95, 0.99, 0.995
and 0.999, indicated by lines with decreasing blackness. $N=480$,
$A=40$, $\alpha=0.2$. Error bars indicate s.e.m. }

\end{figure}

\section{The dynamics of belief propagation during BPD}

Here we explain how we deal with the non-convergence of belief propagation
(BP) during BPD. BP equations are presented in Appendix B in the
main text. At each iteration time step of BP, we calculated Bethe
entropy density 
\begin{equation}
F_{BP}=\frac{1}{N_{\text{unfix}}}\sum_{\mu}\log Z_{\mu}+\frac{1}{N_{\text{unfix}}}\sum_{i}\log Z_{i}-\frac{1}{N_{\text{unfix}}}\sum_{(i,\mu)}\log Z_{(i\mu)},\label{eq:BetheEntropyDensity}
\end{equation}
where 
\begin{equation}
Z_{\mu}=\sum_{\{W_{j}\}_{j\in\partial\mu}}\Theta(s^{\mu}(\sum_{j\in\partial\mu}w_{j}+N_{\partial\mu}^{\text{fix}}-\theta))\prod_{j\in\partial\mu}p_{j\rightarrow\mu}(w_{j})\label{eq:BP_Zmu}
\end{equation}
\begin{equation}
Z_{i}=\sum_{w_{i}}\prod_{\nu\in\partial i}\hat{p}_{\nu\rightarrow i}(w_{i}),\quad Z_{(i\mu)}=\sum_{w_{i}}p_{i\rightarrow\mu}(w_{i})\hat{p}_{\mu\rightarrow i}(w_{i})\label{eq:BP_Zi_Zmui}
\end{equation}
When $|\partial\mu|>7$, we used Gaussian approximation to calculate
eq.\ref{eq:BP_Zmu}, writing $Z_{\mu}$ as $Z_{\mu}=H(s^{\mu}(\frac{\theta-N_{\partial\mu}^{\text{fix}}-a_{\mu}}{\sigma_{\mu}})),$
where $a_{\mu}=\sum_{j\in\partial\mu}\frac{1}{1+e^{-h_{j\rightarrow\mu}}}$,
$\sigma_{\mu}^{2}=\sum_{j\in\partial\mu}\frac{e^{-h_{j\rightarrow\mu}}}{(1+e^{-h_{j\rightarrow\mu}})^{2}}$.
When $|\partial\mu|\le6$, we used exact enumeration to calculate
eq.\ref{eq:BP_Zmu}. 

Convergence of BP was defined as the case when the change of $F_{BP}$
is smaller than $10^{-7}$ in adjacent iteration step. After convergence,
we calculated the marginal probabilities using eq.B3 in the main
text, and fixed the most polarized weight. In the case when BP did
not converge, we used the following heuristic method to evaluate
the marginal probabilities by observing the dynamics of $F_{BP}$. 

The typical shape of $F_{BP}(t)$ (with $t$ indicating iteration
step) in the non-convergence case is shown in \textbf{Supplementary
Fig.\ref{fig:BPDynamics}}: $F_{BP}(t)$ has a inflection point at
$t_{0}$ during decreasing, which is convex downward before $t_{0}$,
and convex upward after $t_{0}$. We presume that the slow dynamics
around $t_{0}$ manifests the adjacency of BP messages around its
fixed point. Therefore, the basic idea of our heuristic method is
to pick $t_{0}$ by observing $F_{BP}(t)$, and then calculate the
marginal probabilities using the BP messages at $t_{0}$. Specifically,
we defined a time step $t$ to be \textit{decreasing-inflection}
if 
\begin{equation}
F_{BP}(t-2)-F_{BP}(t-4)<0,\quad F_{BP}(t)-F_{BP}(t-2)<0\quad F_{BP}(t+2)-F_{BP}(t)<0\label{eq:BP_InflectionDecrease-1}
\end{equation}
and 
\begin{equation}
|F_{BP}(t)-F_{BP}(t-2)|<|F_{BP}(t-2)-F_{BP}(t-4)|,\quad|F_{BP}(t)-F_{BP}(t-2)|<|F_{BP}(t+2)-F_{BP}(t)|,\label{eq:BP_InflectionDecrease-2}
\end{equation}
where eq.\ref{eq:BP_InflectionDecrease-1} is the condition for decrease
and eq.\ref{eq:BP_InflectionDecrease-2} is the condition for inflection.
Note that we used $F_{BP}(t)-F_{BP}(t-2)$ instead of $F_{BP}(t)-F_{BP}(t-1)$
to represent the change of $F_{BP}$ around iteration step $t$,
because $F_{BP}(t)$ sometimes trembled in alternative time steps,
resulting in a zigzag shape. We collected all the decreasing-inflection
points within 4000 iteration steps, then deleted the inflection-decreasing
points at which $F_{BP}<-3$ and $F_{BP}>0.75$, leaving only the
\textit{reasonable }ones: with the motivation that at a realistic
stationary point of BP equations, $F_{BP}$ should not be too negative,
and is unlikely to be larger than its theoretical upper bound $\log2=0.693$.
If there were many reasonable decreasing-inflection points, we chose
the one with minimal $|F_{BP}(t-2)-F_{BP}(t-4)|+|F_{BP}(t)-F_{BP}(t-2)|+|F_{BP}(t+2)-F_{BP}(t)|$
value, which means minimal change in adjacent iteration steps. In
rare cases, there were no reasonable decreasing-inflection points,
then we chose the time points with minimal $|F''(t)|$ (with the
motivation that $F''(t)=0$ is a necessary condition of inflection
point), where $F''(t)$ was numerically estimated as $\frac{1}{2}[(\frac{F_{BP}(t+2)-F_{BP}(t)}{2})-(\frac{F_{BP}(t)-F_{BP}(t-2)}{2})]$. 

\begin{figure}[tbph]
\includegraphics[scale=0.8]{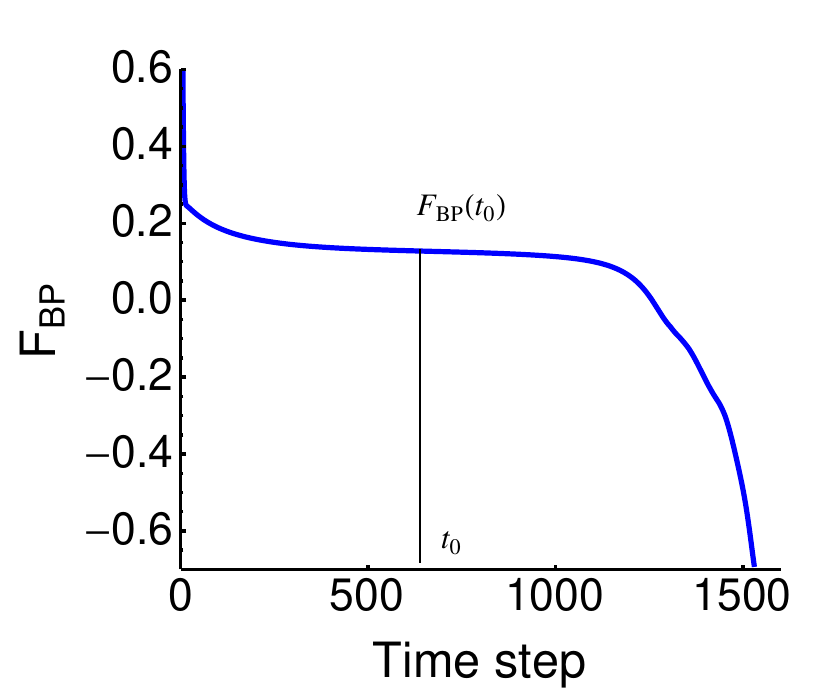}\protect\caption{\label{fig:BPDynamics} An example of $F_{BP}(t)$ when BP does not
converge. $t_{0}$ is an inflection point, which is the point we
look for using heuristic methods. }

\end{figure}

\section{Derivation of eq.9 in the main text}

Eq.9 in the main text relates mean square cross-correlation with
the geometry of pure states. Here we present the key steps when deriving
this equation. 

In the large $N$ limit, mean square cross-correlation can be related
with the variance of overlap distribution as 

\begin{equation}
\frac{1}{N(N-1)}\sum_{i\neq j}[(\langle w_{i}w_{j}\rangle-\langle w_{i}\rangle\langle w_{j}\rangle)^{2}]=\frac{1}{N^{2}}\sum_{i,j}[(\langle w_{i}w_{j}\rangle-\langle w_{i}\rangle\langle w_{j}\rangle)^{2}]=\frac{1}{3}\text{Var}([P_{\xi,\sigma}(q)])\label{eq:3-1}
\end{equation}
where $P_{\xi,\sigma}(q)$ is the distribution of solution overlap
$q$ given a set of IO pairs $\{\xi,\sigma\}$, $[\cdot]$ means
average over the set of of IO pairs. Under the assumption of 1RSB,
\begin{equation}
[P_{\xi,\sigma}(q)]=x_{*}\delta(q-q_{\text{diff}})+(1-x_{*})\delta(q-q_{\text{same}}).\label{eq:3-2}
\end{equation}
After substituting eq.\ref{eq:3-2} into eq.\ref{eq:3-1}, we can
get eq.9 in the main text. So all we need is to prove eq.\ref{eq:3-1}.
\[
\frac{1}{N^{2}}\sum_{i,j}[(\langle w_{i}w_{j}\rangle-\langle w_{i}\rangle\langle w_{j}\rangle)^{2}]
\]
\begin{equation}
=\frac{1}{N^{2}}\sum_{i,j}[\langle w_{i}w_{j}\rangle^{2}]-2\frac{1}{N^{2}}\sum_{i,j}[\langle w_{i}w_{j}\rangle\langle w_{i}\rangle\langle w_{j}\rangle]+\frac{1}{N^{2}}\sum_{i,j}[\langle w_{i}\rangle^{2}\langle w_{j}\rangle^{2}]\label{eq:3-3}
\end{equation}
The first term of eq.\ref{eq:3-3} is 
\[
\frac{1}{N^{2}}\sum_{i,j}[\langle w_{i}w_{j}\rangle^{2}]=\frac{1}{N^{2}}\sum_{i,j}[(\sum_{\alpha}v_{\alpha}\langle w_{i}w_{j}\rangle_{\alpha})^{2}]
\]
\[
=\frac{1}{N^{2}}\sum_{i,j}[\sum_{\alpha\beta}v_{\alpha}v_{\beta}\langle w_{i}\rangle_{\alpha}\langle w_{j}\rangle_{\alpha}\langle w_{i}\rangle_{\beta}\langle w_{j}\rangle_{\beta}]
\]
\begin{equation}
=[\int\mathrm{d}q_{\alpha\beta}q_{\alpha\beta}^{2}P_{\xi,\sigma}(q_{\alpha\beta})]=[\int\mathrm{d}qq^{2}P_{\xi,\sigma}(q)]\label{eq:3-4}
\end{equation}
where $v_{\alpha}$ means the probability that a solution $\mathbf{w}$
lies in the $\alpha$th pure state. In the third step, we used a
property of pure state \cite{Mezard_1987} $\langle w_{i}w_{j}\rangle_{\alpha}=\langle w_{i}\rangle_{\alpha}\langle w_{j}\rangle_{\alpha}$;
in the fourth step, we defined $q_{\alpha\beta}=\frac{1}{N}\sum_{i}\langle w_{i}\rangle_{\alpha}\langle w_{i}\rangle_{\beta}$.
Similarly, the second and third terms of eq.\ref{eq:3-3} are respectively
\[
\frac{1}{N^{2}}\sum_{i,j}[\langle w_{i}w_{j}\rangle\langle w_{i}\rangle\langle w_{j}\rangle]=\frac{1}{N^{2}}\sum_{i,j}[\sum_{\alpha}v_{\alpha}\langle w_{i}\rangle_{\alpha}\langle w_{j}\rangle_{\alpha}\sum_{\beta}\langle w_{i}\rangle_{\beta}\sum_{\gamma}\langle w_{j}\rangle_{\gamma}]
\]
\begin{equation}
=[\int\mathrm{d}q_{\alpha\beta}\mathrm{d}q_{\alpha\gamma}q_{\alpha\beta}q_{\alpha\gamma}P_{\xi,\sigma}(q_{\alpha\beta},q_{\alpha\gamma})]\label{eq:3-5}
\end{equation}
and
\begin{equation}
\frac{1}{N^{2}}\sum_{i,j}[\langle w_{i}\rangle^{2}\langle w_{j}\rangle^{2}]=[\int\mathrm{d}q_{\alpha\beta}\mathrm{d}q_{\gamma\delta}q_{\alpha\beta}q_{\gamma\delta}P_{\xi,\sigma}(q_{\alpha\beta},q_{\gamma\delta})]\label{eq:3-6}
\end{equation}
Under the assumption of ultrametricity of solution overlap, it can
be shown that \cite{Mezard_1984}
\begin{equation}
[P_{\xi,\sigma}(q_{\alpha\beta},q_{\alpha\gamma})]=\frac{1}{2}[P_{\xi,\sigma}(q_{\alpha\beta})]\delta(q_{\alpha\beta}-q_{\alpha\gamma})+\frac{1}{2}[P_{\xi,\sigma}(q_{\alpha\beta})][P_{\xi,\sigma}(q_{\alpha\gamma})]\label{eq:3-7}
\end{equation}
\begin{equation}
[P_{\xi,\sigma}(q_{\alpha\beta},q_{\gamma\delta})]=\frac{1}{3}[P_{\xi,\sigma}(q_{\alpha\beta})]\delta(q_{\alpha\beta}-q_{\gamma\delta})+\frac{2}{3}[P_{\xi,\sigma}(q_{\alpha\beta})][P_{\xi,\sigma}(q_{\gamma\delta})]\label{eq:3-8}
\end{equation}
Substituting eq.\ref{eq:3-7} into eq.\ref{eq:3-5}, and eq.\ref{eq:3-8}
into eq.\ref{eq:3-6}, we have 
\begin{equation}
[\langle w_{i}w_{j}\rangle\langle w_{i}\rangle\langle w_{j}\rangle]=\frac{1}{2}[\int\mathrm{d}qq^{2}P_{\xi,\sigma}(q)]+\frac{1}{2}[\int\mathrm{d}qqP_{\xi,\sigma}(q)]^{2}\label{eq:3-9}
\end{equation}
\begin{equation}
[\langle w_{i}\rangle^{2}\langle w_{j}\rangle^{2}]=\frac{1}{3}[\int\mathrm{d}qq^{2}P_{\xi,\sigma}(q)]+\frac{2}{3}[\int\mathrm{d}qqP_{\xi,\sigma}(q)]^{2}\label{eq:3-10}
\end{equation}
Substituting eq.\ref{eq:3-4} and the two equations above into eq.\ref{eq:3-3},
we have 
\[
\frac{1}{N^{2}}\sum_{i,j}[(\langle w_{i}w_{j}\rangle-\langle w_{i}\rangle\langle w_{j}\rangle)^{2}]=\frac{1}{3}[\int\mathrm{d}qq^{2}P_{\xi,\sigma}(q)]-\frac{1}{3}[\int\mathrm{d}qqP_{\xi,\sigma}(q)]^{2}=\frac{1}{3}\text{Var}([P_{\xi,\sigma}(q)])
\]
which is eq.\ref{eq:3-1}.

\section{Other Supplementary Figures}

See \textbf{Supplementary Figs. \ref{fig:Comparing-the-dynamics}-\ref{fig:The-dynamics-of}}.

\begin{figure}[tbph]
\includegraphics[scale=0.6]{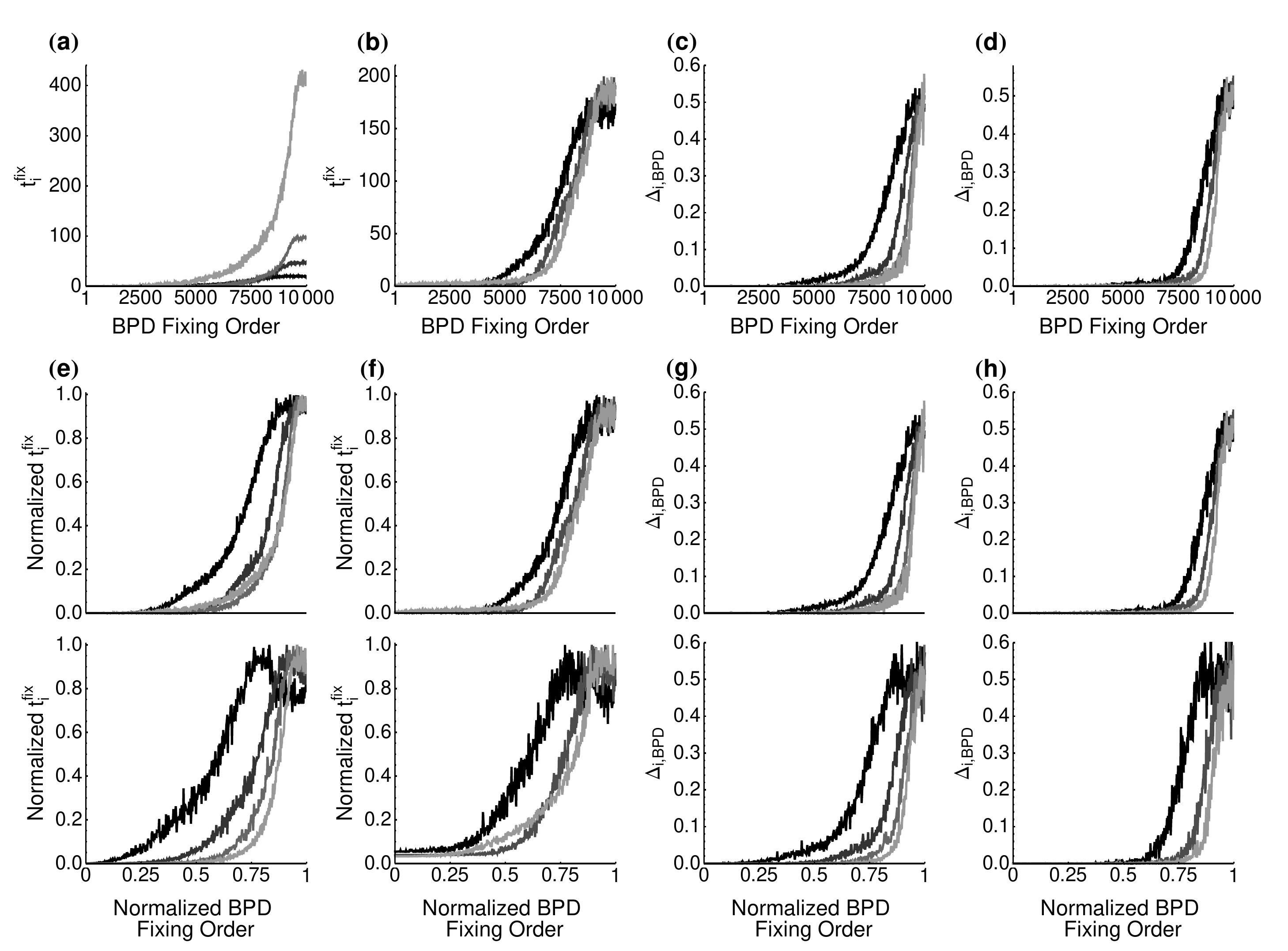}

\protect\caption{\textbf{\label{fig:Comparing-the-dynamics}Comparing the dynamics
of SBPI and rBP with weight-fixing order in BPD.} (a) Weight-fixing
time step $t_{i}^{\text{fix}}$ in SBPI as a function of weight-fixing
order in BPD. $f_{in}=0.1$, 0.2, 0.3, 0.4, represented by curves
with decreasing blackness. (b) The same as (a), but for rBP. $f_{in}=0.1,$
0.2, 0.3. (c) The difference $\Delta_{i,\text{BPD}}$ between the
value of weight $w_{i}$ in a solution found by SBPI from the value
of $w_{i}$ in the solution found by BPD, as a function of BPD-fixing
order of $w_{i}$. (d) The same as (c), but for rBP. In (a-d), $N=10000$.
(e-h) Comparing the cases when $N=10000$ (upper panels) with the
cases when $N=480$ (lower panels). The lower panels for $N=480$
were plotted using the same data as \textbf{Fig.3} of the main text.
(e) Normalized $t_{i}^{\text{fix}}$ as a function of normalized
BPD fixing order, for SBPI. Here, both $t_{i}^{\text{fix}}$ and
BPD fixing order are normalized so that their maximal values are
equal to 1. (f) The same as (e), but for rBP. (g) $\Delta_{i,\text{BPD}}$
as a function of normalized BPD fixing order, for SBPI. Here, we
did not normalize $\Delta_{i,\text{BPD}}$ because its maximal value
is close to 0.5 in all cases. (h) The same as (g), but for rBP. All
panels average over only trials in which SBPI or rBP succeeded to
find solutions. Note that, here, for SBPI, we only plotted the curves
when $f_{in}=0.1$, 0.2, 0.3, 0.4, but did not show the case when
$f_{in}=0.5$ as we did in \textbf{Fig.3} of the main text; for rBP,
we only plotted the curves when $f_{in}=0.1$, 0.2, 0.3, but did
not show the case when $f_{in}=0.4$ as we did in \textbf{Fig.3 }of
the main text. The reason for these choices of $f_{in}$ is that
algorithms may usually fail to find solutions when $f_{in}$ takes
large values. }

\end{figure}

\begin{figure}[tbph]
\includegraphics[scale=0.8]{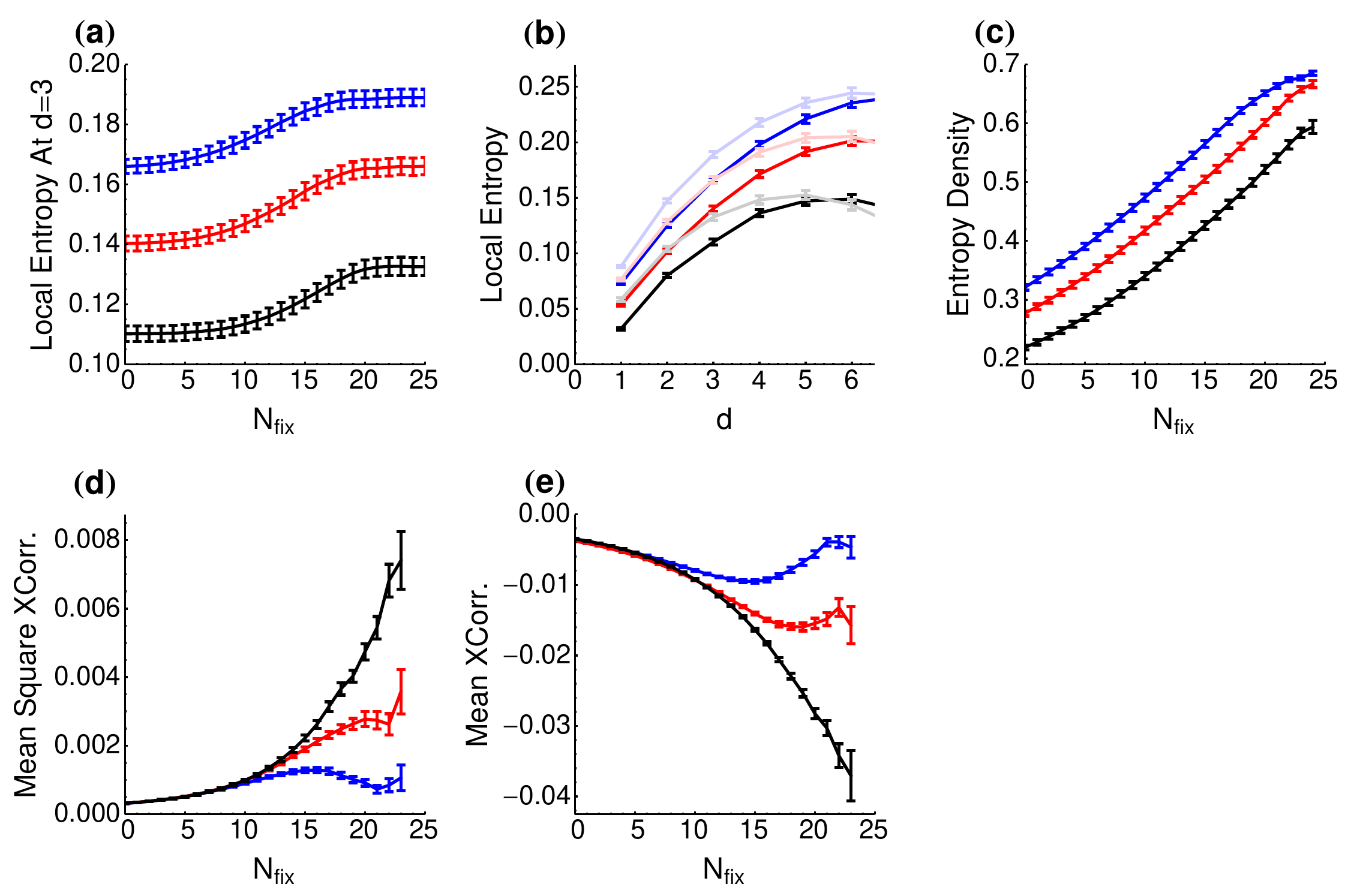}

\protect\caption{\textbf{Supplementary data for exact decimation. }\label{fig:ExactDecimation}(a)
Mean local entropy at Hamming distance $d=3$ from solutions in $\mathcal{S}(N_{\text{fix}})$
during exact decimation, when $f_{in}=0.3$ (blue), 0.4 (red) and
0.5 (black). (b) Local entropy at small Hamming distances average
over from all solutions (untransparent curves), and from the solution
found by exact decimation after fixing all weights (transparent curves).
(c) Solution entropy density in $\mathcal{S}(N_{\text{fix}})$ calculated
by exact enumeration. (d) Mean square cross-correlation $\overline{\text{XCorr}^{2}}$
during exact decimation, when $f_{in}=0.3$ (blue), 0.4 (red), and
0.5 (black). (d) Mean cross-correlation $\overline{\text{XCorr}}$
during exact decimation. Panels (a-c) correspond to \textbf{Fig.4g-i
}in the main text. Panels (d,e) correspond to \textbf{Fig.5b,c }in
the main text. $N=25$, $A=12$, $\alpha=0.36$. }
\end{figure}

\begin{figure}[tbph]
\includegraphics[scale=0.7]{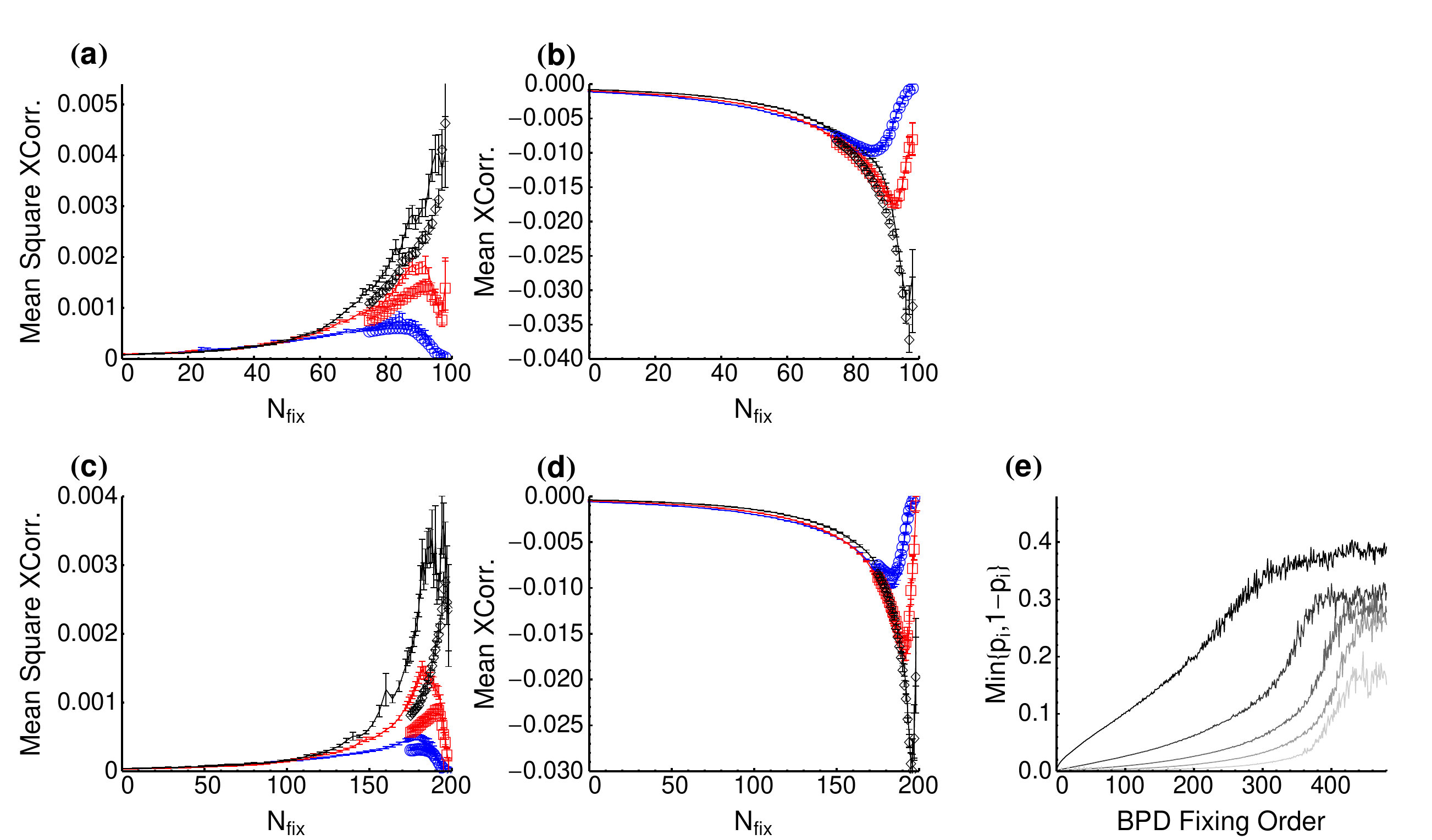}

\protect\caption{\textbf{Supplementary data for BPD. }\label{fig:BPDLargeN}(a) Mean
square cross-correlation $\overline{\text{XCorr}^{2}}$ during BPD,
when $f_{in}=0.3$ (blue), 0.4 (red), and 0.5 (black) in systems
of size $N=100$. Error bars connected by lines represent results
computed by belief propagation; plot markers (circles, squares, diamonds)
represent results computed by exactly enumerated solutions in $\mathcal{S}(N_{\text{fix}})$.
(b) Mean cross-correlation $\overline{\text{XCorr}}$ during BPD,
\textbf{$N=100$}. (c,d) The same as (a,b), except for $N=200$.
(e) $\text{min}\{p_{i},1-p_{i}\}$ (with $p_{i}$ being the probability
that $w_{i}=1$ in full solution space) as a function of the fixing
order of $w_{i}$ in BPD. $f_{in}=0.1$, 0.2, 0.3, 0.4, 0.5, indicated
by lines with decreasing blackness. This panel shows that weights
fixed early in BPD have stronger polarization. Panels (a-d) correspond
to \textbf{Fig.5b-e }in the main text, $A=12$, $\alpha=0.36$. In
(e), $N=480$, $A=40$, $\alpha=0.2$.}
\end{figure}

\begin{figure}[tbph]
\includegraphics[scale=0.6]{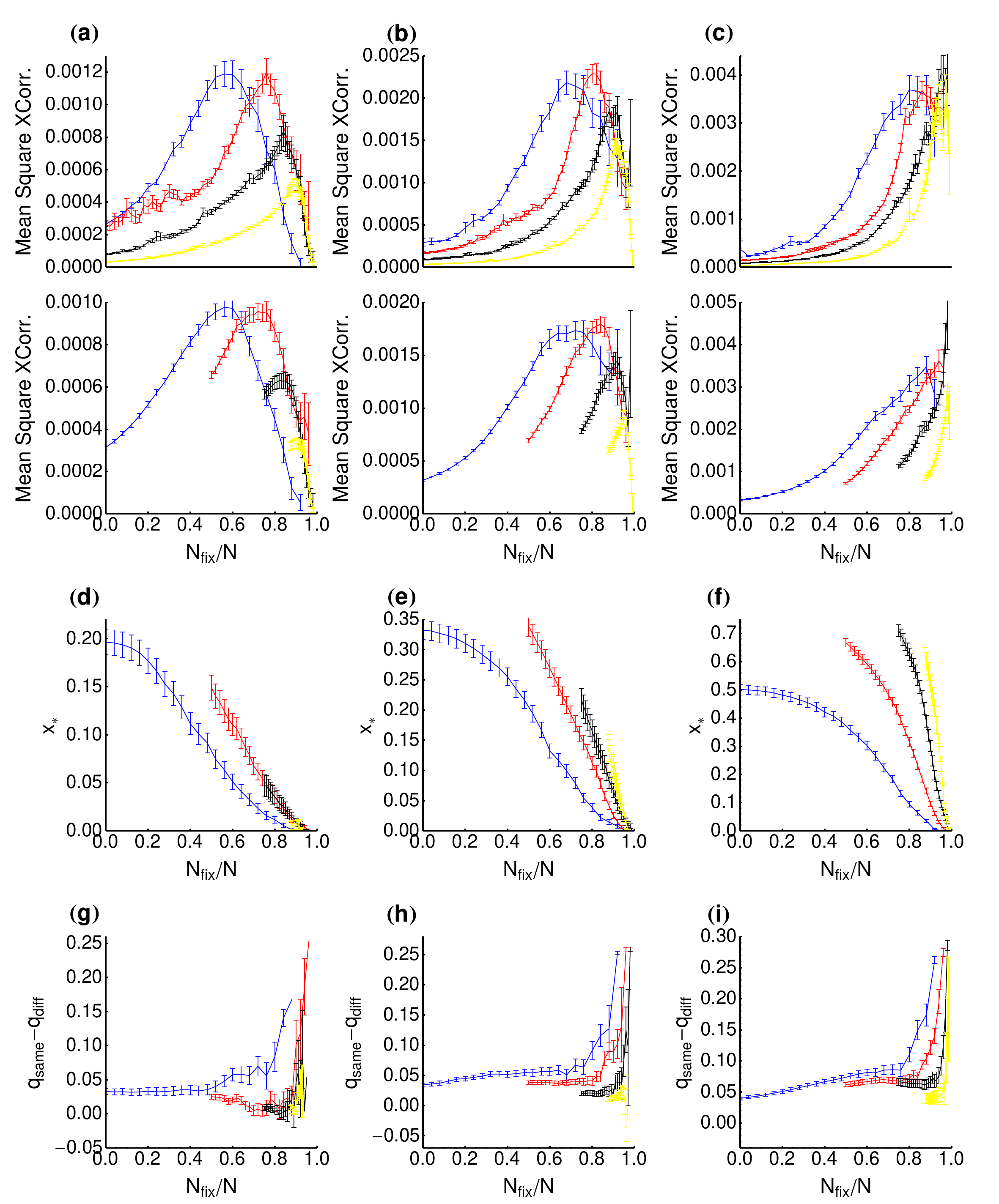}

\protect\caption{\textbf{The scaling of mean square cross-correlation $\overline{\text{XCorr}^{2}}$during
BPD with $N$.} (a) Upper panel: $\overline{\text{XCorr}^{2}}$ during
BPD calculated using belief propagation, when $N=25$ (blue), 50
(red), 100 (black) and 200 (yellow). Lower panel: $\overline{\text{XCorr}^{2}}$
at the late stage of BPD calculated using exact enumeration. $f_{in}=0.3$.
(b) The same as (a), but for $f_{in}=0.4$. (c) The same as (a),
but for $f_{in}=0.5$. Panels (a-c) were plotted using the same data
as those in \textbf{Fig.5b,d} of the main text and in \textbf{Supplementary
Fig.\ref{fig:BPDLargeN}a,c}. (d-f) Parisi parameter $x_{*}$ at
the late stage of BPD calculated using exact enumeration. $f_{in}=0.3$
(panel d), 0.4 (panel e) and 0.5 (panel f). (g-i) $q_{\text{same}}-q_{\text{diff}}$
at the late stage of BPD, at $f_{in}=0.3$ (panel g), 0.4 (panel
h) and 0.5 (panel i). $\alpha=0.36$, $A=12$.\label{fig:The-scaling-of}}

\end{figure}

\begin{figure}[tbph]
\includegraphics[scale=0.6]{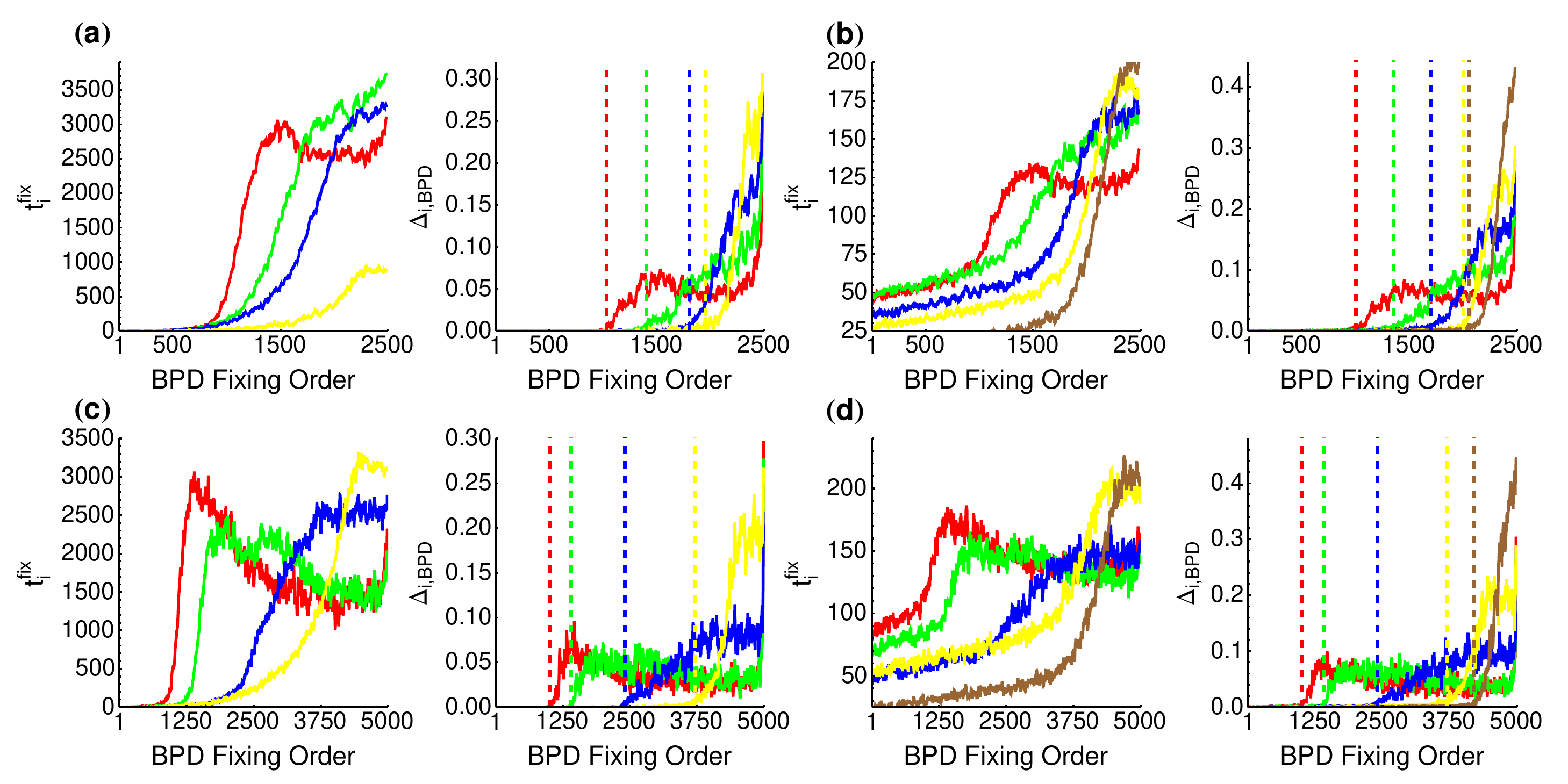}

\protect\caption{\textbf{The dynamics of SBPI and rBP when they fail to find solutions.}
(a) Left: Weight-fixing time step $t_{i}^{\text{fix}}$ in SBPI as
a function of weight-fixing order in BPD, when SBPI fails to find
solutions. Right: The difference $\Delta_{i,\text{BPD}}$ of the
value of weight $w_{i}$ that SBPI ended up from the value of $w_{i}$
that BPD ended up. Colors represent $\alpha$ values indicated in
\textbf{Fig.8b} in the main text. Dashed lines represent the positions
where $\Delta_{i,\text{BPD}}$ start to ramp up. $N=2500$. (b) The
same as (a), but for rBP. (c,d) The same as (a,b), but for $N=5000$.
All panels average over only trials in which SBPI or rBP failed to
find solutions.\label{fig:The-dynamics-of}}

\end{figure}

\bibliographystyle{ieeetr}
\bibliography{referenceSM}